\documentclass[letterpaper,titlepage,11pt]{article}

\usepackage[T1]{fontenc}
\usepackage[utf8]{inputenc}
\usepackage{lmodern}
\usepackage{enumerate}
\usepackage[english]{babel}
\usepackage{comment}
\usepackage[numbers,sort&compress]{natbib}
\usepackage{bm}
\usepackage[usenames,dvipsnames]{xcolor}
\usepackage[a4paper]{geometry}

\usepackage[colorlinks=true,urlcolor=blue,citecolor=magenta]{hyperref}
\usepackage{amsmath}
\usepackage{amsfonts}
\usepackage{amssymb}
\usepackage{graphicx}
\usepackage[dvipsnames]{xcolor}
\usepackage{mathtools}
\usepackage{simplewick}
\usepackage{hyperref}
\usepackage{multirow}
\usepackage{enumitem}
\usepackage{physics}
\usepackage[compat=1.1.0]{tikz-feynman}
\usepackage{subcaption}
\usepackage{booktabs}
\usepackage{xcolor}

\allowdisplaybreaks[1]

\newcommand{\CJ}{\mathcal{J}}
\newcommand{\BCJ}{\bar{\mathcal{J}}}
\newcommand{\CF}{\mathcal{F}}
\newcommand{\CG}{\mathcal{G}}
\newcommand{\CN}{\mathcal{N}}
\newcommand{\CQ}{\mathcal{Q}}

\newcommand{\PSU}{\text{PSU}}
\newcommand{\SU}{\text{SU}}
\newcommand{\U}{\text{U}}
\newcommand{\ds}{\displaystyle}

\newcommand{\spa}{\ , \ \ }
\newcommand{\nn}{\nonumber}
\newcommand{\psd}{\psi^\dagger}
\newcommand{\ra}[1]{\renewcommand{\arraystretch}{#1}}

\newcommand{\beq}{\begin{equation}}
\newcommand{\eeq}{\end{equation}}
\newcommand{\bea}{\begin{eqnarray}}
\newcommand{\eea}{\end{eqnarray}}
\newcommand{\mbf}{\mathbf}

\def\le{\left(}
\def\ri{\right)}

\hypersetup{   colorlinks=true,  citecolor=blue, linkcolor=blue, urlcolor=red }

\begin{document}

\numberwithin{equation}{section}

\begin{titlepage}
\rightline{\vbox{   \phantom{ghost} }}

 \vskip 1.8 cm
\begin{center}
{\LARGE \bf
Nonrelativistic near-BPS corners of 
\\[2mm]
${\cal N} = 4$ super-Yang-Mills with $\SU(1,1)$ symmetry }
\end{center}
\vskip 1 cm

\title{}
\date{\today}
\author{Stefano Baiguera, Troels Harmark, Nico Wintergerst}

\centerline{\large {{\bf Stefano Baiguera, Troels Harmark, Nico Wintergerst}}}

\vskip 1.0cm

\begin{center}
\sl The Niels Bohr Institute, University of Copenhagen\\
Blegdamsvej 17, DK-2100 Copenhagen \O, Denmark
\end{center}

\vskip 1.3cm \centerline{\bf Abstract} \vskip 0.2cm \noindent
We consider limits of $\CN=4$ super Yang-Mills (SYM) theory that approach BPS bounds and for which an $\SU(1,1)$ structure is preserved. 
The resulting near-BPS theories become non-relativistic, with a $\U(1)$ symmetry emerging in the limit that implies the conservation of particle number.
They are obtained by reducing $\CN=4$ SYM on a three-sphere and subsequently integrating out fields that become non-dynamical as the bounds are approached. Upon quantization, and taking into account normal-ordering, they are consistent with taking the appropriate limits of the dilatation operator directly, thereby corresponding to Spin Matrix theories, found previously in the literature. In the particular case of the $\SU(1,1|1)$ near-BPS/Spin Matrix theory, we find a superfield formulation that applies to the full interacting theory. Moreover, for all the theories we find tantalizingly simple semi-local formulations as theories living on a circle. 
Finally, we find positive-definite expressions for the interactions in the classical limit for all the theories, which can be used to explore their strong coupling limits.
This paper will have a companion paper in which we explore BPS bounds for which a $\SU(2,1)$ structure is preserved.

\end{titlepage}
\newpage
\tableofcontents

\section{Introduction}

$\CN=4$ super-Yang-Mills (SYM) theory is conjectured to describe strings and gravity on $\mbox{AdS}_5 \times S^5$ in its strongly coupled limit. Accessing this regime is a challenging task and necessitates looking for limits of $\CN=4$ SYM in which its dynamics simplify. For instance, in its planar limit one achieves a powerful integrability symmetry that enables to solve for the spectrum in the strong coupling limit \cite{Beisert:2010jr}. This can for instance be used to obtain the Hagedorn temperature at any 't Hooft coupling \cite{Harmark:2017yrv,Harmark:2018red}. However, the planar limit corresponds to non-interacting strings and gravitons. Thus, even by including corrections to the planar limit, one would not be able to study phenomena like black holes that involve strong gravity.

Recently a different approach was advocated \cite{Harmark:2019zkn}. The proposal is to consider certain non-relativistic corners that arise as near-BPS limits of $\CN=4$ SYM \cite{Harmark:2014mpa}. In such limits, one can maintain a finite number of colors of $\CN=4$ SYM, and hence strong gravity, but instead the stringy and gravitational dynamics become non-relativistic \cite{Harmark:2008gm,Harmark:2017rpg,Harmark:2018cdl,Harmark:2019upf}. 

One can motivate the interest in the resulting non-relativistic theories from two points of view. One is that they reveal new insights into the dynamics of $\CN=4$ SYM and hence into the AdS/CFT correspondence. Another is that these new theories might provide new non-relativistic realizations of the holographic principle that are important to study on their own. 

In this paper, we continue the investigations of the non-relativistic corners of $\CN=4$ SYM set out in \cite{Harmark:2019zkn}.
Starting with $\CN=4$ SYM on a three-sphere, we consider limits that zoom in close to BPS bounds of the type
\begin{equation}
\label{BPSbound}
E \geq S_1 + \sum_{i=1}^3 \omega_i Q_i \,,
\end{equation}
where $E$ is the energy, $S_1$ one of the angular momenta and $Q_i$, $i=1,2,3$, are the three R-charges of $\CN=4$ SYM on a three-sphere. Moreover, $\omega_i$, $i=1,2,3$, are three constants that characterize the BPS bounds. One can equally well translate these inequalities to bounds on the scaling dimensions for $\CN=4$ SYM on flat space via the state-operator correspondence.

The near-BPS limits we consider send the 't Hooft coupling $\lambda$  to zero while keeping \cite{Harmark:2014mpa}
\begin{equation}
\label{BPSlimit}
\frac{1}{\lambda} (E - S_1 - \sum_{i=1}^3 \omega_i Q_i) \ \ \mbox{fixed} \,.
\end{equation}
Starting with the classical action for $\CN=4$ SYM on a three-sphere, we show using sphere reduction that most of the massive modes on the three-sphere decouple, leaving a subset of dynamical modes that survive the limit. However, some of the non-dynamical modes, which we show includes spherical modes of the gauge field of $\CN=4$ SYM, can still contribute to the effective interaction of the surviving dynamical modes. Using this procedure, with reduction on $S^3$ and integrating out non-dynamical modes, one obtains a classical description of the surviving modes, which is the classical description of the near-BPS theory corresponding to the given BPS bound \eqref{BPSbound}.

These classical near-BPS theories provide the effective description of $\CN=4$ SYM near the BPS bounds \eqref{BPSbound}. We find that all such theories are non-relativistic, in that antiparticles decouple in the limit. Accordingly, one observes the emergence of a $\U(1)$ symmetry corresponding to a conserved number operator.

Upon quantization of the near-BPS theories, they result in  quantum mechanical theories. As part of the quantization one finds self-energy corrections that are easily computable from a normal-ordering prescription.
We show that the quantized near-BPS theories correspond to the Spin Matrix theories \cite{Harmark:2014mpa} that were found previously by considering the same near-BPS limits \eqref{BPSlimit} taken on the quantized theory of  $\CN=4$ SYM on a three-sphere, as described by the dilatation operator \cite{Beisert:2003tq, Beisert:2003jj, Beisert:2004ry}. Indeed, one obtains in  this way  theories with only a subset of the states of $\CN=4$ SYM on a three-sphere, as the rest have decoupled. Moreover, the interaction is directly related to the one-loop dilatation operator of $\CN=4$ SYM. This shows that one 
can consistently quantize the near-BPS theories that we obtain in this paper. 

Due to the particular form of the BPS bound \eqref{BPSbound}, the near-BPS/Spin Matrix theories that we consider here have $\SU(1,1)$ symmetry, possibly as subgroup of a larger global symmetry. 
In the free limit the spectrum 
gives a free energy that goes like temperature squared, indicating that the theories are effectively $(1+1)$-dimensional. Therefore, one would expect to find formulations as non-relativistic $(1+1)$-dimensional quantum field theories. Indeed, such formulations exist, albeit not as fully local quantum field theories and with non-standard features similar to positive energy ghost fields.

In detail we consider four different BPS bounds \eqref{BPSbound} depending on the choice of $(\omega_1,\omega_2,\omega_3)$. In the case $(\omega_1,\omega_2,\omega_3)=(1,0,0)$ one obtains a scalar theory with $\SU(1,1) \times \U(1)$ global symmetry that resemble the positive momentum modes of a scalar field on a circle. Interestingly, the interactions in this case can be viewed as arising from the coupling to a non-dynamical scalar field, resembling a gauge field. 
With $(\omega_1,\omega_2,\omega_3)=(\frac{2}{3},\frac{2}{3},\frac{2}{3})$ one finds instead a theory with fermionic modes with the same global symmetry that can be formulated in terms of the positive momentum modes of a chiral fermion on a circle.

For $(\omega_1,\omega_2,\omega_3)=(1,\frac{1}{2},\frac{1}{2})$ one obtains a non-relativistic theory with $\SU(1,1|1) \times \U(1)$ symmetry that can be regarded as a combination of the two latter theories, with a bosonic and a fermionic field on a circle. This theory is supersymmetric and one can find a superfield formulation in which the interactions arise from integrating out the super-multiplet of a non-dynamical gauge field.  This is the case that we are considering in most detail in this paper, since it is simple to describe but at the same time it contains the bosonic and fermionic cases with $\SU(1,1) \times \U(1)$ global symmetry as subsectors.
Finally, we also consider the maximal case with $(\omega_1,\omega_2,\omega_3)=(1,1,1)$ in which one has a theory with two scalars and two chiral fermions on a circle with $\PSU(1,1|2) \times \U(1)$ global symmetry .

These four near-BPS/Spin Matrix theories are interesting in their own right since they are consistent  limits of $\CN=4$ SYM on a three-sphere that describes the behavior of $\CN=4$ SYM near a BPS bound, or, equivalently, near a zero-temperature critical point if one takes the planar limit \cite{Harmark:2014mpa}. Indeed, it is intriguing that one obtains non-relativistic behavior in such limits.  

Another important reason to study them is that they have holographic duals. One sees this as consequence of the AdS/CFT correspondence, since one can take the same near-BPS limit on the string theory side of the correspondence \cite{Harmark:2006ta,Harmark:2008gm,Harmark:2014mpa,Harmark:2016cjq,Harmark:2017rpg,Harmark:2018cdl,Harmark:2019upf}. The philosophy here is that one can hope to solve this corner of the AdS/CFT correspondence, and then exploit this to learn about the full correspondence. This goal was realized in case of the Hagedorn temperature \cite{Harmark:2006ta,Harmark:2017yrv,Harmark:2018red} and it is also the spirit of the papers \cite{Berkooz:2008gc,Berkooz:2014uwa}. 
 
Alternatively, and even more interestingly, one can view the near-BPS/Spin Matrix theories as fully consistent and self-contained theories that realize the holographic principle. Indeed, this is supported by the fact that Spin Matrix theories in the planar limit reduces to nearest-neighbor spin chains that in a continuum limit are described by sigma-models. Recently, such sigma-models where interpreted as part of a class of non-relativistic sigma-model with a structure that resembles ordinary relativistic string theory, and with a new type of non-relativistic target space geometry called $\U(1)$-Galilean geometry \cite{Harmark:2017rpg,Harmark:2018cdl,Harmark:2019upf}. In this sense one can claim to have shown the emergence of geometry from the Spin Matrix theories.

The missing piece for having a full-fledged realization of the holographic principle is to see the emergence of gravity. In this regard, interesting progress has been made on beta-function calculations \cite{Gomis:2019zyu,Gallegos:2019icg,Bergshoeff:2019pij,Yan:2019xsf} in the related non-relativistic SNC \cite{Andringa:2012uz,Bergshoeff:2018yvt} and TNC \cite{Harmark:2017rpg,Harmark:2018cdl,Harmark:2019upf} string theories, providing the hope that a similar calculation is possible for the string-dual of Spin Matrix theory that indeed possess a Galilean Conformal Algebra as local symmetry.

This paper is organized as follows. In Section \ref{sect-sphere_reduction} we consider the four near-BPS limits of classical $\CN=4$ SYM on a three-sphere. This uses the sphere reduction of  $\CN=4$ SYM of \cite{Ishiki:2006rt} explained in Appendix \ref{app-spher_harmonics_S3} and performed in detail in Appendix \ref{app-quadratic_Hamiltonian_conserved_charges}. In Appendix \ref{app-properties_Clebsch_Gordan_coeff} we exhibit relevant Clebsch-Gordan coefficients and further properties of spherical harmonics. 

In Section \ref{sect-quantization_one_loop_dilatation} we quantize the near-BPS theory with $\SU(1,1|1)$  symmetry and show explicitly that the resulting quantum mechanical theory is the same as the $\SU(1,1|1)$ Spin Matrix theory limit of $\CN=4$ SYM. This means that whether one first quantizes, and then takes the near-BPS limit, yields the same quantum mechanical theory as if one does it in the opposite order. Also, it means that we found a highly efficient way to compute the one-loop dilatation operator of $\CN=4$ SYM. 

In Section \ref{sect-momentum_space_superfields} we find a momentum-space superfield formalism for the $\SU(1,1| 1)$ near-BPS theory, showing manifestly the supersymmetry of this theory. In addition it reveals a very simple formulation of the interactions via a non-dynamical gauge-field multiplet.

In Section \ref{sect-local_formulation} we discuss in detail how to find a local formulation of our four near-BPS/Spin Matrix theories. This reveals intriguing results that shows rather simple formulations of the interactions, at least in the $\SU(1,1|1)$ case and its two $\SU(1,1)$ subsectors. At the same time, the theories are not fully local and exhibit a ghost-like behavior of the dynamical fields.  As part of this we exhibit the algebraic structure of the global symmetry groups in Appendix \ref{app-algebra_osc_repr}.

Finally, we present our conclusions and outlook in Section \ref{sect-conclusions-and-outlook}. 

\section{Classical $S^3$ reduction and near-BPS limits}
\label{sect-sphere_reduction}

In this section we consider the classical Hamiltonian of $\CN=4$ SYM on a three-sphere in the near-BPS limits of the type \eqref{BPSlimit} and show how to derive the classical effective Hamiltonian of the surviving degrees of freedom.
We consider four different limits. In each limit one has fields that decouple and do not contribute to the dynamics. However the gauge field, and in some cases also fermionic fields, can contribute to the resulting dynamics even if they decouple as degrees of freedom and thus they need to be properly integrated out. This happens because the corresponding fields are sourced and can be understood in the same manner as the nondynamical modes of the photon mediating the Coulomb interaction.

\begin{figure}
\centering
\includegraphics[scale=0.5]{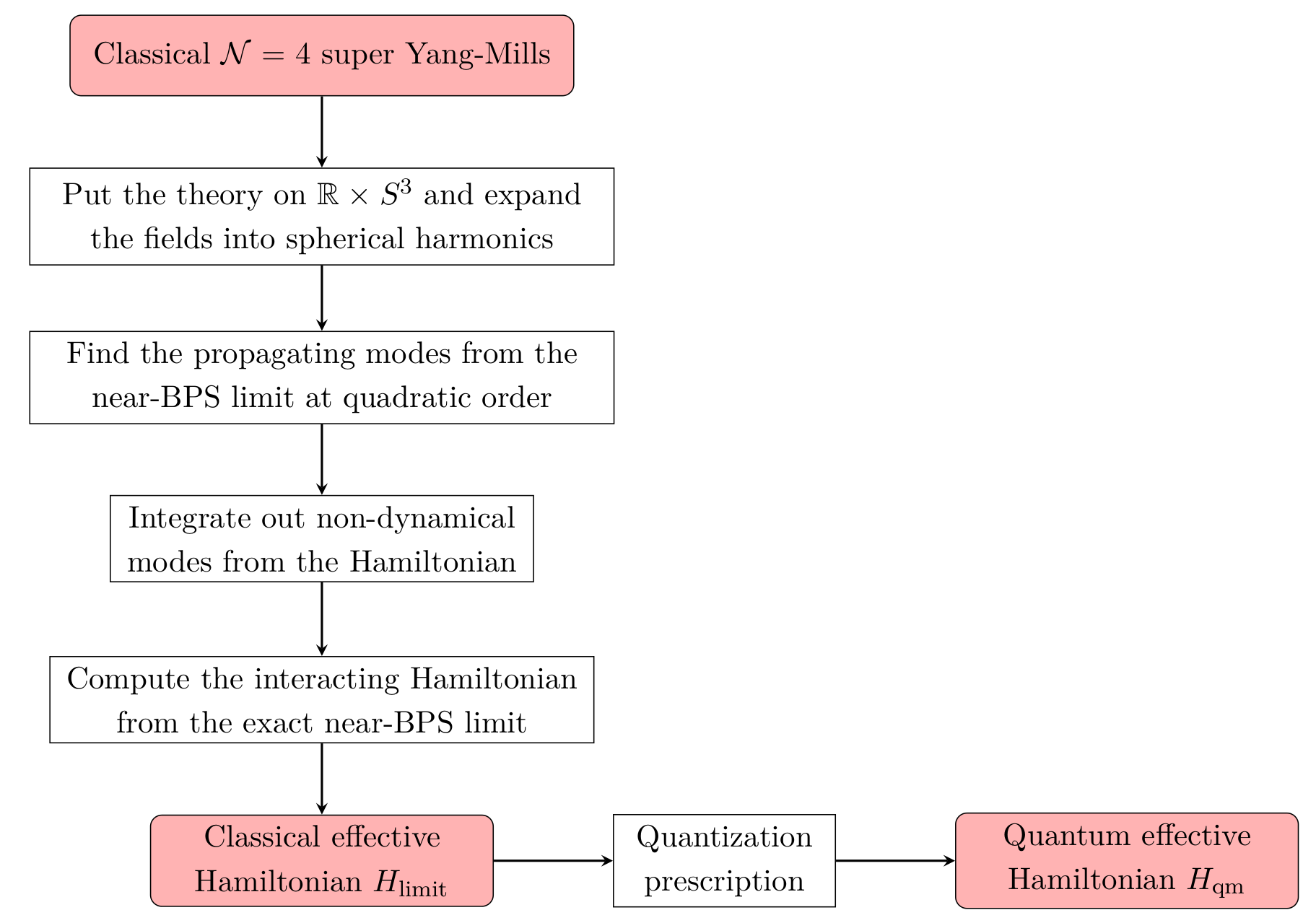}
\caption{\small Pictorial description of the procedure performed in Sections \ref{sect-sphere_reduction} and \ref{sect-quantization_one_loop_dilatation} to find an effective quantum Hamiltonian starting from $\mathcal{N}=4$ supersymmetric Yang-Mills on a three-sphere.}
\label{fig:flow_chart}
\end{figure}

We summarize with the scheme in Fig.~\ref{fig:flow_chart} the main steps of the procedure that we will perform in Sections \ref{sect-sphere_reduction} and \ref{sect-quantization_one_loop_dilatation}.

After setting the stage of the computations in Section \ref{sect-sym_on_s3} 
we first review in detail in Section \ref{sec:bos_su11} the limit given only by bosonic modes with a global $\SU(1,1)$ symmetry. This case was considered previously in \cite{Harmark:2019zkn}. Then we proceed in Section \ref{sec:su111} with the limit that adds fermionic modes to this, providing a theory with $\SU(1,1|1)$ symmetry which we show explicitly in Section \ref{sect-momentum_space_superfields} to be supersymmetric. We proceed with a subcase of this with only fermionic degrees of freedom in Section \ref{sec:ferm_su11}. Finally, in Section \ref{sec:su112} we consider the limit giving the maximally possible amount of bosonic and fermionic modes, which has a global $\PSU(1,1|2)$ symmetry and hence has extended supersymmetry.

In Section \ref{sect-quantization_one_loop_dilatation} we show that when one quantizes the four classical near-BPS theories that we obtain in Sections \ref{sec:bos_su11}-\ref{sec:su112}, one obtains bosonic $\SU(1,1)$ Spin Matrix theory (SMT), $\SU(1,1|1)$ SMT,  fermionic $\SU(1,1)$ SMT and $\PSU(1,1|2)$ SMT.

\subsection{${\cal N} = 4$ SYM on $S^3$}
\label{sect-sym_on_s3}

Our starting point is the classical action of ${\cal N} = 4$ super-Yang-Mills theory compactified on a three-sphere
\begin{multline}
\label{theS}
S = \int_{\mathbb{R} \times S^3} \sqrt{-\mathrm{det} \, g_{\mu\nu}} \, \tr \left\lbrace - \frac{1}{4} F_{\mu\nu}^2 - |D_{\mu} \Phi_a|^2 - |\Phi_a|^2 - i \psi^{\dagger}_a \bar{\sigma}^{\mu} D_{\mu} \psi^A + g \sum_{A,B,a} C_{AB}^{a} \psi^A [\Phi_a, \psi^B]
  \right. \\
\left. + g \sum_{A,B,a} \bar{C}^{aAB} \psi^{\dagger}_A [\Phi^{\dagger}_a, \psi^{\dagger}_B]
- \frac{g^2}{2} \sum_{a,b} \le |[\Phi_a, \Phi_b]|^2 + |[\Phi_a, \Phi^{\dagger}_b]|^2 \ri
 \right\rbrace \, .
\end{multline}
From this one can straightforwardly obtain the classical Hamiltonian $H$ of $\CN=4$ SYM on $S^3$ by a Legendre transform.
In the action \eqref{theS} $g$ is the Yang-Mills coupling constant, and we introduced complex combinations of the real scalar fields transforming in the $\mathbf{6}$ representation of the R-symmetry group $ \mathrm{SO}(6)\simeq \SU(4),$ defined as $\Phi_a = \frac{1}{\sqrt{2}} ( \phi_{2a-1} + i \phi_{2a})$ with $a \in \lbrace 1,2,3 \rbrace . $ 
The Weyl fermions $\psi^A$ with $A \in \lbrace 1,2,3,4 \rbrace$ transform in the representation $\mathbf{4}$ of  $\SU(4) .$
The action is canonically normalized on the $\mathbb{R} \times S^3$ background with the radius of the three-sphere set to unity.
The field strength is defined as
\beq
F_{\mu\nu} = \partial_{\mu} A_{\nu} - \partial_{\nu} A_{\mu} + i g [A_{\mu} , A_{\nu}] \, ,
\eeq
and the covariant derivatives $D_{\mu}$ as
\bea
& D_{\mu} \Phi_a = \partial_{\mu} \Phi_a + i g [A_{\mu}, \Phi_a] \, , & \\
& D_{\mu} \psi^A = \nabla_{\mu} \psi^A + i g [A_{\mu}, \psi^A] \, , &
\eea
where $\nabla_{\mu}$ is the covariant derivative on the three-sphere, {\sl i.e.} it contains the spin connection contribution for the fermions.
The $C^{a}_{AB}$ are Clebsch-Gordan coefficients coupling two $\mathbf{4}$  representations and one $\mathbf{6}$ representation of the R-symmetry group $\SU(4).$
All the fields in the action transform in the adjoint representation of the gauge group $\SU(N)$.

One can now decompose all the fields into spherical harmonics on $S^3$. For this, we follow the procedure and conventions of \cite{Ishiki:2006rt}. We have given the relevant details of this in Appendix \ref{app-spher_harmonics_S3} and \ref{app-quadratic_Hamiltonian_conserved_charges}.

Before we turn to the individual limits we first discuss the gauge field. 
In all four limits, the gauge field degrees of freedom will decouple on-shell. However, it contributes to the dynamics exactly like an off-shell longitudinal photon does in QED and i ntegrating it out gives rise to an effective interaction of the surviving mode at order $g^2$. Since this is a feature that all four sectors share, we make a few remarks about it here.

We will work in Coulomb gauge, corresponding to imposing
\begin{equation}
\nabla_i A^i = 0 \, .
\end{equation}
In our analysis below, it proves useful to first integrate out, {\sl i.e.} solve for, all auxiliary degrees of freedom, here the temporal and longitudinal components of the gauge field. This procedure is standard, but since it is central to our arguments we display it here in some detail.

To this end, we focus on the quadratic action for the gauge field, but we also include a generic source to keep track of the correct constraint structure.  
We have
\begin{equation}
S_A = \int_{\mathbb{R}\times S^3} \sqrt{-\mathrm{det} \, g_{\mu\nu}} \, \tr(-\frac{1}{4} F_{\mu\nu}^2 - A^\mu j_\mu)
\,.
\end{equation}
The canonical momenta are
\begin{equation}
\Pi_0 = \frac{1}{\sqrt{-\mathrm{det} \, g_{\mu\nu}}} \frac{\delta S_A}{\delta \dot{A}_0} = 0 \, , \qquad
  \Pi_i = \frac{1}{\sqrt{-\mathrm{det} \, g_{\mu\nu}}} \frac{\delta S_A}{\delta \dot{A}_i} = F_{0i}\,, \qquad
\end{equation}
yielding the Hamiltonian
\begin{equation}
H_A = \int_{\mathbb{R} \times S^3} \sqrt{- \mathrm{\det} \, g_{\mu\nu}} \,  \tr( \frac{1}{2} \Pi_i^2 + \frac{1}{4} F_{ij}^2  - A_0(\nabla_i \Pi^i + j_0) + A^i j_i + \eta \nabla_i A^i ) \,,
\label{eq:Hamiltonian_gauge_field_plus_currents}
\end{equation}
where we have introduced a Lagrange multiplier $\eta$ to enforce Coulomb gauge. 
We obtain the constraints
\begin{equation}
\label{cons1}
\nabla_i \Pi^i + j_0 = 0\,,\quad \nabla_i A^i=0\,.
\end{equation}
We have chosen to treat $A_0$ as a Lagrange multiplier that enforces the Gauss' law, and no longer as one of the dynamical variables\footnote{This is possible because this field has no dynamics (the canonical momentum is vanishing), and the non-trivial spatial dependence is encoded into the momentum $\Pi_i.$}.
Thus, we have two second class constraints, enough to eliminate the remaining unphysical degrees of freedom.

In order to solve the constraints \eqref{cons1}, it proves useful to decompose all the fields into spherical harmonics on $S^3$ (see Appendix \ref{app-spher_harmonics_S3}). 
Inserting all the decompositions into 
the Hamiltonian \eqref{eq:Hamiltonian_gauge_field_plus_currents}, we find 
\beq
\begin{aligned}
\hspace*{-2mm}
H_A =  \tr\sum_{J,m,\tilde{m}} & \left\lbrace  \sum_{\rho = -1}^1 \frac{1}{2} |\Pi_{(\rho)}^{Jm\tilde{m}}|^2 + \sum_{\rho = \pm1} \frac{1}{2} \omega_{A,J}^2 |A_{(\rho)}^{Jm\tilde{m}}|^2  - \chi^{Jm\tilde{m}}\left(2i\sqrt{J(J+1)}\Pi_{(0)}^{Jm\tilde{m}} +  j_0^{\dagger \, Jm\tilde{m}}\right) \right. \\
& \left. + \sum_{\rho = -1}^1 A_{(\rho)}^{Jm\tilde{m}} j_{(\rho)}^{\dagger\,Jm\tilde{m}}  - 2i\sqrt{J(J+1)}\eta^{\dagger \, Jm\tilde{m}}A_{(0)}^{Jm\tilde{m}} \right\rbrace \,,
\end{aligned}
\eeq
while the constraints \eqref{cons1} become
\begin{equation}
2i\sqrt{J(J+1)}\Pi_{(0)}^{Jm\tilde{m}} +  j_0^{\dagger J m\tilde{m}} = 0\,,\quad A_{(0)}^{Jm\tilde{m}}=0\,.
\end{equation}
Since we can directly solve the constraints for $ A_{(0)}^{Jm\tilde{m}}$ and its symplectic partner $\Pi_{(0)}^{Jm\tilde{m}}$, we can insert the solution into the Hamiltonian without changing the Poisson bracket. We thus obtain the unconstrained Hamiltonian
\begin{equation}
\label{eq:Ham_freeYM}
H_A = \tr\sum_{J,m,\tilde{m}} \left[ \sum_{\rho = \pm1}\left( \frac{1}{2}|\Pi_{(\rho)}^{Jm\tilde{m}}|^2 + \frac{1}{2} \omega_{A,J}^2 |A_{(\rho)}^{Jm\tilde{m}}|^2+ A_{(\rho)}^{Jm\tilde{m}} j_{(\rho)}^{\dagger\,Jm\tilde{m}}\right) + \frac{1}{8J(J+1)} |j_0^{Jm\tilde{m}}|^2 \right] \,.
\end{equation}
The form of the currents can now straightforwardly be 
reconstructed from the full ${\cal N} = 4$ Hamiltonian, and all further interactions can be restored. Instead of doing so in full generality, we will consider the near-BPS limit individually and reconstruct the interactions case by case, where they simplify considerably.

We proceed now by considering the four near-BPS limits individually in the following subsections \ref{sec:bos_su11}-\ref{sec:su112}. 
In each case we will employ the following procedure
\begin{enumerate}
	\item Isolate the propagating modes in a given near-BPS limit from the quadratic classical Hamiltonian.
	\item Derive the form of the currents that couple to the gauge fields.
	\item Integrate out additional non-dynamical modes that give rise to effective interactions in a given near-BPS limit.
	\item Derive the interacting Hamiltonian by taking the limit.
\end{enumerate}
In all of the four near-BPS limits the single angular momentum $S_1$ is turned on, corresponding to BPS bounds of the form  $H \geq S_1 + \sum_{i=1}^3 \omega_i Q_i$ where $H$ is the Hamiltonian, $S_1$ is one of the angular momenta and $Q_i$, $i=1,2,3$, are the three R-charges  of $\CN=4$ SYM on $S^3$. The coefficients $\omega_i$ in front of the R-charges are given in the Table \ref{tab:BPS_bounds}. A derivation of these coefficients can be found in \cite{Harmark:2007px}. 
For each case, the near-BPS limit is 
\begin{equation}
\label{newnearBPSlimit}
g \rightarrow 0 \quad \mbox{with} \quad \frac{H - S_1 - \sum_{i=1}^3 \omega_i Q_i}{g^2} \quad \mbox{fixed} \,.
\end{equation}
Note that $N$ is held fixed in this limit while $g\rightarrow 0$. We find that the surviving degrees of freedom  are described by a  Hamiltonian $H_{\rm limit}$ of the form
\begin{equation}
\label{full_ham_gen}
H_{\rm limit} = L_0 + \tilde{g}^2 H_{\rm int} \spa H_{\rm int} = \lim_{g \rightarrow 0} \frac{H - S_1 - \sum_{i=1}^3 \omega_i Q_i}{g^2 N} \,,
\end{equation}
where $L_0$ is the Cartan charge of $\SU(1,1)$, $H_{\rm int}$ is the part of the Hamiltonian that describes the interactions and $\tilde{g}$ is the coupling constant of the resulting non-relativistic theory.

\begin{table}\centering
	\ra{1.3}
	\begin{tabular}{@{}|c|c|c|c|c|@{}}\toprule[0.1em]
Sectors		& $\SU(1,1)$ bosonic & $\SU(1,1)$ fermionic & $\SU(1,1|1)$ & $\PSU(1,1|2)$ \\ \midrule[0.05em]
		$\sum_{i=1}^3 \omega_i Q_i$  & $Q_1$ & $\frac{2}{3}(Q_1+Q_2+Q_3)$ & $Q_1+ \frac{1}{2}(Q_2+Q_3)$ & $Q_1+Q_2$ \\
		\bottomrule[0.1em]
	\end{tabular}
	\caption{List of the combinations of the R-charges  defining the limits of $\mathcal{N}=4$ SYM theory towards BPS bounds $H \geq S_1+ \sum_{i=1}^3 \omega_i Q_i. $}
	\label{tab:BPS_bounds}
\end{table}

\subsection{Bosonic $\SU(1,1)$ limit - The simplest case}
\label{sec:bos_su11}

The first BPS bound we consider is $H \geq S_1 + Q_1$. As we shall see, the dynamical theory that one obtains from the near-BPS limit \eqref{newnearBPSlimit} has a global $SU(1,1) \times U(1)$ symmetry of the interactions. 

\subsubsection*{Free Hamiltonian and reduction of degrees of freedom}

We start from the quadratic Hamiltonian $H_0$, in which interaction terms are omitted, and also the R-charge $Q_1$ and the angular momentum $S_1$  of ${\cal N} = 4$ SYM on $S^3$. These are all given in Appendix \ref{app-quadratic_Hamiltonian_conserved_charges}. The propagating degrees of freedom can be extracted by considering the near-BPS limit to lowest order in the coupling, which means we should set $H_0 - Q_1 - S_1 = 0$. The left hand side reads
\begin{multline}
H_0 - S_1 - Q_1 = \sum_{J,m,\tilde{m}}  \tr \Bigg\lbrace
|\Pi_a^{Jm\tilde{m}} + i(\delta^1_a+\tilde{m}-m)\Phi_a^{\dagger Jm\tilde{m}}|^2 + (\omega_J^2 - (\delta^1_a+\tilde{m}-m)^2) |\Phi_a^{Jm\tilde{m}}|^2 
 \\
 + \sum_{\kappa=\pm 1} \left( \sum_{A=1,4}  \le \omega_J^{\psi} + m - \tilde{m} - \frac{\kappa}{2} \ri \psd_{JM,\kappa, A} \psi^A_{JM,\kappa} 
+ \sum_{A=2,3} \le \omega_J^{\psi} + m - \tilde{m} + \frac{\kappa}{2} \ri \psd_{JM,\kappa, A} \psi^A_{JM,\kappa}
 \right) \\
  +  \sum_{\rho=-1,1} \frac{1}{2} \left( |\Pi_{(\rho)}^{Jm\tilde{m}} - i(m-\tilde{m}) A_{(\rho)}^{\dagger\,Jm\tilde{m}}|^2+ (\omega_{A,J}^2 - (m-\tilde{m})^2) |A_{(\rho)}^{Jm\tilde{m}}|^2 \right)
\Bigg\rbrace \,,
\end{multline}
with $\omega_J = 2J+1$,  $\omega_J^{\psi} = 2J + \frac{3}{2}$ and $\omega_{A,J} \equiv 2J + 2$. 
Equating this expression to zero now yields a set of conditions on the fields. First of all, since for the gauge field $|m-\tilde{m}| \leq 2J+1$, one finds 
\begin{equation}
\label{eq:APi_zero}
A_{(\rho)}^{Jm\tilde{m}} = {\cal O}(g)\,,\quad\Pi_{(\rho)}^{Jm\tilde{m}}-i(m-\tilde{m})A_{(\rho)}^{\dagger\,Jm\tilde{m}} = {\cal O}(g)\,.
\end{equation}
Second, for the scalar field $\Phi_1$ we find for $J = -m = \tilde{m}$
\begin{equation}
\label{eq:Phi_const}
\Pi_1^{J,-J,J} + i\omega_J \Phi_1^{\dagger \, J,-J,J} = {\cal O}(g)\,,
\end{equation}
and for all other eigenvalues of momentum $(m, \tilde{m})$
\begin{equation}
\label{eq:Phizero_const}
\Phi_{1}^{Jm\tilde{m}} = {\cal O}(g)\,,\quad \Pi_1^{Jm\tilde{m}}  = {\cal O}(g)\,.
\end{equation}
The other two scalar fields satisfy for all possible values of $(m, \tilde{m})$ the conditions
\begin{equation}
\Phi_{2}^{Jm\tilde{m}} =  \Pi_2^{Jm\tilde{m}} = \Phi_{3}^{Jm\tilde{m}} =  \Pi_3^{Jm\tilde{m}}  = {\cal O}(g)\,.
\label{eq:Phi_2and3_zero_constraint}
\end{equation}
For the fermions, non-trivial degrees of freedom would arise when we are able to make the prefactor of the quadratic terms in the fields to vanish.
However, when $\kappa=1$
\beq
\omega_J^{\psi} = 2J+ \frac{3}{2} \, , \qquad
|m| \leq J + \frac{1}{2} \, , \qquad
|\tilde{m}| \leq J \, ,
\label{eq:choice_m_m_tilde_fermionic_SU(1,1)}
\eeq 
there is no way to make the $J$--independent constant to vanish.
The same phenomenon happens with $\kappa=-1,$ with the roles of $(m, \tilde{m})$ exchanged.
This tells us that in the bosonic $\SU(1,1)$ sector we have for all choices of the indices $(A,\kappa)$ the condition
\beq
\psi^A_{JM,\kappa} = \mathcal{O}(g) \, .
\label{eq:vanishing_fermions_su11bos_limit}
\eeq
It is clear that each of the above constraints eliminates a dynamical degree of freedom from the theory, as one forfeits the choice of freely choosing initial conditions. Instead, the corresponding fields are entirely determined by the remaining degrees of the freedom, as encoded by the right hand sides of the above relations. We can make these explicit by demanding compatibility with Hamiltonian evolution.
It is simple to see that Eq.~\eqref{eq:Phizero_const} weakly commutes, {\sl i.e.} commutes on the constraint surface, with $H$, since no linear term in $\Phi$ and $\Pi$ are present. The same holds for the first constraint in Eq.~\eqref{eq:APi_zero}, for the scalars in \eqref{eq:Phi_2and3_zero_constraint} and for the fermionic field in \eqref{eq:vanishing_fermions_su11bos_limit}. 
On the other hand, the gauge field does appear linearly, namely through its coupling to the sources, as outlined in Eq.~\eqref{eq:Ham_freeYM}. Therefore,
\begin{equation}
\{H,\Pi_{(\rho)}^{Jm\tilde{m}}-i(m-\tilde{m})A_{(\rho)}^{\dagger\,Jm\tilde{m}}\} \approx  (\omega_{A,J}^2 - (m-\tilde{m})^2) A_{(\rho)}^{\dagger\,Jm\tilde{m}} + j_{(\rho)}^{\dagger\,Jm\tilde{m}} \,.
\end{equation}
Hence we impose the RHS side as a constraint to have a consistent Hamiltonian evolution. 
Finally, one can check that Eq.~\eqref{eq:Phi_const} does not generate additional requirements. 
We thus obtain the set of constraints
\begin{align}
\label{eq:constraints_gauge_fields_su11bos}
&A_{(\rho)}^{Jm\tilde{m}} = -\frac{1}{\omega_{A,J}^2 - (m-\tilde{m})^2} j^{ Jm\tilde{m}}_{(\rho)} \,,\quad\Pi_{(\rho)}^{Jm\tilde{m}} = 0\,,\\
\label{eq:constraints_scalars1_su11bos}
&\Phi_{a=2,3}^{Jm\tilde{m}} = 0\,,\quad\Pi_{a=2,3}^{Jm\tilde{m}}  = 0\,,\\[1mm]
\label{eq:constraints_scalars2_su11bos}
&\Phi_{1}^{Jm\tilde{m}} = 0\,,\quad\Pi_{1}^{Jm\tilde{m}}  = 0 \quad (\mbox{except when}\ \ J=-m=\tilde{m}) \,,
\\[1mm]
&\Pi_1^{J,-J,J} + i\omega_J \Phi_1^{\dagger \, J,-J,J} = 0\,.
\label{eq:constraint_dynamical_scalars_su11bos}
\end{align}
Thus, the only dynamical degrees of freedom left are the modes $\Phi_1^{J,-J,J}$ that obey the constraint \eqref{eq:constraint_dynamical_scalars_su11bos}. Essentially, \eqref{eq:constraint_dynamical_scalars_su11bos} is responsible for making the limiting theory non-relativistic as it decouples the anti-particles. 
Indeed, this condition relates the momentum with the complex conjugate of the field, implying that at the quantum level the field $\Phi_n$ will annihilate a particle and the hermitian conjugate $\Phi^{\dagger}_n$ will create it.
As we explain below, this goes in hand with a $\U(1)$ global symmetry responsible for the conservation of particle number.
This behavior is standard in the non-relativistic low-momentum limit of QFTs \cite{Bergman:1991hf}. Here we see that the same phenomenon happens when focusing on a near-BPS limit of $\mathcal{N}=4$ SYM.

Before we turn to the interactions, we consider the free part of the resulting Hamiltonian.
The quadratic piece is simply obtained by inserting the constraint \eqref{eq:constraint_dynamical_scalars_su11bos} into the quadratic Hamiltonian \eqref{H0_sphere}. Before doing so, however, we note that \eqref{eq:constraint_dynamical_scalars_su11bos} implies a change of brackets, since $\Phi_1^{J,-J,J}$ and $(\Phi_1^{J,-J,J})^\dagger$ no longer commute on the constraint surface. The Dirac brackets can be straightforwardly worked out, yielding (with matrix indices suppressed) 
\begin{equation}
\{\Phi_1^{J,-J,J}, (\Phi_1^{J',-J',J'})^\dagger \}_D = \frac{i}{2\omega_J}\delta_{JJ'}
\label{eq:Dirac_bracket_scalar}
\end{equation}
We make the redefinition
\begin{equation}
\label{newphidef}
\Phi_{2J} = \sqrt{2\omega_{J}} \Phi_1^{J,-J,J} \,,
\end{equation}
in order to have a canonical normalization and to take into account that $J$ both takes integer and half-integer values. The Dirac bracket \eqref{eq:Dirac_bracket_scalar} then becomes canonical
\begin{equation}
\label{eq:Dirac_bracket_scalar2}
\Big\{ (\Phi_n)^i {}_j , (\Phi_{n'}^\dagger)^k {}_l \Big\}_D = i \delta_{n,n'} \delta^i {}_l \delta^k {}_j
\end{equation}
With this, we obtain for the quadratic Hamiltonian
\begin{equation}
H_0 = S_1+ Q_1=  \sum_{n=0}^\infty (n+1) \tr |\Phi_n|^2\,.
\end{equation}
It is important to see how the $\SU(1,1)$ symmetry emerges. Consider
\begin{equation}
\label{bossu11_su11gens}
L_0 =  \sum_{n=0}^{\infty}  \le n + \frac{1}{2} \ri \tr |\Phi_{n} |^2  \spa
L_+ = (L_-)^{\dagger} = \sum_{n=0}^{\infty} \le n+1 \ri \tr ( \Phi^{\dagger}_{n+1} \Phi_n ) \, . 
\end{equation}
Using the bracket \eqref{eq:Dirac_bracket_scalar2} one finds that these charges obey the $\SU(1,1)$ brackets $\lbrace L_0 , L_{\pm} \rbrace_D = \pm i L_{\pm}, $ $ \lbrace L_+, L_- \rbrace_D = -2i L_0 $. The interactions that we find below have vanishing brackets with $L_0$ and $L_\pm$ which means that the interactions have a global $\SU(1,1)$ symmetry.

The difference between $H_0$ and $L_0$ is 
\begin{equation}
\label{numberop_su11}
H_0 = L_0 + \frac{1}{2} \hat{N} \spa \hat{N} \equiv \sum_{n=0}^\infty  \tr |\Phi_n|^2
\end{equation}
We notice that $\hat{N}$ commutes with $H_0$, $L_0$ and $L_\pm$ as well as the interaction terms with respect to the bracket \eqref{eq:Dirac_bracket_scalar2}. This means in particular that $\hat{N}$ is a conserved charge. Indeed, $\hat{N}$ is the number operator when quantizing this theory, and we recognize the fact that the conservation of the number operator is a hallmark of a non-relativistic theory. 
This in turn means we are allowed to switch the free part of the Hamiltonian to be $L_0$ instead of $H_0$ since they differ by a conserved quantity (one can view this switch as a time-dependent redefinition of the fields). This will turn out to be a natural choice for all of the four limits.

\subsubsection*{Interactions}

We now exhibit the interacting part of the Hamiltonian $H_{\rm int}$ that arise in the near-BPS limit.
Since $H_0 - S_1 - Q_1 = 0$ by construction, we can define the interacting Hamiltonian as
\begin{equation}
\label{Hintdef_bossu11}
H_\text{int} =  \lim_{g \to 0} \frac{H - S_1 - Q_1}{g^2 N}\,.
\end{equation}
Non-trivial contributions to $H_\text{int}$ arise from integrating out the gauge field. 
On the surface defined by the constraints \eqref{eq:constraints_gauge_fields_su11bos} we find that the contributions to $H - S_1 - Q_1$ amount to
\begin{equation}
\label{eq:gauge_terms}
\sum_{J,m,\tilde{m}} \tr \left(\frac{1}{8J(J+1)} |j_0^{Jm\tilde{m}}|^2 - \sum_{\rho = \pm1} \frac{1}{2 ( \omega_{A,J}^2 - (m-\tilde{m})^2 )} |j_{(\rho)}^{Jm\tilde{m}}|^2\right) \,.
\end{equation}
To these terms we should add contributions from the scalar sector that we will derive now.
To this end, we add the entire scalar sector and work out the form of the currents. The relevant interaction terms involving scalars in the Hamiltonian $H$ of $\CN=4$ SYM, Eq.~\eqref{eq:app_full_interacting_N=4SYM_Hamiltonian}, are
\begin{eqnarray}
&& \sum_{J,m,\tilde{m}} \tr \left\lbrace  \frac{g^2}{2}{\cal C}^{\CJ_2}_{\CJ_1,J M}{\cal C}^{\CJ_3}_{\CJ_4,J M}[\Phi_1^{\CJ_1},\Phi_1^{\CJ_2}{}^\dagger][\Phi_1^{\CJ_3},\Phi_1^{\CJ_4}{}^\dagger] + i g {\cal C}^{\CJ_2}_{\CJ_1,JM} \chi^{JM}\left([\Phi_1^{\CJ_2}{}^\dagger,\Pi_1^{\CJ_1}{}^\dagger] + [\Phi_1^{\CJ_1},\Pi_1^{\CJ_2}]\right)  \right. \nn \\ &&
\left. - 4g\sqrt{J_1(J_1+1)} {\cal D}^{\CJ_2}_{\CJ_1, JM\rho}A^{JM}_{(\rho)}[\Phi_1^{\CJ_1},\Phi_1^{\CJ_2}{}^\dagger]  \right\rbrace \,.
\label{eq:relevant_Hamiltonian_su11bos}
\end{eqnarray}
where we used the short-hand notation 
\begin{equation}
\CJ = (J,-J,J) \,, 
\label{eq:short-hand_notation_momenta}
\end{equation}
{\sl i.e.} $\Phi_1^{\CJ}=\Phi_1^{J,-J,J}$, since we can restrict ourselves to the surviving scalar modes.
In this expression the quantities $\mathcal{C}, \mathcal{D}$ are Clebsch-Gordan coefficients that couple respectively three scalars or two scalars and one vector harmonics. We define them and show some of their properties in Appendix \ref{app-properties_Clebsch_Gordan_coeff}.
From \eqref{eq:relevant_Hamiltonian_su11bos}, we can directly read off the currents. We have
\begin{equation}
j_0^{\dagger\,Jm\tilde{m}} = i g {\cal C}^{\CJ_2}_{\CJ_1,JM} \left([\Phi_1^{\CJ_2}{}^\dagger,\Pi_1^{\CJ_1}{}^\dagger] + [\Phi_1^{\CJ_1},\Pi_1^{\CJ_2}]\right) = 2 g (1+J_1+J_2) {\cal C}^{\CJ_2}_{\CJ_1,JM} [\Phi_1^{\CJ_1},\Phi_1^{\CJ_2} {}^\dagger]\,,
\end{equation}
where the latter equality holds on the constraint surface. Furthermore
\begin{equation}
j_{(\rho)}^{\dagger\,Jm\tilde{m}} =- 4g\sqrt{J_1(J_1+1)} {\cal D}^{\CJ_2}_{\CJ_1, JM\rho}
[\Phi_1^{\CJ_1},\Phi_1^{\CJ_2} {}^\dagger] \,.
\end{equation}
We can now proceed to find the interaction Hamiltonian \eqref{Hintdef_bossu11}. Employing \eqref{Hintdef_bossu11} we obtain
\begin{multline}
H_\text{int} = \frac{1}{4N}\sum_{J,m,\tilde{m}}\sum_{J_1,J_2,J_3,J_4} \left(\prod_{i=1}^4 \frac{1}{\sqrt{\omega_{J_i}}}\right)
\tr \Bigg(\frac{1}{2J(J+1)} (1+J_1+J_2)(1+J_3+J_4) {\cal C}^{\CJ_2}_{\CJ_1,JM} {\cal C}^{\CJ_3}_{\CJ_4,JM} \\
- \sum_{\rho = \pm1}\frac{8}{\omega_{A,J}^2 - (m-\tilde{m})^2} \sqrt{J_1(J_1+1)}\sqrt{J_4(J_4+1)} {\cal D}^{\CJ_2}_{\CJ_1, JM\rho}{\cal \bar D}^{\CJ_3}_{\CJ_4, JM\rho}\\
+ \frac{1}{2}{\cal C}^{\CJ_2}_{\CJ_1,J M}{\cal C}^{\CJ_3}_{\CJ_4,J M} \Bigg)[\Phi_{2J_1},\Phi^\dagger_{2J_2}][\Phi_{2J_3},\Phi^\dagger_{2J_4}]\,.
\end{multline}
where we used the redefinition \eqref{newphidef}.
It is clear that the only nontrivial contributions arise from $\tilde{m} = -m$. For notational convenience, we consider $M = (-m,m)$. Inserting this and making $\omega_{A,J}$ explicit yields
\begin{multline}
H_\text{int} = \frac{1}{8N}\sum_{J,m}\sum_{J_1,J_2,J_3,J_4} \left(\prod_{i=1}^4 \frac{1}{\sqrt{\omega_{J_i}}}\right) \tr \Bigg(\left(\frac{(1+J_1+J_2)(1+J_3+J_4) }{J(J+1)}+1\right) {\cal C}^{\CJ_2}_{\CJ_1,JM} {\cal C}^{\CJ_3}_{\CJ_4,JM} \\
- \sum_{\rho = \pm1}\frac{4}{(J+1)^2 - m^2} \sqrt{J_1(J_1+1)}\sqrt{J_4(J_4+1)} {\cal D}^{\CJ_2}_{\CJ_1, JM\rho}{\cal \bar D}^{\CJ_3}_{\CJ_4, JM\rho}\Bigg)[\Phi_{2J_1},\Phi^\dagger_{2J_2}][\Phi_{2J_3},\Phi^\dagger_{2J_4}]\,.
\label{eq:H_int_bosonicSU(1,1)_before_sum}
\end{multline}
We can now directly use the crossing relations \eqref{eq:CDrelation}  to see that upon a shift $J \to J-1$ in the contribution with $\rho = 1$, all terms in the above sum cancel except for a nontrivial remainder from the lower boundary of summation. We distinguish between $\Delta{J} \equiv J_2 - J_1 \neq 0$ and $\Delta{J} = 0$ and moreover choose $J_2 > J_1$ without loss of generality, accounting for the converse with a factor of $2$. In this way we obtain
\begin{equation}
\label{Hint_bossu11}
H_\text{int}^{(J_1 \neq J_2)} = \frac{1}{4N}\tr \sum_{J_1,J_4 \geq 0} \sum_{\Delta{J}>0}\frac{1}{\Delta{J}} [\Phi_{2J_1},\Phi^\dagger_{2J_1 +2 \Delta{J}}][\Phi_{2J_4+2\Delta{J}},\Phi^\dagger_{2J_4}]
=  \frac{1}{2N} \sum_{l=1}^{\infty}\frac{1}{l}
\tr \le q^{\dagger}_l q_l \ri 
\,,
\end{equation}
where we defined the $\SU(N)$ charge density 
\beq
q_l= \sum_{n =0}^{\infty} [\Phi^{\dagger}_{n}, \Phi_{n+l}] \, . 
\label{eq:charge_density_scalar_su11bos}
\eeq
Next, let us consider the case $J_1 = J_2$. Here, the abovementioned trick to shift part of the expression by $J \rightarrow J-1$ in order to receive contributions only from the lowest boundary of summation still works.
However, the additional subtlety is that the first term in \eqref{eq:H_int_bosonicSU(1,1)_before_sum} is singular in $J=0,$ and then it is summed over $J>0.$ Collecting everything, we get
\begin{multline}
\label{Hint_bossu11_extra}
H_\text{int}^{(J_1 = J_2)} = \frac{1}{8N}\tr
\sum_{J_1,J_3 \geq 0}\Bigg[\frac{1-4J_1J_3}{(1+2J_1)(1+2J_3)} + \sum_{J > 0} \Bigg(\frac{1}{J} - \frac{(2J_1-J)(2J_3-J)}{(J+1)(2J_1+J+1)(2J_3+J+1)}\Bigg)\\
\times\frac{(2J_1)!(2J_1)!}{(2J_1-J)!(2J_1+J)!}\frac{(2J_3)!(2J_3)!}{(2J_3-J)!(2J_3+J)!}\Bigg][\Phi_{2J_1},\Phi^\dagger_{2J_1}][\Phi_{2J_3},\Phi^\dagger_{2J_3}]\,,
\end{multline}
which resums into 
\begin{equation}
H_\text{int}^{(J_1 = J_2)} 
=\frac{1}{8N}\sum_{n= 0}^\infty \frac{1-n}{1+n}\tr( q_0 [\Phi^\dagger_{n},\Phi_{n}])\,,
\end{equation}
where $q_0 = \sum_{n =0}^{\infty} [\Phi^{\dagger}_{n}, \Phi_{n}]$ is the $\SU(N)$ charge. 
The Gauss law on the three-sphere implies that $q_0=0$ and hence $H_\text{int}^{(J_1 = J_2)}$ is zero when taking this into account.

The full Hamiltonian \eqref{full_ham_gen} then becomes
\begin{equation}
\label{fullH_bossu11}
H_\text{limit} = L_0 +   \frac{\tilde{g}^2}{2N} \sum_{l=1}^{\infty}\frac{1}{l}
\tr \le q^{\dagger}_l q_l \ri 
\end{equation}
taking into account that all physical configurations have zero $\SU(N)$ charge $q_0=0$ due to the Gauss law on the three-sphere and with $L_0$ given in Eq.~\eqref{bossu11_su11gens}. 
This is the interacting Hamiltonian describing the effective dynamics of $\mathcal{N}=4$ SYM near the $\SU(1,1)$ bosonic BPS bound.
It is a non-relativistic theory because of Eq.~\eqref{eq:constraint_dynamical_scalars_su11bos}, which relates the canonical momentum to the complex conjugate field, as it happens for this class of quantum field theories.
In addition, the non-relativistic nature of the system is also clear from the conservation of the number operator $\hat{N}$ defined in \eqref{numberop_su11} corresponding to a further $\U(1)$ symmetry in addition to the global $\SU(1,1)$.
Finally, it is straightforward to show that \eqref{Hint_bossu11} commutes with the $\SU(1,1)$ charges $L_0$ and $L_\pm$ \eqref{bossu11_su11gens} under the brackets \eqref{eq:Dirac_bracket_scalar2}. This shows that the interaction term of \eqref{fullH_bossu11} is invariant under a global $\SU(1,1)$ symmetry. Upon quantization, we show in Section \ref{sect-quantization_one_loop_dilatation} that the Hamiltonian \eqref{fullH_bossu11} is equivalent to $\SU(1,1)$ Spin Matrix theory.

\subsection{$\SU(1,1|1)$ limit - A first glance at SUSY}
\label{sec:su111}

We turn now to the BPS bound $H \geq S_1 + Q_1+ \tfrac{1}{2}(Q_2 + Q_3)$. In this case, the theory that emerges from the limit \eqref{newnearBPSlimit} has a $\SU(1,1|1) \times \mathrm{U}(1)$ symmetry of its interactions. As we shall see, the additional symmetry compared to the $\SU(1,1)$ case of Section \ref{sec:bos_su11} is related to the fact that one has fermionic modes in addition to the bosonic modes of the $\SU(1,1)$ case. In Sections \ref{sect-quantization_one_loop_dilatation}, \ref{sect-momentum_space_superfields} and \ref{sect-local_formulation} we study this theory further, and show among other things that it is supersymmetric.

\subsubsection*{Free Hamiltonian and reduction of degrees of freedom}

We follow the same procedure as in Section \ref{sec:bos_su11}. Using Appendix \ref{app-quadratic_Hamiltonian_conserved_charges} we find that the quadratic terms in the left-hand side of the BPS bound $H - S_1 - Q_1- \tfrac{1}{2}(Q_2 + Q_3) \geq 0$ are given by 
\begin{equation}
\label{H0_su111}
\begin{array}{l} \ds
 H_0 - S_1 - Q_1 - \frac{1}{2} (Q_2+Q_3)  =  \sum_{JM} \tr \Bigg\{ \sum_{\kappa=\pm 1} \Bigg[ (\omega^{\psi}_J +m- \tilde{m}-\kappa) \psd_{JM,\kappa,1} \psi^1_{JM,\kappa} 
 \\[5mm] \ds 
+ \sum_{A=2,3}\le \omega^{\psi}_J +m- \tilde{m}  + \frac{\kappa}{2} \ri \psd_{JM,\kappa,A} \psi^A_{JM,\kappa}  + (\omega^{\psi}_J +m- \tilde{m}) \psd_{JM,\kappa,4} \psi^4_{JM,\kappa}  \Bigg]
 \\[5mm] \ds
	 + \left|\Pi_1^{JM} + i(1+\tilde{m}-m) \Phi_1^{\dagger\,JM} \right|^2 + (\omega_J^2 - (1+\tilde{m}-m)^2) |\Phi_1^{JM} |^2
	 \\[3mm] \ds
	 + \sum_{a = 2,3} \left( \left|\Pi_a^{JM} + i \le \frac{1}{2}+\tilde{m}-m \ri (\Phi_a^\dagger)^{JM} \right|^2 + \le \omega_J^2 - \le \frac{1}{2}+\tilde{m}-m \ri^2 \ri |\Phi_a^{JM}|^2 \right)
	  \\[5mm] \ds
+ 
 \sum_{\rho=-1,1} \frac{1}{2} \left(|\Pi_{(\rho)}^{Jm\tilde{m}} - i(m-\tilde{m})A_{(\rho)}^{\dagger\,Jm\tilde{m}}|^2+ (\omega_{A,J}^2 - (m-\tilde{m})^2) |A_{(\rho)}^{Jm\tilde{m}}|^2\right)   
 \Bigg\}
  \, .
\end{array}
\end{equation}
Imposing that this expression is zero gives a set of constraints.
While the combination of R-charges is different from the bosonic $\SU(1,1)$   case of Section \ref{sec:bos_su11}, the constraints for the scalars are given by the same set \eqref{eq:Phi_const}, \eqref{eq:Phizero_const} and \eqref{eq:Phi_2and3_zero_constraint} because of the inequalities $|m| \leq J$ and $ |\tilde{m}| \leq J . $
On the other hand, we have now surviving fermionic modes, corresponding to the conditions 
\beq
A=1 \, , \qquad
\kappa = 1 \, , \qquad
m = -J - \frac{1}{2} \, , \qquad
\tilde{m} =  J \, .
\label{eq:solution_m_mtilde_SU_1_1_ferm}
\eeq
All the other fermionic modes, {\sl i.e.}~modes with different $\SU(4)$ index $A$, different value of $\kappa$, or with a different choice of momenta $m, \tilde{m}$, decouple in the $g\rightarrow 0$ limit.
When requiring the compatibility of the constraints with Hamiltonian evolution, we do not obtain additional constraints from the fermionic terms.
In fact the Hamiltonian is always at least quadratic in both the scalars and the fermions, and then weakly commutes with their constraints.
Finally, one sees that the gauge field appears the same way as in Section \ref{sec:bos_su11}, namely that it decouples and only mediates an effective interaction, since it satisfies the constraint \eqref{eq:constraints_gauge_fields_su11bos}.
We summarize here all the constraints:
\begin{align}
\label{eq:constraint1_su111}
&A_{(\rho)}^{Jm\tilde{m}} = -\frac{1}{\omega_{A,J}^2 - (m-\tilde{m})^2} j^{ Jm\tilde{m}}_{(\rho)} \,,\quad\Pi_{(\rho)}^{Jm\tilde{m}} = 0\,,\\
\label{eq:constraint2_su111}
&\Phi_{a=2,3}^{Jm\tilde{m}} = 0\,,\quad\Pi_{a=2,3}^{Jm\tilde{m}}  = 0\,,\\
\label{eq:constraint3_su111}
&\Phi_{1}^{(J,m \ne -J, \tilde{m} \ne J)} = 0\,,\quad\Pi_{1}^{(J,m \ne -J, \tilde{m} \ne J)}  = 0\,,\\
\label{eq:constraint4_su111}
&\Pi_1^{J,-J,J} + i\omega_J \Phi_1^{\dagger \, J,-J,J} = 0 \,, \\
\label{eq:constraint5_su111}
&\psi^{A=1}_{(J,m\ne -J-\frac{1}{2},\tilde{m}\ne J); \kappa = 1} = 0 \spa
\psi^{A=1}_{J,m,\tilde{m}, \kappa =- 1} = 0 \spa
\psi^{A=2,3,4}_{Jm\tilde{m}, \kappa} = 0 
\,.
\end{align}
We notice again that \eqref{eq:constraint4_su111} induces a change of the Dirac brackets for the bosonic modes as in Eq.~\eqref{eq:Dirac_bracket_scalar}, while this does not happen for the fermionic modes.
For this reason we again use the redefinition of the scalar modes \eqref{newphidef}. According to this, we define
\beq
\Phi_{2J} \equiv \sqrt{2 \omega_J} \Phi_1^{J,-J,J} \,, \qquad
\psi_{2J} \equiv \psi^{A=1}_{J,-J-\frac{1}{2},J;\kappa=1} \, ,
\label{eq:abbreviation_fields_su111}
\eeq
which correspond to the surviving degrees of freedom in the $g\rightarrow 0$ limit. The Dirac anti-brackets for the fermionic modes are 
\begin{equation}
\label{eq:Dirac_bracket_fermion}
\Big\{ (\psi_n)^i{}_j , (\psi_n^\dagger)^k{}_l \Big\}_D = i \delta_{n,n'} \delta^i{}_l \delta^k{}_j
\end{equation}
Evaluating now the free Hamiltonian $H_0$ of Eq.~\eqref{H0_su111} on the constraints \eqref{eq:constraint1_su111}-\eqref{eq:constraint5_su111} we find
\begin{equation}
H_0 = \tr \sum_{n=0}^{\infty} \left[ \le n + 1 \ri |\Phi_{n} |^2 + \le n+\frac{3}{2} \ri  |\psi_n|^2 \right] \, , 
\label{eq:free_Hamiltonian_sphere_red_su111}
\end{equation}
We record also the $\SU(1,1)$ charges
\begin{eqnarray}
\label{eq:representation_L0_on_modes_su111}
 L_0 &=& \tr \sum_{n=0}^{\infty} \left[ \le n + \frac{1}{2} \ri |\Phi_{n} |^2 + \le n+1 \ri  |\psi_n|^2 \right] \, ,   \\ \ds
 L_+ &=& (L_-)^{\dagger} = \tr \sum_{n=0}^{\infty} \left[ \le n+1 \ri \Phi^{\dagger}_{n+1} \Phi_n + \sqrt{(n+1)(n+2)} \psi^{\dagger}_{n+1} \psi_n \right] \, . 
\label{eq:representation_L+-_on_modes_su111}
\end{eqnarray}
We notice that $H_0$ is related to $L_0$ as
\begin{equation}
\label{numberop_su111}
H_0 = L_0 + \frac{1}{2} \hat{N} \spa \hat{N} \equiv \tr \sum_{n=0}^\infty (  |\Phi_n|^2 + |\psi_n|^2 )
\end{equation}
$\hat{N}$ commutes with $H_0$, $L_0$ and $L_\pm$ as well as the interaction terms in terms of the brackets \eqref{eq:Dirac_bracket_scalar2} and \eqref{eq:Dirac_bracket_fermion} thus $\hat{N}$ is a conserved charge. As for the bosonic $\SU(1,1)$ case this corresponds to the conservation of particle number and it gives an extra $\U(1)$ symmetry which is a signature of non-relativistic theories. 
Moreover, we can again switch the free part of the Hamiltonian to be $L_0$ instead of $H_0$ since they differ by a conserved quantity.

\subsubsection*{Interactions}

Following the general steps given in Section \ref{sect-sym_on_s3} we need to derive the currents which couple to the matter fields and to integrate out non-dynamical modes which give non-vanishing quartic effective interactions. 
The interacting Hamiltonian in this limit is defined by
\beq
\label{Hint_def_su111}
H_{\rm int} = \lim_{g \rightarrow 0} \frac{H-S_1-Q_1-\frac{1}{2} (Q_2+Q_3)}{g^2 N} \, .
\eeq
The contribution to $H-S_1-Q_1-\frac{1}{2} (Q_2+Q_3) $ of the currents that couple to the gauge field is again given by Eq.~\eqref{eq:gauge_terms}. 
In addition to this, one has the interaction terms recorded in Eq.~\eqref{eq:relevant_Hamiltonian_su11bos}, and from Eq.~\eqref{eq:app_full_interacting_N=4SYM_Hamiltonian},
one sees that one has interactions between the surviving fermionic modes and the modes of the gauge field as well
\beq
  \sum_{JM} \sum_{J_1,J_2} \tr \le  g \mathcal{F}^{\BCJ_1}_{\BCJ_2, JM}  \chi^{JM}
\lbrace \psd_{2J_2} , \psi_{2J_1} \rbrace + g \mathcal{G}^{\BCJ_1}_{\BCJ_2, JM \rho} A_{(\rho)}^{JM}
\lbrace \psd_{2J_2} , \psi_{2J_1} \rbrace \ri \, ,
\label{eq:new_interaction_SU_1_1_ferm}
\eeq
where we introduced the short-hand notation%
\footnote{The $\CF$ and $\CG$ Clebsch-Gordan coefficients are evaluated on the momenta corresponding to the surviving dynamical degrees of freedom of the sector. However, due to the redefinition \eqref{eq:redefinition_fermions} of the fermionic modes with $\kappa=1$, which exchanges a field with its hermitian conjugate, the momenta on which the Clebsch-Gordan coefficients \eqref{FGcoef_modes} are evaluated have opposite signs than the momenta of the modes \eqref{eq:abbreviation_fields_su111}.  Thanks to this modification, we observe that the conditions on momenta coming from triangle inequalities of the Clebsch-Gordan coefficients $\mathcal{F}, \mathcal{G}$ are consistent with momentum conservation as evaluated directly from the creation or annihilation of particles dictated by the field content of the interactions.}
\beq
\label{FGcoef_modes}
\BCJ = ( J,J+\frac{1}{2},-J,\kappa=1)
\eeq
See Appendix \ref{app-properties_Clebsch_Gordan_coeff} for the definitions of the Clebsch-Gordan coefficients $\CF$ and $\CG$.
Finally, there are Yukawa-type terms that couple fermions and the scalar fields 
\begin{multline} 
\label{eq:su111_int}
 -i \frac{g}{\sqrt{2}} \tr \sum_{J,J_1,J_2}  \left\lbrace (-1)^{m-\tilde{m}  + m_2 - \tilde{m}_2 + \tfrac{\kappa_2}{2}} {\cal F}^{J_1,M_1,\kappa_1}_{J_2,-M_2,\kappa_2;J,-M} \left(\psi^\dagger_{J_1,M_1,\kappa_1,4}[\Phi_1^{\dagger\,JM},\psi_{J_2,-M_2,\kappa_2,1}]  \right)  \right. \\
\left. - (-1)^{-m_1 + \tilde{m}_1 + \tfrac{\kappa_1}{2}} {\cal F}^{J_1,-M_1,\kappa_1}_{J_2,M_2,\kappa_2;J,M} \left( \psi^{4}_{J_1,M_1,\kappa_1}[\Phi_1^{JM},\psd_{J_2,-M_2,\kappa_2,1}] \right) - (J_1 \leftrightarrow J_2) \right\rbrace \, ,
\end{multline}
where the antisymmetrization $J_1 \leftrightarrow J_2$ in the last line is referred to all the terms in the interaction. Using the properties of $\CF$ written in  \eqref{eq:property_complex_conjugation_F_coefficients} and \eqref{eq:properties_F_coefficients}  one finds 
\beq
\label{yukawa_terms}
 - i \frac{g}{\sqrt{2}} \tr \sum_{J_1,J_2} \sum_{J,M,\kappa}  2 (-1)^{2 J_1} \mathcal{F}^{\CJ_1}_{J M \kappa ; \CJ_2}  \left\lbrace 
- \psi^\dagger_{J,M,\kappa,4}[\Phi^{\CJ_2 \dagger}_1,\psi_{J_1}] 
 +   \psi^4_{J,M,\kappa}[\Phi^{\CJ_2}_1,\psd_{J_1}] 
 \right\rbrace  \, .
\eeq
With this, we recorded all the interaction terms in the Hamiltonian of $\CN=4$ SYM on $S^3$ which are relevant for this particular limit\footnote{The full interacting Hamiltonian of $\mathcal{N}=4$ SYM action after reduction on the three-sphere is given in Eq.~\eqref{eq:app_full_interacting_N=4SYM_Hamiltonian}.}.

From the terms \eqref{eq:new_interaction_SU_1_1_ferm} which couple to the gauge field we extract the currents 
\bea
\label{eq:bosonic_gauge_current_su111}
& j_0^{\dagger\,Jm\tilde{m}}  = 2 g (1+J_1+J_2) {\cal C}^{\CJ_2}_{\CJ_1,JM} [\Phi^{\CJ_1}_1,\Phi^{\CJ_2 \dagger}_1] + g \mathcal{F}^{\BCJ_1}_{\BCJ_2, JM}  
\lbrace \psi_{2J_1} , \psd_{2J_2} \rbrace \, , & \\
& j_{(\rho)}^{\dagger\,Jm\tilde{m}} =- 4g\sqrt{J_1(J_1+1)} {\cal D}^{\CJ_2}_{\CJ_1, JM\rho}[\Phi^{\CJ_1}_1,\Phi^{\CJ_2 \dagger}_1] + g \mathcal{G}^{\BCJ_1}_{\BCJ_2, JM \rho}
\lbrace \psi_{2J_1} , \psd_{2J_2} \rbrace \, . &
\label{eq:fermionic_gauge_current_su111}
\eea
We can now use this in Eq.~\eqref{eq:gauge_terms} to find the explicit contributions from integrating out the gauge field.
The purely bosonic part, which combines with the quartic scalar self-interaction in Eq.~\eqref{eq:relevant_Hamiltonian_su11bos}, gives as a result the Eq.~\eqref{eq:H_int_bosonicSU(1,1)_before_sum}, and after solving the sum over $J$ one finds Eqs.~\eqref{Hint_bossu11} and \eqref{Hint_bossu11_extra}.

The purely fermionic part of Eq.~\eqref{eq:gauge_terms} combined with \eqref{eq:bosonic_gauge_current_su111}-\eqref{eq:fermionic_gauge_current_su111} gives instead
\beq
\begin{aligned}
\frac{1}{2N} \sum_{JM} \sum_{J_1,J_2,J_3,J_4} & \tr \le - \sum_{\rho = \pm 1} \frac{1}{\omega^2_{A,J}- (m - \tilde{m})^2} \mathcal{G}^{\BCJ_1}_{\BCJ_2, JM \rho} \bar{\mathcal{G}}^{\BCJ_4}_{\BCJ_3, JM \rho}  \right. \\
& \left.  + \frac{1}{4J(J+1)} \mathcal{F}^{\BCJ_1}_{\BCJ_2, JM} \mathcal{F}^{\BCJ_4}_{\BCJ_3, JM}  \ri 
 \lbrace \psd_{2J_2} , \psi_{2J_1} \rbrace \lbrace \psd_{2J_4} , \psi_{2J_3} \rbrace \, ,
\end{aligned}\label{eq:interacting_Hamiltonian_fermSU(1,1)}
\eeq
where $\bar{\mathcal{G}}$ denotes the complex conjugate of $\mathcal{G}$ and we divided with $g^2 N$ as in \eqref{Hint_def_su111}. To evaluate this we use the results of Appendix \ref{app-properties_Clebsch_Gordan_coeff} where we expressed all the above terms with Clebsch-Gordan coefficients using only the $\mathcal{C}$ Clebsch-Gordan coefficient. It is important in this to keep properly track of the constraints on the momenta. All terms in the sum should have 
\beq
m = - \tilde{m} = J_2 - J_1 = J_3 - J_4 \equiv \Delta J \, .
\label{eq:constraints_angular_momenta_ferm_SU(1,1)}
\eeq
On the other hand, the triangle inequalities fix the range of summation of the momentum $J$ and slightly differ for the various terms.
The quadratic combinations in $\mathcal{F}$ impose $|J_1-J_2| \leq J \leq \mathrm{min} (J_1+J_2, J_3 + J_4)$
while the quadratic expressions in $\mathcal{G}$ impose the same conditions when $\rho=1,$ and the different constraints
$|J_1-J_2| \leq J \leq \mathrm{min} (J_1+J_2-1, J_3 + J_4-1)$
when $\rho= - 1$. A remarkable simplification comes now from applying Eq.~\eqref{eq:FGrelation_saturated_momenta}, which upon a shift $J \rightarrow J-1$ in the contribution $\mathcal{G} \bar{\mathcal{G}}$ with $\rho=-1,$ allows to cancel all the terms in the sum except for a single contribution coming from the lower boundary of summation.
This allows to explicitly compute the sum over $J,$ the only remaining term coming exactly from $\mathcal{G} \bar{\mathcal{G}}$ with $\rho=-1$ evaluated at the boundary value $\Delta J$. 
In conclusion, \eqref{eq:interacting_Hamiltonian_fermSU(1,1)} gives the following four-fermion contributions to $H_\text{int}$
\beq
\frac{1}{2N}  \sum_{l=1}^{\infty} \frac{1}{l} \,\, \mathrm{tr} (  \tilde{q}^{\dagger}_l \tilde{q}_l ) 
- \frac{1}{8N} \sum_{n=0}^{\infty} \tr \le \lbrace \psi^{\dagger}_{n} , \psi_{n} \rbrace\tilde{q}_0 \ri  \, ,
\label{eq:interacting_Hamiltonian_su11ferm}
\eeq
where we defined the $\SU(N)$ charge density 
\beq
\label{eq:fermsu11charge}
\tilde{q}_l =  \sum_{n = 0}^\infty \frac{\sqrt{n+1}}{\sqrt{n+l+1}} \lbrace \psd_n , \psi_{n+l} \rbrace \, .
\eeq
Note that the first term in \eqref{eq:interacting_Hamiltonian_su11ferm} arise from $\Delta J \ne 0$. Instead the second term arise from  $\Delta J =0$ and one sees that it is zero on singlet states and therefore does not contribute. Note also that an important difference from the case $\Delta J \ne 0$ is that the $\mathcal{F} \mathcal{F}$ term in \eqref{eq:interacting_Hamiltonian_fermSU(1,1)} has a singular prefactor when $J=0,$ which means that the only remaining contribution from the entire interaction comes from the $\mathcal{G} \bar{\mathcal{G}}$ term with $\rho=1$ evaluated at $J=0$.

Finally, the mixed bosonic-fermionic part of Eq.~\eqref{eq:gauge_terms} combined with \eqref{eq:bosonic_gauge_current_su111}-\eqref{eq:fermionic_gauge_current_su111} gives the following contribution to $H_\text{int}$
\beq
\begin{aligned}
 & \frac{1}{2N} \sum_{JM} \sum_{J_1,J_2,J_3,J_4}  \left\lbrace  \sum_{\rho = \pm 1}  \frac{2 \le \sqrt{J_1(J_1+1)} \mathcal{D}^{\CJ_2}_{\CJ_1, JM \rho} \bar{\mathcal{G}}^{\BCJ_4}_{\BCJ_3, JM \rho} + \sqrt{J_2(J_2+1)}
 \bar{\mathcal{D}}^{\CJ_1}_{\CJ_2, JM \rho} \mathcal{G}^{\BCJ_3}_{\BCJ_4, JM \rho}  \ri}{\omega^2_{A,J}- (m - \tilde{m})^2}   \right. \\
& \left.  + \frac{J_1+J_2+1}{4J(J+1)} \le \mathcal{C}^{\CJ_2}_{\CJ_1, JM} \mathcal{F}^{\BCJ_4}_{\BCJ_3, JM} + \mathcal{C}^{\CJ_1}_{\CJ_2, JM} \mathcal{F}^{\BCJ_3}_{\BCJ_4, JM} \ri  \right\rbrace 
\frac{1}{\sqrt{ \omega_{J_1} \omega_{J_2}}}
\tr \le  [ \Phi_{2J_1} , \Phi^\dagger_{2J_2} ] \lbrace \psi_{2J_3} , \psd_{2J_4} \rbrace \ri \, .
\end{aligned}
\label{eq:interaction_gauge_current_su111}
\eeq
Notice that the two pieces mediated by the gauge field come in pairs, with the constraints 
\beq
m=- \tilde{m} = \Delta J \equiv J_1 - J_2 = J_4 - J_3  \quad \vee   \quad
m=- \tilde{m} =- \Delta J \equiv J_2 - J_1 = J_3 - J_4 \, .
\eeq
For this reason, it is possible to split the result in two parts. 
For both of them, the sum over $J$ can be analytically computed with a similar trick as in the previous cases: we shift the terms with $\rho=-1$ in the sums over $\mathcal{D} \mathcal{G}$ to find that the sum vanishes due to \eqref{eq:app_crossing_relation_gauge_current_su111}, and then we conclude that there is only a contribution from the lower extremum of summation.
Collecting these results and adding the quartic bosonic terms Eqs.~\eqref{Hint_bossu11} and \eqref{Hint_bossu11_extra} and quartic fermionic terms \eqref{eq:interacting_Hamiltonian_su11ferm} one finds the simple result 
 \beq
  \frac{1}{2N} \sum_{l=1}^\infty \frac{1}{l} \tr \le \hat{q}_{l}^{\dagger} \hat{q}_{l} \ri  
 - \frac{1}{8N} \tr (\hat{q}_{0}^2) + \frac{1}{4N} \sum_{n=0}^\infty \frac{1}{n+1}  \tr \le  [\Phi^\dagger_{n},\Phi_{n}] \hat{q}_{0} \ri  
 \, ,
 \label{eq:terms_from_current_su111}
\eeq
where we defined the total $\SU(N)$ current density $\hat{q}_{l}= q_l + \tilde{q}_l$.
This expression is the quartic scalar self-interaction of Eq.~\eqref{eq:relevant_Hamiltonian_su11bos}  plus 
Eq.~\eqref{eq:gauge_terms} with the current given by \eqref{eq:bosonic_gauge_current_su111}-\eqref{eq:fermionic_gauge_current_su111}. We see that only the first term in \eqref{eq:terms_from_current_su111} contributes to the interactions since the Gauss law on the three-sphere means that $\hat{q}_{0}$ is zero.

Finally, we should consider the Yukawa-type terms \eqref{yukawa_terms}. The presence of these interactions imply that the field $\psi_4$ is sourced, and should thus be integrated out.
After doing that, we find the following further contribution to $H_{\rm int}$ 
\beq
 \frac{1}{2N}  \sum_{J,M,\kappa} \sum_{J_1,J_2,J_3,J_4}  \frac{2 (-1)^{2J_4 - 2J_1} \mathcal{F}^{\BCJ_1}_{JM\kappa; \CJ_2} \mathcal{F}^{\BCJ_4}_{JM\kappa ; \CJ_3} }{\sqrt{\omega_{J_2} \omega_{J_3}}(\kappa \omega^{\psi}_J - (m - \tilde{m}))} 
 \tr \le [\Phi_{2J_3},\psd_{2J_4}] [\psi_{2J_1} ,\Phi^\dagger_{2J_2}] \ri \,  .
 \label{eq:interaction_from_integrating_out_psi_4_su_1_1_1}
\eeq
We consider the sum over $J$ by splitting between the cases $J_1 < J_2, J_1 \geq J_2 .$
It turns out that the argument of the sum vanishes once we shift $J \rightarrow J-\frac{1}{2}$ in the term with $\kappa=-1 .$
Then the result reduces to a boundary term when $J_1 \geq J_2, $ while it vanishes when $J_1 < J_2,$ since in the two cases the extremes of summation change. 
The result is%
\footnote{We observe that the physical consequence of having different extremes of summation is that the interaction only contains a particular assignment of momenta: in the quantized theory we annihilate a boson and create a fermion with higher momentum, and at the same time we annihilate a fermion to create a boson with lower momentum.
The reverse possibility is forbidden.}
\beq
\frac{1}{2N} \sum_{J_2, J_3, \Delta J \geq 0} \frac{\tr \le [\Phi_{2J_3},\psd_{2J_3+2\Delta J}] [\psi_{2J_2+2\Delta J},\Phi^\dagger_{2J_2}] \ri}{\sqrt{(2(J_2+\Delta J)+1)(2(J_3+\Delta J)+1)}}  \, ,
\label{eq:sum_over_J_FFterms_su_1_1_1}
\eeq
where we defined $\Delta J \equiv J_1 - J_2 = J_4 - J_3$.

In summary, the effective Hamiltonian in the $g\rightarrow 0$ near-BPS limit towards the BPS bound $H \geq S_1+Q_1+\frac{1}{2}(Q_2+Q_3)$ is 
\begin{equation}
\label{eq:final_ham_su111}
H_{\rm limit} = L_0 + \tilde{g}^2 H_{\rm int}
\end{equation}
with the interaction Hamiltonian \eqref{Hint_def_su111} given by 
\begin{equation}
H_{\rm int}  = \frac{1}{2N} \sum_{l=1}^\infty  \frac{1}{l} \tr \le \hat{q}_{l}^{\dagger} \hat{q}_{l} \ri  
 + \frac{1}{2N} \sum_{l = 0}^{\infty}  \tr ( F_l^\dagger F_l ) \, .
 \label{eq:final_interaction_sphere_reduction_su111}
 \end{equation}
where we defined
\begin{equation}
\label{Fldef}
F_l =  \sum_{m=0}^\infty \frac{ [\psi_{m+l},\Phi_m^\dagger] }{\sqrt{m+l+1}} \,.
\end{equation} 
In Eq.~\eqref{eq:final_interaction_sphere_reduction_su111}
we took into account that all physical configurations have zero $\SU(N)$ charge $\hat{q}_{0}=0$ due to the Gauss law on the three-sphere. Note that $H_{\rm int}$ is manifestly positive.

One can check that the interacting Hamiltonian $H_{\rm int}$ commutes with the number operator $\hat{N}$ of Eq.~\eqref{numberop_su111} as well as the $\SU(1,1)$ charges $L_0$ and $L_\pm$ in \eqref{eq:representation_L0_on_modes_su111}-\eqref{eq:representation_L+-_on_modes_su111} with respect to the Dirac brackets \eqref{eq:Dirac_bracket_scalar2} and \eqref{eq:Dirac_bracket_fermion}. This means the theory has a global $\SU(1,1) \times \U(1)$ invariance. However, this can be enhanced to $\SU(1,1|1) \times \U(1)$ by considering the conserved supercharges. We define
\begin{equation}
\label{eq:CQdef}
\CQ = \sum_{n=0}^\infty \sqrt{\frac{n+1}{2}} \tr \left( \psi_n^\dagger \Phi_{n+1} +\Phi_n^\dagger \psi_n \right) \,.
\end{equation}
One can now show
\begin{equation}
\label{eq:CQprops}
L_0 = \{ \CQ , \CQ^\dagger \}_D  \spa \{ H_{\rm int} , \CQ \}_D = 0 \,,
\end{equation}
using the Dirac brackets \eqref{eq:Dirac_bracket_scalar2} and \eqref{eq:Dirac_bracket_fermion}.
This reveals that the near-BPS theory is supersymmetric.

The non-relativistic nature of the near-BPS theory is apparent from the the conservation of the number operator $\hat{N}$, which is related to the decoupling of anti-particles in the limit as one can see from the constraint Eq.~\eqref{eq:constraint4_su111}. In addition, it is seen by the fact that the surviving dynamical fermion appears with only a fixed choice for the chirality $\kappa=1,$ thus giving a description in terms of a single Grassmann-valued field.
This phenomenon also happens when considering the non-relativistic limit of the Dirac equation in $3+1$ dimensions, since after sending $c \rightarrow \infty$ one of the Weyl spinors composing the Dirac fermion becomes heavy and decouples from the theory, leaving only a single Weyl spinor entering the Schroedinger-Pauli equation.
Such a result can be found by requiring Galilean invariance from first principles and can be generalized to other dimensions \cite{LevyLeblond:1967zz}.
It also applies in the context of null reduction, a procedure that allows to find Bargmann-invariant theories starting from relativistic systems in one higher dimension \cite{Duval:1984cj}. This mechanism works naturally also for non-relativistic supersymmetric theories built from null reduction \cite{Auzzi:2019kdd}. 

In Section \ref{sect-quantization_one_loop_dilatation} we quantize this theory and find that it is equivalent to $\SU(1,1|1)$ Spin Matrix theory. 
In Section \ref{sect-momentum_space_superfields} we show that the natural presence of the supercharge $\CQ$ is related to the fact that one can formulate it in terms of a momentum-space superfield formalism. Finally, in Section \ref{sect-local_formulation} we consider a local formulation of this near-BPS theory and comment on this.

\subsection{Fermionic $\SU(1,1)$ limit - A subcase of $\SU(1,1|1)$}
\label{sec:ferm_su11}

For completeness we consider here briefly the BPS bound $H \geq S_1+\frac{2}{3}(Q_1+Q_2+Q_3)$. The near-BPS limit gives in this case a subsector of the $\SU(1,1|1)$ near-BPS limit in which only the fermionic modes survive. The global symmetry of this theory is $\SU(1,1)\times \U(1)$.

Considering the quadratic terms on the left-hand side of the BPS bound $H - S_1-\frac{2}{3}(Q_1+Q_2+Q_3) \geq 0$ we find
\beq
\begin{aligned}
& H_0 - S_1 - \frac{2}{3} (Q_1+Q_2+Q_3) = \\
& = \sum_{J,m,\tilde{m}}  \tr \Bigg\lbrace
\left|\Pi_a^{Jm\tilde{m}} + i \le \frac{2}{3}+\tilde{m}-m \ri \Phi_a^{\dagger Jm\tilde{m}} \right|^2 + \le \omega_J^2 - \le \frac{2}{3}+\tilde{m}-m \ri^2 \ri | \Phi_a^{Jm\tilde{m}}|^2 
 \\
& + \sum_{\kappa=\pm 1} \left( \le \omega_J^{\psi} + m - \tilde{m} - \kappa \ri \psd_{JM,\kappa, 1} \psi^1_{JM,\kappa} 
+ \sum_{A=2,3,4} \le \omega_J^{\psi} + m - \tilde{m} + \frac{\kappa}{3} \ri \psd_{JM,\kappa, A} \psi^A_{JM,\kappa}
 \right) \\
&  +  \sum_{\rho=-1,1} \frac{1}{2} \left(|\Pi_{(\rho)}^{Jm\tilde{m}} - i(m-\tilde{m})A_{(\rho)}^{\dagger\,Jm\tilde{m}}|^2+ (\omega_{A,J}^2 - (m-\tilde{m})^2) |A_{(\rho)}^{Jm\tilde{m}}|^2\right)   
\Bigg\rbrace \,.  
\end{aligned}
\eeq
As in Sections \ref{sec:bos_su11} and \ref{sec:su111} this provides a set of constraints. Comparing to Section \ref{sec:su111} we find that the constraints are \eqref{eq:constraint1_su111}, \eqref{eq:constraint2_su111} and \eqref{eq:constraint5_su111} with the additional constraints
\beq
\Phi_{1}^{Jm\tilde{m}} =  \Pi_1^{Jm\tilde{m}}  = 0 \,,
\eeq
which means that all scalar fields decouple. The only surviving modes are thus $\psi_{2J} \equiv \psi^{A=1}_{J,-J-\frac{1}{2},J;\kappa=1}$ with the Dirac anti-bracket given by  \eqref{eq:Dirac_bracket_fermion}. Since the gauge field enters in the same way as in the $\SU(1,1|1)$ case of Section \ref{sec:su111} the terms that one obtains from integrating out the gauge field are the same. Thus, one can obtain the interacting Hamiltonian $H_{\rm int}$ of this near-BPS limit simply by be setting the modes $\Phi_{n}=0$ in the $\SU(1,1|1)$ case. We find therefore $H_{\rm limit} = L_0 + \tilde{g}^2 H_{\rm int}$ with
\begin{equation}
H_{\rm int}  = \frac{1}{2N} \sum_{l=1}^\infty \frac{1}{l} \tr \le \tilde{q}_l^{\dagger} \tilde{q}_l\ri  
 \, .
 \label{eq:final_interaction_sphere_reduction_fermsu11}
 \end{equation}
where the $\SU(N)$ charge density is defined by \eqref{eq:fermsu11charge} and 
we took into account that all physical configurations have zero $\SU(N)$ charge $\tilde{q}_0=0$ due to the Gauss law on the three-sphere. The properties of this theory are now inherited from the $\SU(1,1|1)$ case. In particular, $H_{\rm int}$ has the global symmetry with respect to $\SU(1,1)$. Moreover, the number operator $\hat{N}$ is conserved which again is in accordance with this being a non-relativistic theory.

\subsection{$\PSU(1,1|2)$ limit - The maximal case}
\label{sec:su112}

The last BPS bound that we consider is $H \geq S_1 + Q_1 + Q_2$.
The theory emerging from the limit \eqref{newnearBPSlimit} contains interactions with global invariance $ \PSU{(1,1|2)} \times \U(1).$ In particular, it is supersymmetric and includes a $\SU(2)$ residue of the original R-symmetry: this means that we will find again both bosonic and fermionic modes, but now both of them will transform as a doublet under this group. 
In Section \ref{sect-quantization_one_loop_dilatation} we quantize this theory, while in Section \ref{sect-local_formulation} we show that it can be described in terms of local fields.

\subsubsection*{Free Hamiltonian and reduction of the degrees of freedom}

Given the free Hamiltonian $H_0$ and the Cartan charges derived in Appendix \ref{app-quadratic_Hamiltonian_conserved_charges}, we consider the near-BPS bound at lowest order in the coupling, {\sl i.e.} we impose $H_0 - S_1 - Q_1 - Q_2 =0.$
The left hand side reads
\begin{multline}
 H_0 - S_1 - Q_1 - Q_2  = \\
\tr \sum_{JM} \Bigg\{ \sum_{\kappa=\pm 1}  \left[ \sum_{A=1}^4  ( \omega^{\psi}_J + m - \tilde{m}  ) (\psi^A_{JM,\kappa})^\dagger \psi^A_{JM,\kappa} -  \kappa (\psi^{A=1}_{JM,\kappa})^\dagger \psi^{A=1}_{JM,\kappa} + \kappa (\psi^{A=2}_{JM,\kappa})^\dagger \psi^{A=2}_{JM,\kappa}  \right]\\
+ \sum_{a = 1,2}|\Pi_a^{JM} + i(1+\tilde{m}-m)(\Phi_a^\dagger)^{JM}|^2 + (\omega_J^2 - (1+\tilde{m}-m)^2) |\Phi_a^{JM}|^2\\
+ |\Pi_3^{JM} + i(\tilde{m}-m) \Phi_3^{\dagger\,JM}|^2 + (\omega_J^2 - (\tilde{m}-m)^2) |\Phi_3^{JM}|^2 \\
+  \sum_{\rho=-1,1} \frac{1}{2} \left(|\Pi_{(\rho)}^{Jm\tilde{m}} - i(m-\tilde{m})A_{(\rho)}^{\dagger\,Jm\tilde{m}}|^2+ (\omega_{A,J}^2 - (m-\tilde{m})^2) |A_{(\rho)}^{Jm\tilde{m}}|^2\right)   \Bigg\} \, .
\label{eq:quadratic_BPS_limit_su112}
\end{multline}
The vanishing of this expression gives a set of constraints.
The common feature with the other cases is that the gauge field is non-dynamical, since it appears again with the same combination as in Section \ref{sec:bos_su11}, and then gives rise to the same constraints \eqref{eq:constraints_gauge_fields_su11bos}.
On the other hand, now there is more space for scalars and fermions, indeed we find the generalization of Eq.~\eqref{eq:constraint_dynamical_scalars_su11bos}
\beq
\Pi_a^{J,-J,J} + i \omega_J \Phi^{\dagger \, J, -J, J}_a = 0 \, \qquad (a=1,2) \, ,
\eeq
and there are no constraints on the fermionic modes with
\bea
\label{eq:dynamical_ferm_su112_1}
& A=1 \, , \qquad  \kappa=1 \, , \qquad m=-J-\frac{1}{2} \, , \qquad  \tilde{m} = J \, , & \\
& A=2 \, , \qquad  \kappa=-1 \, , \qquad m=-J \, , \qquad  \tilde{m} = J + \frac{1}{2} \, . &
\label{eq:dynamical_ferm_su112_2}
\eea
All the other scalars and fermionic modes decouple in the $g \rightarrow 0$ limit.
In addition, the compatibility with Hamiltonian evolution works in the same way as in the previous cases, {\sl i.e.} no additional constraints are generated, except for the non-trivial Dirac bracket involving the new scalar surviving the limit, in complete analogy with Eq.~\eqref{eq:Dirac_bracket_scalar}
\begin{equation}
\{\Phi_2^{J,-J,J}, (\Phi_2^{J',-J',J'})^\dagger \}_D = \frac{i}{2\omega_J}\delta_{JJ'} \, .
\end{equation}
The entire set of constraints is given by \eqref{eq:constraint1_su111}-\eqref{eq:constraint5_su111}, the only difference being that we need to apply all the previous identities involving the scalar $\Phi_1$ and the fermion $\psi_1$ to the new dynamical modes $\Phi_2, \psi_2 , $ too\footnote{Strictly speaking, the identities involving the two fermions are not the same, because the dynamical modes differ. However, here we mean that all the fermionic modes vanish except for the cases selected by \eqref{eq:dynamical_ferm_su112_1} and \eqref{eq:dynamical_ferm_su112_2}.}.
Indeed, the dynamical bosons and fermions form a doublet under the residual $\SU{(2)}$ R-symmetry.
We remark this explicitly, and we canonically normalize the Dirac brackets of the scalar fields, by introducing the notation
\begin{align}
& \Phi_a^{2J} \equiv \Big( \sqrt{2\omega_J}\Phi_1^{J,-J,J} , \sqrt{2\omega_J}\Phi_2^{J,-J,J} \Big) \, ,  & \\
& \psi^a_{2J} \equiv  \Big( \psi^{A=1}_{J,-J-\tfrac{1}{2},J,\kappa = 1} , \psi^{A=2}_{J,-J,J+\tfrac{1}{2},\kappa = -1} \Big)  \, . & 
\label{eq:dynamical_modes_su_1_1_2}
\end{align}
They will be the dynamical modes entering all the interactions of the sector, with brackets given by Eq.~\eqref{eq:Dirac_bracket_scalar2} and \eqref{eq:Dirac_bracket_fermion} for all the fields in each doublet.

The evaluation of the free Hamiltonian $H_0$ in Eq.~\eqref{eq:quadratic_BPS_limit_su112} on the constraints gives
\beq
H_0 = \tr \sum_{n=0}^{\infty} \left[\le n+ 1 \ri |\Phi^a_n|^2 + \le n+\frac{3}{2} \ri |\psi^a_n|^2 \right] \, ,
\eeq
which is the natural generalization of the quadratic Hamiltonian of the $\SU{(1,1|1)}$ sector.
The $\SU(1,1)$ generators similarly generalize with a $\SU{(2)}$ structure and read
\beq
L_0 = \tr \sum_{n=0}^{\infty}  \left[\le n+ \frac{1}{2} \ri |\Phi^a_n|^2 + \le n+1 \ri |\psi^a_n|^2 \right] \, ,
\eeq
\beq
L_+ = (L_-)^{\dagger} = \tr \sum_{n=0}^{\infty} \left[ (n+1) (\Phi^{\dagger}_a)_{n+1} \Phi^a_n 
+ \sqrt{(n+1)(n+2)} (\psi^{\dagger}_a)_{n+1} \psi^a_n \right] \, .
\eeq
This shows that the free Hamiltonian and $L_0$ are related with a shift by a number operator $\hat{N}$ such that
\beq
H_0 = L_0 + \frac{1}{2} \hat{N} \, , \qquad
\hat{N} \equiv \tr \sum_{n=0}^{\infty} \le |\Phi^a_n|^2 + |\psi^a_n|^2 \ri \, .
\eeq
The number operator $\hat{N}$ is a conserved charge because it commutes with $H_0, L_0 , L_{\pm}$ and the interactions, due to the brackets \eqref{eq:Dirac_bracket_scalar2} and \eqref{eq:Dirac_bracket_fermion}. 
Hence we can define the free part of the Hamiltonian to be $L_0 ;$ the charge $\hat{N}$ and the corresponding invariance of the Hamiltonian correspond to the particle number symmetry typical of non-relativistic theories.

\subsubsection*{Interactions}

The interacting Hamiltonian in this sector is defined by
\beq
\label{Hint_def_su112}
H_{\rm int} = \lim_{g \rightarrow 0} \frac{H-S_1-Q_1-Q_2}{g^2 N} \, .
\eeq
Following the general strategy outlined in Section \ref{sect-sym_on_s3}, we identify the following interactions:
\begin{itemize}
\item Contribution of the currents for the coupling to the non-dynamical gauge field.
\item Quartic scalar self-interaction.
\item Yukawa term, which gives rise to effective quartic interactions after integrating out one of the non-dynamical fields.
\end{itemize}
In principle these possibilities are the same allowed for the $\SU(1,1|1)$ sector, the difference being that from a technical point of view there are more possibilities among the non-dynamical fields to integrate out, and the interactions have an additional $\SU{(2)}$ structure.

We start from the generalization of the currents \eqref{eq:bosonic_gauge_current_su111} and \eqref{eq:fermionic_gauge_current_su111}, which now read
\beq
\label{eq:current_j0_su112}
\begin{aligned}
\hspace{-2mm} j_0^{\dagger\,Jm\tilde{m}}  = & g \,  \frac{J_1+J_2+1}{\sqrt{\omega_{J_1} \omega_{J_2}}} {\cal C}^{\CJ_2}_{\CJ_1,JM} [\Phi^a_{2J_1},(\Phi_a)^\dagger_{2J_2}] \\
& + g \mathcal{F}^{\BCJ_1}_{\BCJ_2, JM}  
\lbrace \psi^1_{2J_1} , (\psi_1)^{\dagger}_{2J_2} \rbrace 
 + g \mathcal{F}^{\BCJ_2}_{\BCJ_1, JM}  
\lbrace \psi^2_{2J_1} , (\psi_2)^{\dagger}_{2J_2} \rbrace  \, , 
\end{aligned}
\eeq
\beq
\label{eq:current_jrho_su112}
\begin{aligned}
\hspace{-2mm} j_{(\rho)}^{\dagger\,Jm\tilde{m}}  = &- 4g \sqrt{\frac{J_1(J_1+1)}{\omega_{J_1} \omega_{J_2}}} {\cal D}^{\CJ_2}_{\CJ_1, JM\rho}[\Phi^a_{2J_1},(\Phi_a)^\dagger_{2J_2}] \\
&  + g \mathcal{G}^{\BCJ_1}_{\BCJ_2, JM \rho}
\lbrace \psi^1_{2J_1} , (\psi_1)^{\dagger}_{2J_2} \rbrace 
- g \mathcal{G}^{\BCJ_2}_{\BCJ_1, JM, -\rho}
\lbrace \psi^2_{2J_1} , (\psi_2)^{\dagger}_{2J_2} \rbrace  \, .
\end{aligned}
\eeq
Here we used Eq.~\eqref{eq:app_relation_between_twoprescriptions_F} and \eqref{eq:relation_between_two_cases_G} to express the result only in terms of the short-hand Clebsch-Gordan coefficients introduced in Eq.~\eqref{eq:short-hand_notation_momenta} and \eqref{FGcoef_modes}, and we immediately rescaled the scalar fields according to the definition \eqref{eq:dynamical_modes_su_1_1_2}.

These currents are singlets under $\SU(2)$ and contributes to the interactions via Eq.~\eqref{eq:gauge_terms}.
Using techniques analog to the method explained in Section \ref{sec:su111} for the $\SU(1,1|1)$ sector by means of the identities given in Appendix \ref{app-properties_Clebsch_Gordan_coeff}, we reduce all the sums over intermediate momenta $J$ to a boundary term.

In order to perform this method for the purely scalar part, however, we also need to include the quartic bosonic self-interaction, which partially contribute to this result.
The corresponding term in the general $\mathcal{N}=4$ SYM Hamiltonian is
\begin{equation}
 \frac{g^2}{2} \tr(\frac{1}{2}([\Phi_1,\Phi_1^\dagger]^2+[\Phi_2,\Phi_2^\dagger]^2) + |[\Phi_1,\Phi_2]|^2 + |[\Phi_1,\Phi_2^\dagger]|^2)\,,
\end{equation}
which we can equivalently write as
\begin{equation}
 \frac{g^2}{2} \tr( \frac{1}{2} |[\Phi_a,(\Phi_a^\dagger)]|^2 + |[\Phi_a,\Phi_b]|^2)
\label{eq:quartic_self_interaction_su112}
\end{equation}
The first term contributes to the effective interactions mediated by the gluons, having the structure of a product of $\SU(2)$ singlets, {\sl i.e.} it has a $\SU(2)$ double trace structure. 
Combining such a term with the formula \eqref{eq:gauge_terms} with currents \eqref{eq:current_j0_su112} and \eqref{eq:current_jrho_su112}, we obtain
\beq
\frac{1}{2N}  \sum_{l=1}^{\infty} \frac{1}{l} \tr \le \hat{q}_l^{\dagger} \hat{q}_l \ri \, ,
\label{eq:sphere_reduction_su112_term_mediated_gluon}
\eeq
where the charge densities are $\hat{q}_l = q_l + \tilde{q}_l$ with
\beq
 q_s \equiv \sum_{n=0}^{\infty} \sum_{a=1,2} [(\Phi_a^{\dagger})_n , (\Phi_a)^{n+l}] \, , \qquad
\tilde{q}_l \equiv \sum_{n=0}^{\infty} \sum_{a=1,2} \frac{\sqrt{n+1}}{\sqrt{n+l+1}}  \lbrace (\psi_a^{\dagger})_n , (\psi_a)^{n+l} \rbrace  \, ,  
\eeq
The other term included in the quartic scalar self-interaction \eqref{eq:quartic_self_interaction_su112} requires some additional care. 
It is a $\SU(2)$ single trace operator, and as such it cannot can be mediated by a $\SU(2)$ singlet. Consequently, it gives rise to a genuinely new interaction of the form
\begin{equation}
\begin{aligned}
 & \frac{1}{4N} \sum_{JM,J_i} \frac{{\cal C}^{JM}_{\CJ_1;\CJ_2}{\cal C}^{J,-M}_{J_3,J_3,-J_3;J_4,J_4,-J_4}}{\sqrt{(2J_1+1)(2J_2+1)(2J_3+1)(2J_4+1)}}\tr([\Phi_a^{2J_1},\Phi_b^{2J_2}][(\Phi_b^\dagger)^{2J_3},(\Phi_a^\dagger)^{2J_4}])\\
&= \frac{1}{2N} \sum_{l,m,n=0}^{\infty} \frac{1}{m+n+l+1}\tr([\Phi_a^{m+l},\Phi_b^{n}][(\Phi_b^\dagger)^{n+l},(\Phi_a^\dagger)^{m}])\,.
\end{aligned}
 \label{eq:purely_bosonic_term_su_1_1_2}
\end{equation}
In this case the sum over $J$ in the first line is trivial because the conditions on momenta saturate the triangle inequalities, and this fixes $J=J_1+J_2 = J_3+J_4.$ 
The second line is then obtained with straightforward shifts and rescalings of the labels.

The remaining interactions of the sector all arise from the Yukawa cubic term in the $\mathcal{N}=4$ action, and they generalize Eq.~\eqref{eq:su111_int}.
Due to the broader field content of the sector with respect to the previous cases, it is now possible to obtain effective quartic interactions which survive the limit by integrating out three different non-dynamical fields: $\psi^3, \psi^4$ or $\Phi_3 .$

We start by integrating out the fermion fields $\psi^3, \psi^4,$ which works conceptually in the same way as for the $\SU(1,1|1)$ sector and brings to an effective interaction analog to Eq.~\eqref{eq:interaction_from_integrating_out_psi_4_su_1_1_1}. 
Since there is an additional $\SU{(2)}$ structure, we find more possible combinations of the fields.
The sum over intermediate momenta $J$ can be performed with a shift $J \rightarrow J-\frac{1}{2}$ in appropriate terms, leading again to a contribution coming from the boundary of summation.
We find
\beq
\begin{aligned}
 &   \frac{1}{2N} \sum_{m, n, l = 0}^{\infty} \frac{\tr \le [(\Phi^\dagger_a)_{m},(\psi_b)_{m+l}] [(\psi_b)^{\dagger}_{n+l},(\Phi_a)_{n}] \ri }{\sqrt{(m+l+1)(n+l+1)}} \\
& -\frac{1}{2N} \sum_{m,n, l = 0}^{\infty} \sqrt{\frac{m +1}{n + l+1}} 
\frac{ \epsilon_{ac}\epsilon_{bd}\tr \le [(\psi_a^{\dagger})_{m} , (\Phi_b)_{m+ l + 1}] [(\psi_c^{\dagger})_{n+ l} , (\Phi_d)_{n} ] \ri }{m+n+ l +2}  \\
& -\frac{1}{2N} \sum_{m,n,l=0}^{\infty} \sqrt{\frac{m+1}{n+ l+1}}  
\frac{ \epsilon_{ac}\epsilon_{bd} \tr \le [(\Phi_a^{\dagger})_{m+ l + 1} , (\psi_b)_{m}] [(\Phi_c^{\dagger})_{n} , (\psi_d)_{n+l} ] \ri }{m+n+ l+2} \, .
\end{aligned}
\label{eq:final_mixed_interaction_su_1_1_2}
\eeq
The last interaction comes from integrating out the non-dynamical scalar $\Phi_3$ from the Yukawa term.
This gives rise to a new quartic combination of purely fermionic fields, whose explicit expression can be worked out by similar manipulations as above, giving
\beq
 \frac{1}{2N} \sum_{m, n, l = 0}^{\infty} \sqrt{\frac{(m +1)(n +1)}{(m  + l +1)(n  +l +1)}}
\frac{\tr \le \lbrace (\psi_a)_{m+ l}, (\psi_b)_{n}   \rbrace 
		\lbrace (\psd_b)_{n + l} , (\psd_a)_{m}  \rbrace  \ri}{m + n + l +2}  \, .
\eeq 
This concludes the treatment of the interacting Hamiltonian of the $\PSU(1,1|2)$ theory.
The full Hamiltonian of the system in the near-BPS limit $g \rightarrow 0$ with bound $H \geq S_1+Q_1+Q_2$ is
\beq
H_{\rm limit} = L_0 + \tilde{g}^2 H_{\rm int} \, .
\label{eq:full_Hamiltonian_su112}
\eeq
The interacting Hamiltonian is obtained by using the definition \eqref{Hint_def_su112} and collecting all the previous terms.
Remarkably, the final expression can be written in a convenient form showing that it is manifestly positive definite by means of the property
\beq
\tr \le \lbrace \psi_1 , \psi_2 \rbrace [\Phi_1^{\dagger}, \Phi_2^{\dagger}] \ri = 
\tr \le [\psi_1 , \Phi_1^{\dagger}] [\psi_2 , \Phi_2^{\dagger}] \ri
- \tr \le [\psi_1, \Phi_2^{\dagger}] [\psi_2, \Phi_1^{\dagger}] \ri \, .
\eeq
Thus we obtain
\beq
\begin{aligned}
	H_{\rm int}  = \frac{1}{2N} \sum_{l=1}^{\infty}  \frac{1}{l} \tr \le \hat{q}_l^{\dagger} \, \hat{q}_l  \ri
	+ \frac{1}{2N} \sum_{l = 0}^{\infty}  \tr ( (F_{ab})_l^\dagger (F_{ab})_l )
+ \frac{1}{2N}	\sum_{l=1}^{\infty} 
 \tr ( (G_{ab})_l^\dagger (G_{ab})_l )
	 \, ,
	\end{aligned}
\label{eq:sphere_reduction_su112_result_refined}
\eeq
where we defined
\begin{equation}
(F_{ab})_l =  \sum_{m=0}^\infty \frac{ [(\psi_a)_{m+l},(\Phi_b)_m^\dagger] }{\sqrt{m+l+1}} \, , \qquad
(G_{ab})_l =   \frac{1}{\sqrt{l}}  (j_{ab})_l  \, ,
\end{equation} 
\beq
(j_{ab})_{l} =
 \sum_{m, n=0}^{\infty} \le  \sqrt{\frac{m+1}{n+1}} \lbrace (\psi_a)_{m}, (\psi_b)_{n}  \rbrace \delta (m + n +2 -l)  + [(\Phi_a)_{m} , (\Phi_b)_{n}] \delta (m + n +1 -l) \ri \, .
 \label{eq:current_su112}
\eeq
The expression \eqref{eq:sphere_reduction_su112_result_refined} is invariant under the global group $\SU{(1,1)} \times \U{(1)},$ as can be checked explicitly by computing the commutators with the particle number operator $\hat{N}$ and with the charges $L_0, L_{\pm} .$
In addition, it is also invariant under extended supersymmetry, with supercharges
\beq
\begin{aligned}
\label{eq:CQdef2}
& \mathcal{Q}_1 = \sum_{a=1,2} \sum_{n=0}^{\infty} \sqrt{\frac{n+1}{2}} \tr \le (\psi^{\dagger}_a)_n (\Phi^a)_{n+1} 
+ (\Phi^{\dagger}_a)_n (\psi^a)_n \ri \, ,  & \\
& \mathcal{Q}_2 = \sum_{a,b=1,2} \sum_{n=0}^{\infty} \sqrt{\frac{n+1}{2}} \epsilon^{ab} \tr \le (\psi^{\dagger}_a)_n (\Phi_b)_{n+1} 
+ (\Phi^{\dagger}_b)_n (\psi_a)_n \ri \, . &
\end{aligned}
\eeq
satisfying
\beq
L_0 = \lbrace \CQ_1, \CQ^{\dagger}_1 \rbrace_D =
\lbrace \CQ_2, \CQ^{\dagger}_2 \rbrace_D  \, , \qquad
\lbrace H_{\rm int} , \CQ_1 \rbrace_D = \lbrace H_{\rm int} , \CQ_2 \rbrace_D = 0 \, .
\eeq
This can be shown to be true by using the Dirac brackets \eqref{eq:Dirac_bracket_scalar2} and \eqref{eq:Dirac_bracket_fermion} for all the copies of the fields.
The same comments given in Section \ref{sec:su111} about the non-relativistic nature of the model are true.
In addition, we observe that the broader field content of this near-BPS limit allows for a set of new interactions in the last two lines of Eq.~\eqref{eq:sphere_reduction_su112_result_refined}, where the distribution of momenta between the bosonic and fermionic degrees of freedom is shifted by unity.
This aspect is related to the fact that the scalars transform under the $j=1/2$ representation of $\SU(1,1),$ while the fermionic field under the $j=1$ representation.
We will investigate in more details the consequences of this observation in Section \ref{sect-local_formulation}, where this will play an important role to determine the local description of the sector.


\section{Quantization of near-BPS theories}
\label{sect-quantization_one_loop_dilatation}

In Section \ref{sect-sphere_reduction} we found non-relativistic theories that describe the effective dynamics of $\CN=4$ SYM near BPS bound, when taking the near-BPS limit \eqref{newnearBPSlimit}. These theories are classical, as they arise from limits of the classical Hamiltonian of $\mathcal{N}=4$ SYM on a three-sphere.
In this section we consider the quantization of the near-BPS theories that we have obtained.

In Section \ref{sect-quantize_near_BPS} we quantize the $\SU(1,1|1)$ near-BPS theory and find its full quantum mechanical Hamiltonian and the Hilbert space on which it acts. We show that the quantized theory includes normal-ordering effects that can be viewed as self-energy corrections. With these effects included, we review in Section \ref{sect-quantize_near_BPS} that the quantized $\SU(1,1|1)$ near-BPS theory is equivalent to $\SU(1,1|1)$ Spin Matrix theory \cite{Harmark:2014mpa}. As we explain in Section \ref{sect-quantization_vs_nearBPS}, this means that taking the near-BPS limit \eqref{newnearBPSlimit} on the level of the classical Hamiltonian of $\CN=4$ SYM on a three-sphere, as we did in Section \ref{sect-sphere_reduction}, and then quantizing the resulting near-BPS theory, is equivalent to first quantizing $\CN=4$ SYM on a three-sphere, and then taking the near-BPS limit \eqref{newnearBPSlimit} of the quantum Hamiltonian, which is equivalent to the dilatation operator of $\CN=4$ SYM.
Thus, one gets the same result whether one first quantizes, and then takes the near-BPS limit, or if one first takes the near-BPS limit, and then quantizes.

We stress that since one can view the bosonic and fermionic $\SU(1,1)$ near-BPS theories as truncations of the $\SU(1,1|1)$ near-BPS theory, the conclusions we draw for the $\SU(1,1|1)$ case will hold for these cases as well. We furthermore comment on the extension to the $\PSU(1,1|2)$ case.

\subsection{Quantization of $\SU(1,1|1)$ near-BPS theory}
\label{sect-quantize_near_BPS}

We perform now the complete quantization procedure for the $\SU(1,1|1)$ near-BPS theory. This theory is rich enough to show the appearance of non-trivial contributions from the normal ordering of both the bosonic and fermionic terms in the Hamiltonian, and it is also the simplest case where supersymmetry arises.
The procedure can be straightforwardly generalized to include the new interactions of the $\PSU(1,1|2)$ sector as no additional subtleties arise.

First of all, we replace all the Dirac brackets with (anti)commutators 
\beq
\lbrace \cdot, \cdot \rbrace_D \rightarrow i [ \cdot , \cdot \rbrace  \, , 
\eeq
where we denoted with $\lbrace \rbrace_D $ in the LHS the classical brackets and in the RHS the notation stresses that the symmetry depends from the bosonic or fermionic nature of the fields involved.
Then we introduce raising and lowering operators obeying the canonical commutation relations
\beq
[ (a_r)^i_{\,\, j} , (a^{\dagger}_s)^{k}_{\,\, l}  ] = \delta^i_{\,\, l} \delta^k_{\,\, j} \delta_{rs} \, , \qquad
\lbrace (b_r)^i_{\,\, j} , (b^{\dagger}_s)^{k}_{\,\, l}  \rbrace = \delta^i_{\,\, l} \delta^k_{\,\, j} \delta_{rs} \, , 
\label{eq:commutation_relations_ladder_operators_SMT}
\eeq
where $ a_s \equiv \Phi_s , \, a^{\dagger}_s \equiv \Phi^{\dagger}_s $ are bosonic, and $ b_s \equiv \psi_s , \, b^{\dagger}_s \equiv \psi^{\dagger}_s  $ are fermionic.
These oscillators carry indices $i,j$ for the internal $\SU(N)$ symmetry and an index $s$ corresponding to a representation of the 
spin group $\SU(1,1|1) .$

Using this dictionary, we directly promote the classical result \eqref{eq:final_interaction_sphere_reduction_su111} to a quantum-mechanical Hamiltonian 
\beq
\begin{aligned}
H_{\rm qm} & = \tr \le \sum_{s=0}^{\infty} \le s + \frac{1}{2} \ri a^{\dagger}_s a_s + \sum_{s=0}^{\infty} (s+1) b^{\dagger}_s b_s  
 + \frac{\tilde{g}^2}{2N} \sum_{s=1}^{\infty} \frac{1}{s}  (q_s^{\rm tot})^{\dagger} q_s^{\rm tot}  \ri    \\
&  + \frac{\tilde{g}^2}{2N} \sum_{s,s_1, s_2=0}^{\infty}  \frac{1}{\sqrt{(s_1+s+1)(s_2+s+1)}} \tr \le [a_{s_2},b^{\dagger}_{s_2+s}] [b_{s_1+s},a^\dagger_{s_1}] \ri \, ,
\end{aligned}
\eeq
where we defined the quantum version of the charge densities as
\beq
q_l \equiv \sum_{s=0}^{\infty} :[a^\dagger_s , a_{s+l}]: \, , \qquad
\tilde{q}_l = \sum_{s=0}^{\infty} \frac{\sqrt{s+1}}{\sqrt{s+l+1}} :\lbrace b^{\dagger}_s , b_{s+l} \rbrace: \, , \qquad
\hat{q}_{l} = q_l + \tilde{q}_l \, .
\label{eq:definition_charge_densities_SMT_Hamiltonian_su111}
\eeq
At the classical level, the zero mode of the total current $\hat{q}_0$ vanishes due to the Gauss law on the three-sphere.
At the quantum-mechanical level, $\hat{q}_0$ is zero when acting on physical states 
\beq
\hat{q}_0 | \mathrm{phys} \rangle = 0 \, .
\eeq
Hence, the Hilbert space of the quantum theory corresponds to the states which are singlets with respect to the $\SU(N)$ symmetry.
Now we show that normal ordering is responsible for the appearance of the self-energy corrections.
The result for the $\SU(1,1)$ bosonic sector was derived in \cite{Harmark:2019zkn}, but here we review and generalize the procedure including also the fermionic partner.

For the bosonic part, the following result can be obtained by using the commutation relations \eqref{eq:commutation_relations_ladder_operators_SMT} in the explicit evaluation of the normal ordered interaction:
\beq
\sum_{s=1}^{\infty} \frac{1}{s} \tr \le q^{\dagger}_s q_s \ri = 
\sum_{l=1}^{\infty} \frac{1}{s} \tr \le :q^{\dagger}_s q_s: \ri 
+ 2 N \sum_{s=0}^{\infty} h(s) \tr \le a^{\dagger}_s a_s \ri - 2 \sum_{s=0}^{\infty} h(s) \tr ( a^{\dagger}_s ) \tr (a_s) \, .
\label{eq:self_energy_corrections_bosons}
\eeq
Here we defined the harmonic numbers as
$
h(s) = \sum_{k=1}^s \frac{1}{k} \, .
$
An analog computation applied to the fermionic part of the Hamiltonian gives the similar relation
\beq
\sum_{s=1}^{\infty} \frac{1}{s} \tr \le \tilde{q}^{\dagger}_s \tilde{q}_s \ri = 
\sum_{l=1}^{\infty} \frac{1}{s} \tr \le : \tilde{q}^{\dagger}_s \tilde{q}_s: \ri 
+ 2 N \sum_{s=0}^{\infty} h(s+1) \tr \le b^{\dagger}_s b_s \ri + 2 \sum_{s=0}^{\infty} h(s+1) \tr ( b^{\dagger}_s ) \tr (b_s) \, .
\label{eq:self_energy_corrections_fermions}
\eeq
In this case, the different argument of the harmonic numbers comes from the identity
\beq
\sum_{l=1}^{\infty} \frac{1}{l} \frac{s+1}{s+l+1} = h(s+1) \, ,
\eeq
which in turn arises from the normalization of the fermionic ladder operators.
In the $\SU(1,1|1)$ sector there are also mixed bosonic-fermionic interactions, but they do not contribute to self-energy corrections, as can be checked explicitly: 
\beq
\sum_{s=1}^{\infty} \frac{1}{s} \tr \le q^{\dagger}_s \tilde{q}_s + \tilde{q}^{\dagger}_s q_s \ri =
\sum_{s=1}^{\infty} \frac{1}{s} \tr \le : q^{\dagger}_s \tilde{q}_s + \tilde{q}^{\dagger}_s q_s : \ri \, .
\label{eq:self_energy_corrections_mixed}
\eeq
To proceed further, we need to work out the implication of the $\SU(N)$ singlet constraint, which implicitly enters the Hamiltonian as the term with $s=0$ in the quartic interactions mediated by the non-dynamical gauge field. 
For the bosonic case, we need to use the identity\footnote{Strictly speaking, the singlet constraint involves the total charge density $\hat{q}_0$ and not the single terms $q_0, \tilde{q}_0 .$ However the mixed terms do not have normal ordering issues because the bosonic operators commute with the fermionic ones, hence it is not restrictive to consider only the diagonal terms in the computation of the self-energy corrections.}
\beq
\begin{aligned}
\sum_{m=0}^{\infty} h(m) \tr \le :[a^{\dagger}_m , a_m]: q_0 \ri = &
\sum_{m,n = 0}^\infty h(m) \tr(:\left[a_{m}^\dagger,a^m\right] \left[a_{n}^\dagger, a^n\right]:) \\
& + 2 N \sum_{m=0}^\infty h(m) \tr(a_{m}^\dagger a^m) - 2 \sum_{m=0}^\infty h(m) \tr(a_{m}^\dagger)\tr(a^m)\,,
\end{aligned}
\label{eq:singlet_condition_boson_su111}
\eeq
while for the fermionic case the analog expression is
\begin{multline}
\sum_{m=0}^{\infty} h(m+1) \tr(:\left\{b_{m}^\dagger,b^m\right\}: \tilde{q}_0) = \sum_{m,n=0}^{\infty} h(m+1) \tr(:\left\{b_{m}^\dagger,b^m\right\}\left\{b_{n}^\dagger, b^n\right\}:) \\ 
 + 2 N \sum_{m=0}^{\infty} h(m+1) \tr(b_m^\dagger b^m) + 2 \sum_{m=0}^{\infty} h(m+1) \tr(b_m^\dagger) \tr(b^m)\,.
 \label{eq:singlet_condition_fermion_su111}
\end{multline}
No additional self-energy corrections arise instead from the mixed bosonic-fermionic interaction.
The crucial observation is that all the self-energy terms cancel when summing the right-hand side of
Eqs.~\eqref{eq:self_energy_corrections_bosons}, \eqref{eq:self_energy_corrections_fermions}, \eqref{eq:self_energy_corrections_mixed} with Eqs.~\eqref{eq:singlet_condition_boson_su111} and \eqref{eq:singlet_condition_fermion_su111}. 

The quartic interaction which was mediated by the non-dynamical fermionic field (as explained in Section \ref{sec:su111}) is already normal-ordered.
In this way the quantum Hamiltonian of the near-BPS $\SU(1,1|1)$ theory becomes
\beq
\begin{aligned}
 H_{\rm qm} & =  \sum_{s=0}^{\infty} \le s + \frac{1}{2} \ri \tr \le a^{\dagger}_s a_s \ri + \sum_{s=0}^{\infty} (s+1) \tr \le b^{\dagger}_s b_s  \ri
 + \frac{\tilde{g}^2}{2N} \sum_{s=1}^{\infty} \frac{1}{s} \tr \le : \hat{q}_s^{\dagger} \hat{q}_s : \ri    \\
& + \frac{\tilde{g}^2}{2N} \sum_{s_1,s_2 = 0}^\infty h(s_1) \tr(:\left[a_{s_1}^\dagger,a^{s_1}\right] \left[a_{s_2}^\dagger, a^{s_2}\right]:) 
 + \frac{\tilde{g}^2}{2N} \sum_{s_1,s_2=0}^{\infty} h(s_1+1) \tr(:\left\{b_{s_1}^\dagger,b^{s_1}\right\}\left\{b_{s_2}^\dagger, b^{s_2}\right\}:)   \\
 & + \frac{\tilde{g}^2}{4N} \sum_{s_1, s_2=0}^{\infty}   h(s_1+1)  \tr \le :\lbrace b^{\dagger}_{s_1} , b^{s_1} \rbrace [a^{\dagger}_{s_2}, a^{s_2}]: \ri 
 + \frac{\tilde{g}^2}{4N} \sum_{s_1, s_2=0}^{\infty}   h(s_1)  \tr \le :[a^{\dagger}_{s_1}, a^{s_1}] \lbrace b^{\dagger}_{s_2} , b^{s_2} \rbrace : \ri  \\
&  + \frac{\tilde{g}^2}{2N} \sum_{s,s_1, s_2=0}^{\infty}  \frac{1}{\sqrt{(s_1+s+1)(s_2+s+1)}} \tr \le :[a_{s_2},b^{\dagger}_{s_2+s}] [b_{s_1+s},a^\dagger_{s_1}]: \ri \, .
\end{aligned}
\label{eq:quantized_normal_ordered_H_su111}
\eeq
We will see in the following that this is equivalent to $\SU(1,1|1)$ SMT \cite{Harmark:2014mpa}.

\subsection{Quantization vs near-BPS limit}
\label{sect-quantization_vs_nearBPS}

\begin{figure}
\centering
\includegraphics[scale=0.4]{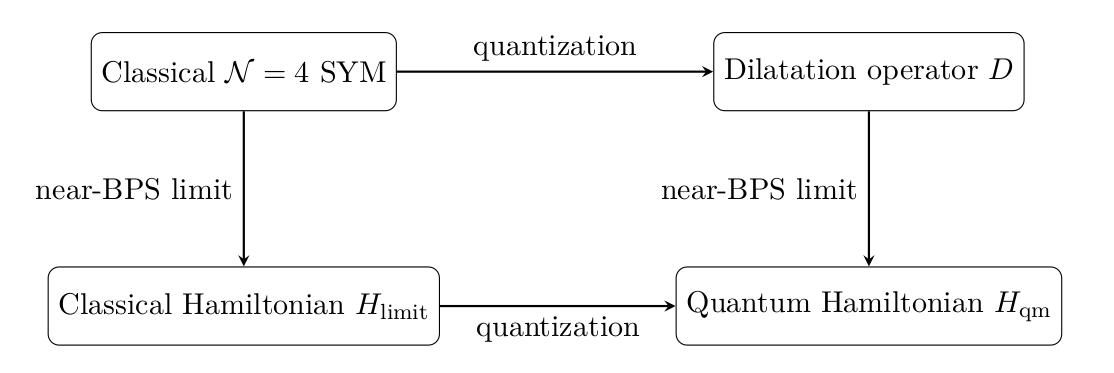}
\caption{\small Commutative diagram representing two different ways to obtain the same quantum theory. One path is to first take the near-BPS limit classically, and then to quantize the theory. That is the path of this paper performed in Sections \ref{sect-sphere_reduction} and \ref{sect-quantize_near_BPS}. The other path is to first quantize $\CN=4$ SYM and then take the near-BPS limit. This corresponds to the Spin Matrix theory limit of \cite{Harmark:2014mpa}. Both paths yield the same result. }
\label{fig:commutative_diagram}
\end{figure}

Above in Section \ref{sect-sphere_reduction} we have taken near-BPS limits \eqref{newnearBPSlimit} of classical $\mathcal{N}=4$ SYM on a three-sphere, to obtain a classical description of the near-BPS dynamics close to certain BPS bounds. Subsequently we quantized the resulting near-BPS theory, specifically in the $\SU(1,1|1)$ case, to obtain the quantum Hamiltonian \eqref{eq:quantized_normal_ordered_H_su111} in Section \ref{sect-quantize_near_BPS}. This was done by using a standard normalordering prescription. 
In this way we found a quantum Hamiltonian that effectively describes a lower-dimensional theory with non-relativistic symmetries. The route to obtain this result is illustrated in one of the paths in the diagram of Figure \ref{fig:commutative_diagram}.

As we shall see in this section, there is another route to the same result, also illustrated in Figure \ref{fig:commutative_diagram}. In this case, we start by quantizing $\mathcal{N}=4$ SYM on a three-sphere. The quantum Hamiltonian is then given by the full dilatation operator $D$ of $\mathcal{N}=4$ SYM on $\mathbb{R}^4$ \cite{Beisert:2003tq, Beisert:2003jj, Beisert:2004ry} (by the state/operator correspondence). One can then subsequently take the same near-BPS limit \eqref{newnearBPSlimit} as for the classical description. Amazingly, as we show in detail below, this will reveal the exact same quantum theory. Thus, in short, quantizing and taking near-BPS limit commute with each other. 

The alternative route, with quantizing first and then taking the near-BPS limit, has been previously explored in \cite{Harmark:2014mpa} and in references therein. In these works, the near-BPS limit is  known as the Spin Matrix theory (SMT) limit and the resulting quantum theory as Spin Matrix theory (SMT). Thus, we show in this paper a different route to obtain SMT.

That quantization and near-BPS limit commutes, as illustrated in Figure \ref{fig:commutative_diagram}, is not a priori evident.
The commutativity of the limits is particularly non-trivial for non-relativistic theories: in fact it is known that procedures like the $c \rightarrow \infty$ limit or null reduction do not commute a priori with the quantization of the theory, or with other generic limits that one can perform in such systems.
For our near-BPS limits, however, we show that the diagram in Fig. \ref{fig:commutative_diagram} is commutative, {\sl i.e.} the two prescriptions lead to the same result.
A posteriori, this matching justifies the prescription given for the quantization of the classical result coming from the sphere reduction.

To exhibit the connection to the SMT/near-BPS limits of the full dilatation operator $D$ of $\CN=4$ SYM we focus on the $\SU(1,1|1)$ case for which we found the quantum Hamiltonian \eqref{eq:quantized_normal_ordered_H_su111}. 
The focus here is on the interacting part  which should be compared to the $\SU(1,1|1)$ sector of the one-loop contribution to the full dilatation operator $D$. 
Writing
\begin{equation}
H_{\rm qm} = \sum_{s=0}^{\infty} \le s + \frac{1}{2} \ri \tr \le a^{\dagger}_s a_s \ri + \sum_{s=0}^{\infty} (s+1) \tr \le b^{\dagger}_s b_s \ri + \tilde{g}^2 H_{\rm qm,int}
\end{equation}
we are interested in the quantum interacting Hamiltonian  $H_{\rm qm,int}$. 
It is convenient to write it in terms of renormalized four-point vertices,
\beq
\begin{aligned}
\label{oneloop}
H_{\rm qm,int} & = \frac{1}{4N} \sum_{s=0}^{\infty} \sum_{s_1,s_2=0}^{s} \tr \le :[a^{\dagger}_{s_1},a_{s_2}] [a^{\dagger}_{s_1-s},a_{s_2-s}]: \ri
\le \delta_{s_1,s_2} \le h(s_1) + h(s_2-s) \ri -  \frac{ 1-\delta_{s_1,s_2}}{|s_1-s_2|} \ri \\
& + \frac{\tilde{g}^2}{4N} \sum_{s=0}^{\infty} \sum_{s_1,s_2=0}^{s} \sqrt{\frac{(s_1+1)(s_2-s+1)}{(s_2+1)(s_1-s+1)}}
 \tr \le :\lbrace b^{\dagger}_{s_1},b_{s_2}\rbrace \lbrace b^{\dagger}_{s_1-s},b_{s_2-s}\rbrace: \ri \\
 & \times \le \delta_{s_1,s_2} \le h(s_1+1) + h(s_2-s+1) \ri -  \frac{ 1-\delta_{s_1,s_2}}{|s_1-s_2|} \ri \\
& +\frac{\tilde{g}^2}{4N} \sum_{s=0}^{\infty} \sum_{s_1,s_2=0}^{s} \sqrt{\frac{s_1-s+1}{s_2-s+1}} \tr \le :[a^{\dagger}_{s_1},a_{s_2}] \lbrace b^{\dagger}_{s_1-s},b_{s_2-s}\rbrace: \ri
\le \delta_{s_1,s_2} h(s_1) - \frac{ 1-\delta_{s_1,s_2}}{|s_1-s_2|} \ri \\
& +\frac{\tilde{g}^2}{4N} \sum_{s=0}^{\infty} \sum_{s_1,s_2=0}^{s} \sqrt{\frac{s_2+1}{s_1+1}} \tr \le :\lbrace b^{\dagger}_{s_1},b_{s_2}\rbrace [a^{\dagger}_{s_1-s},a_{s_2-s}]: \ri
\le \delta_{s_1,s_2} h(s_1+1) - \frac{ 1-\delta_{s_1,s_2}}{|s_1-s_2|} \ri \\
&  + \frac{\tilde{g}^2}{2N} \sum_{s,s_1, s_2=0}^{\infty}  \frac{1}{\sqrt{(s_1+s+1)(s_2+s+1)}} \tr \le :[a_{s_2},b^{\dagger}_{s_2+s}] [b_{s_1+s},a^\dagger_{s_1}]: \ri \, .
\end{aligned}
\eeq
The terms in the right parenthesis for each vertex correspond exactly to the one-loop dilatation operator in the $\SU(1,1|1)$ sector \cite{Beisert:2003jj}. 
Thus, we have shown the commutativity in the diagram of Figure \ref{fig:commutative_diagram}. This means that the quantum Hamiltonian \eqref{eq:quantized_normal_ordered_H_su111} indeed is that of $\SU(1,1|1)$ SMT. 

It is interesting to note that the computation of \eqref{oneloop}, from point of view of the one-loop dilatation operator of $\CN=4$ SYM, is quite involved. One has to compute divergent Feynman diagrams for $\mathcal{N}=4$ SYM and perform dimensional regularization for two-point functions. Instead, we have found the same result from a classical computation, {\sl i.e.} the near-BPS limit of the sphere reduction of the action on $\mathbb{R} \times S^3 $, along with a simple normal-ordering prescription to obtain the Hamiltonian at the quantum level.

\subsubsection*{Generalization to the $\PSU(1,1|2)$ sector}

Here we comment on the generalization of the result to the $\PSU(1,1|2)$ sector.
In this case, the bosonic and fermionic fields are both supplemented by an additional $\SU(2)$ index due to the residual R-symmetry of the system.
We then define the ladder operators as
\beq
(a_a)_s \equiv (\Phi_a)_s \, , \qquad
(a_a^{\dagger})_s \equiv (\Phi^{\dagger}_a)_s \, , \qquad
(b_a)_s \equiv (\psi_a)_s \, , \qquad
(b^{\dagger}_a)_s \equiv (\psi^{\dagger}_a)_s \, .
\eeq
The prescription to quantize the Hamiltonian coming from the sphere reduction is still to directly promote the result at quantum level, without further changes.
This implies that the interacting part is given by
\beq
\begin{aligned}
	H_{\rm qm, int}  = \frac{1}{2N} \sum_{l=1}^{\infty}  \frac{1}{l} \tr \le \hat{q}_l^{\dagger} \, \hat{q}_l  \ri
	+ \frac{1}{2N} \sum_{l = 0}^{\infty}  \tr ( (F_{ab})_l^\dagger (F_{ab})_l )
+ \frac{1}{2N}	\sum_{l=1}^{\infty} 
 \tr ( (G_{ab})_l^\dagger (G_{ab})_l )
	 \, ,
	\end{aligned}
\eeq
where now we have
\beq
 q_l \equiv \sum_{a=1,2} \sum_{n=0}^{\infty} :[(a_a^{\dagger})_n , (a_a)^{n+l}]: \, , \quad
\tilde{q}_l \equiv \sum_{a=1,2} \sum_{n=0}^{\infty} \frac{\sqrt{n+1}}{\sqrt{n+l+1}}  :\lbrace (b_a^{\dagger})_n , (b_a)^{n+l} \rbrace:   \, ,
\eeq
\begin{equation}
(F_{ab})_l =  \sum_{m=0}^\infty \frac{ [(b_a)_{m+l},(a_b)_m^\dagger] }{\sqrt{m+l+1}} \, , \qquad
(G_{ab})_l =   \frac{1}{\sqrt{l}}  (j_{ab})_l  \, ,
\end{equation} 
\beq
(j_{ab})_{l} =
 \sum_{m, n=0}^{\infty} \le  \sqrt{\frac{m+1}{n+1}} \lbrace (b_a)_{m}, (b_b)_{n}  \rbrace \delta (m + n +2 -l)  + [(a_a)_{m} , (a_b)_{n}] \delta (m + n +1 -l) \ri \, .
\eeq
Working out the $\SU(N)$ singlet condition and writing all the expressions in terms of normal ordered quantities allows to recast the result in a form where the vertices are renormalized in the same way as computed from the one-loop corrections to the dilatation operator in this sector.
The procedure is completely analog to the $\SU(1,1|1)$ case, and we simply need to complement the result with the additional $\SU(2)$ structure.


\section{Momentum-space superfield formalism}
\label{sect-momentum_space_superfields}

The spin group of the SMT Hamiltonian in the $\SU(1,1|1)$ limit is supersymmetric, i.e. it admits the existence of a complex supercharge relating the bosonic and fermionic dynamical degrees of freedom surviving the near-BPS limit of $\mathcal{N}=4$ SYM. 
It is then reasonable to expect that there exists a suitable superspace formulation which makes this invariance manifest and allows to reproduce the field content and the Hamiltonian in terms of superfields.

Indeed, we now show in detail that this is possible.
We stress that while we will give a semi-local description of this model in section \ref{sect-local_formulation}, it should be considered as a complementary way to describe the system, but not as a necessary step.
In fact, all the expressions that we are going to introduce in this section can be considered independently as a way to obtain the classical Hamiltonian \eqref{eq:final_ham_su111}.
 
Following the discussion of Section \ref{sec:su111}, it is convenient to use $L_0$ as the free part of the Hamiltonian; this reads
\beq
L_0 = \int dt \, \sum_{s=0}^{\infty} \tr  \le \le s+\frac{1}{2} \ri \Phi^{\dagger}_s \Phi_s + (s+1) \psi^{\dagger}_s \psi_s \ri  \, .
\label{eq:free_Hamiltonian_momentum_space_components_su111}
\eeq
The eigenvalues are explicitly given by $s+1-R,$ with $R=(\frac{1}{2},0)$ being the $U(1)_R$ charge of bosons and fermions\footnote{See Appendix \ref{app-algebra_osc_repr} for more details on the R-charge and the generators in the oscillator representation.}, respectively.

Since the sector contains only a single complex supercharge, a corresponding superspace formulation accordingly requires the introduction of a single complex Grassmannian coordinate $(\theta, \theta^{\dagger}).$
Moreover, the requirement that the anticommutator of supercharges closes on $L_0$ fixes their expressions to be
\beq
\CQ = \frac{\partial}{\partial \theta} + \frac{1}{2} \theta^{\dagger} (s+1-R) \, , \qquad
\CQ^{\dagger} = \frac{\partial}{\partial \theta^{\dagger}} + \frac{1}{2} \theta (s+1-R) \, ,
\label{eq:supercharge_momentum_space}
\eeq
which indeed satisfy
\beq
\lbrace  \CQ, \CQ^{\dagger} \rbrace = s+1-R = L_0 \, .
\eeq
The most general superfield that we can define in a superspace with one complex Grassmann coordinate is given by
\beq
X_s (t, \theta, \theta^{\dagger}) = A_s (t) + \theta B_s (t) + \theta^{\dagger} C_s (t)  + \theta \theta^{\dagger} D_s (t) \, .
\eeq
The component modes appearing in the definition of the superfield can have a priori both bosonic or fermionic statistics.
In particular, in two dimensions both choices are allowed. We will distinguish the two possibilities by calling the superfields either bosonic or fermionic depending from the behaviour of the lowest component.
In fact, fixing the statistics of $A_s (t)$ is sufficient to fix the statistics of all the other component fields in the expansion.

Given the general expression of the superfield, it turns out that the number of component fields in the multiplet is too big, and we need some constraints in order to find an irreducible representation.
This task can be achieved by defining the covariant derivatives 
\beq
D = i \frac{\partial}{\partial \theta} - \frac{i}{2} \theta^{\dagger} (s+1-R) \, , \qquad
D^{\dagger} = -i \frac{\partial}{\partial \theta^{\dagger}} + \frac{i}{2} \theta (s+1-R) \, ,
\label{eq:susy_covariant_derivatives_momentum_space}
\eeq
which satisfy the following commutation relations:
\beq
 \lbrace D, \CQ \rbrace = \lbrace D^{\dagger} , \CQ^{\dagger} \rbrace  = \lbrace D, \CQ^{\dagger} \rbrace = \lbrace D^{\dagger} , \CQ \rbrace = 0 \, , \qquad
  \lbrace D, D^{\dagger} \rbrace = -L_0 \, . 
\eeq
In this way we define the notion of chiral $\Sigma_s$ and anti-chiral $\Sigma^{\dagger}_s$ superfields by requiring the conditions
\beq
D^{\dagger} \Sigma_s = 0 \, , \qquad
D \Sigma^{\dagger}_s = 0 \, .
\eeq
We will show that the only matter field needed to build the Hamiltonian in superfield language is a chiral fermionic superfield $\Psi$ plus its hermitian conjugate $\Psi^{\dagger}.$ 
For this reason, we directly consider the case where the bottom component of the supermultiplet is a complex fermion and we impose the (anti)chirality constraints to get
\bea
\label{eq:Fourier_modes_superfields_su111}
& \Psi_s (t, \theta, \theta^{\dagger}) = \frac{\psi_s (t)}{\sqrt{s+1}} + \theta \Phi_s (t) - \frac{1}{2} \theta \theta^{\dagger} \sqrt{s+1} \psi_s (t) \, , & \\
& \Psi^{\dagger}_s (t, \theta, \theta^{\dagger}) = \frac{\psi^{\dagger}_s (t)}{\sqrt{s+1}} + \theta^{\dagger} \Phi^{\dagger}_s (t) - \frac{1}{2} \theta \theta^{\dagger} \sqrt{s+1} \psi^{\dagger}_s (t) \, . &
\label{eq:Fourier_modes2_superfields_su111}
\eea
Notice that the particular normalization of the fermionic components reflects the definition of the charge densities, see Eqs.~\eqref{eq:charge_density_scalar_su11bos} and \eqref{eq:fermsu11charge}.
This will play an important role to determine the correct form of the interactions.
The constrained superfield gives an irreducible matter supermultiplet, since it only contains a single complex scalar and the fermionic partner, which are the surviving degrees of freedom of the near-BPS limit. No auxiliary fields are needed.

The supersymmetry transformations of all the modes can be found by computing
\beq
\delta \Psi_s = \le \epsilon \CQ + \epsilon^{\dagger} \CQ^{\dagger} \ri \Psi_s \, ,
\eeq
and then projecting the result on the various components.
We find that the free Hamiltonian $L_0$ can be written as
\beq
L_0 = - \int dt \int d \theta d \theta^{\dagger} \sum_{s=0}^{\infty} \tr \le  \Psi^{\dagger}_s (s+1-R) \Psi_s \ri \, . 
\eeq
Working out the rules of Berezin integration, this is easily shown to correspond in component formalism to Eq.~\eqref{eq:free_Hamiltonian_momentum_space_components_su111}.

Now we move to the interacting part of the Hamiltonian.
Having at disposal the fermionic superfield containing all the dynamical fields of the theory, in principle we can build higher-order terms with appropriate combinations of the superfield.
However, it turns out that the choice of the fermion as the lowest component of the supermultiplet and the Grassmannian nature of the superspace coordinates are responsible for the identities
\beq
\Psi^2_s = 0 \, , \qquad
(\Psi^{\dagger}_s)^2 = 0 \, ,
\eeq
which are a natural supersymmetric generalization of the concept of Grassmann variable.
In particular, this fact rules out the construction of a superpotential, {\sl i.e.} an expression holomorphic in the superfields, which is the natural candidate for renormalizable interactions in standard relativistic theories.
While this fact forbids to build the $\SU(1,1|1)$ Hamiltonian only in terms of the fermionic superfield, on the other hand it shows that another kind of supermultiplet is required to specify the theory, and in fact we will need to add a bosonic (anti)chiral superfield.
The necessity to integrate in a new field also arises at the level of components, as we will show.
This is justified by the fact that the theory still contains remnants of the original gauge symmetry of the $\mathcal{N}=4$ SYM action, which in the near-BPS limit are non-dynamical and mediate the interactions via the currents associated to the matter fields.

Using the component formalism, we define the modes of the current to be 
\beq
j_l = \sum_{s=0}^{\infty} \left(\sqrt{\frac{s+ 1}{s + l + 1}} \left\{\psi_{s}^\dagger,\psi_{s + l}\right\}+  \left[\Phi^\dagger_s,\Phi_{s+l}\right]\right)\,.
\eeq
Notice that this definition is exactly the total current\footnote{We change here the notation of the current as $ j_l$ instead of $\hat{q}_l$ to avoid confusion with supersymmetry charges and to stress that it plays the role of a current in a QFT coupling to a mediator gauge field.} $\hat{q}_{l}= q_l + \tilde{q}_l$ written in terms of bosonic and fermionic currents of Eqs.~\eqref{eq:charge_density_scalar_su11bos} and \eqref{eq:fermsu11charge}.
Now we introduce a \emph{gauge} contribution to the Hamiltonian given by a kinetic term for a complex scalar mode $A_s(t)$ and the minimal coupling between such field and the current:
\beq
H \supset \int dt \, \le -\sum_{s=0}^{\infty} s \tr \le A_s^\dagger A_s \ri +  \tilde{g}  \sum_{s=0}^{\infty} \tr \left(A^\dagger_s j_s + A_s j^\dagger_s\right) \ri \, .
\eeq 
The equation of motion for the constrained gauge field $A_s$ in Fourier space is
\begin{equation}
s A_s - \tilde{g} j_s = 0\,,
\label{eq:integrating_out_A_s_momentum_space}
\end{equation}
and after integrating it out, the term added to the Hamiltonian becomes
\begin{equation}
 \int dt \, \sum_{s=1}^{\infty}  \frac{1}{s} \tr \le j_s^\dagger j_s \ri \,.
\label{eq:local_effective_interaction_su111}
\end{equation}
This shows that by integrating in an auxiliary gauge mediator we get precisely this term entering the Hamiltonian.
The quartic interaction between two scalars and two fermions in Eq.~\eqref{eq:final_interaction_sphere_reduction_su111} can be obtained in component formalism by simply combining the fields as
\begin{equation}
H \supset \int dt \, \sum_{s_1, s_2,l=0}^{\infty}  \frac{1}{\sqrt{(s_1+l+1)(s_2+l+1)}} \tr \le [\Phi_{s_2},\psd_{s_2+l}] [\psi_{s_1+l},\Phi^\dagger_{s_1}] \ri
\Bigg) \,.
\label{eq:quartic_interaction_momentum_space_su111}
\end{equation}

How is possible to obtain this term from the superfield perspective if we cannot build holomorphic combinations of fermionic superfields?
It turns out that the supersymmetrization of the gauge mediator will solve the problem at once, accounting for both the term containing the currents and the quartic mixed interaction.

In fact, we define the following bosonic (anti)chiral superfield
\bea
\label{eq:momentum_space_gauge_superfield1}
& \mathcal{A}_s (t,\theta, \theta^{\dagger}) = A_s (t) + \theta \frac{\lambda_s (t)}{\sqrt{s+1}}  - \frac{1}{2} \theta \theta^{\dagger} s A_s (t) \, ,  & \\
& \mathcal{A}^{\dagger}_s (t,\theta, \theta^{\dagger}) = A^{\dagger}_s (t) - \theta^{\dagger} \frac{\lambda^{\dagger}_s (t)}{\sqrt{s+1}}  - \frac{1}{2} \theta \theta^{\dagger} s A^{\dagger}_s (t) \, . &
\label{eq:momentum_space_gauge_superfield2}
\eea
Since the theory is supersymmetric, we had to introduce in the definition a complex fermion $\lambda_s ,$ which we will interpret as a residual gaugino mediating another interaction.
In this way, we can write the complete Hamiltonian of the sector as
\beq
\begin{aligned}
H =& \int dt \int d \theta d \theta^{\dagger} \, \tr  \left\lbrace \sum_{s=0}^{\infty} \le  \mathcal{A}^{\dagger}_s \mathcal{A}_s  - \Psi^{\dagger}_s \le s+1-R  \ri \Psi_s \ri \right. \\
& \left.  +  \tilde{g} \sum_{s_1,s_2=0}^{\infty} \Psi^{\dagger}_{s_1} \Big[ \mathcal{A}^{\dagger}_{s_2-s_1+R-\frac{1}{2}} , \Psi_{s_2} \Big] 
+ \tilde{g} \sum_{s_1,s_2=0}^{\infty} \Psi^{\dagger}_{s_1} \Big[ \mathcal{A}_{s_1-s_2+R-\frac{1}{2}} , \Psi_{s_2} \Big]  \right\rbrace \, . 
\end{aligned}
\eeq
Although this is not manifest, the terms in the second line can be interpreted as a covariant derivative $i \mathcal{D}_x $ written in momentum space, as we will see with a local formulation in section \ref{sec:localsu111}.
The previous expression in component formalism is
\beq
\begin{aligned}
H= \int dt \, & \sum_{s=0}^{\infty}  \tr  \le \le s+\frac{1}{2} \ri \Phi^{\dagger}_s \Phi_s + (s+1) \psi^{\dagger}_s \psi_s 
-\sum_{s=0}^{\infty} s  A_s^\dagger A_s \right. \\ 
& \left. +  \tilde{g}  \sum_{s=0}^{\infty} \le A^\dagger_s j_s + A_s j^\dagger_s \ri
- \tilde{g} \sum_{s,s_1,s_2=0}^{\infty} \lambda_s [\psi_{s_1} , \Phi_{s_2}^{\dagger}] + \tilde{g} \sum_{s,s_1,s_2=0}^{\infty} \lambda^{\dagger}_s [\Phi_{s_1}, \psi^{\dagger}_{s_2}]  \ri \, .
\end{aligned}
\eeq
The first line contains all the kinetic terms, while the second line all the couplings with the currents.
The remarkable fact is that the gaugino $\lambda_s$ is not dynamical, and can be easily integrated out giving a quartic interaction which corresponds exactly to Eq.~\eqref{eq:quartic_interaction_momentum_space_su111}.
Then we see how the superfield formulation solves the problem: the fermionic partner of the remnant gauge field allows to build a term using minimal coupling without resorting to any (anti)holomorphic superpotential.
At this point we observe that the field $A_s$ is also non-dynamical, and following the step in Eq.~\eqref{eq:integrating_out_A_s_momentum_space} we integrate out this field to get the quartic interaction  \eqref{eq:local_effective_interaction_su111}.


\section{Local formulations}
\label{sect-local_formulation}

In Section \ref{sect-sphere_reduction} we have presented non-relativistic near-BPS theories that arise from limits of classical $\CN=4$ SYM on a three-sphere. In Section \ref{sect-quantization_one_loop_dilatation} we have quantized these theories employing a normal-ordering prescription. 
The quantized theories are the Spin Matrix theories (SMTs) considered in \cite{Harmark:2014mpa} and references therein, that also can be obtained directly from limits of quantized $\CN=4$ SYM. 

In this section we find local formulations of the quantized near-BPS theories/SMTs. 
Our main focus is on  $\SU(1,1|1)$ SMT, but we also comment on the other cases with $\SU(1,1)$ symmetry as well. 
In particular, we initially present the bosonic $\SU(1,1)$ sector as a simple setting to introduce the procedure, and we finally comment on the $\mathrm{P}\SU(1,1|2)$ case, being the one with richest structure.

\subsection{Local representations on fields}

In all of the four cases that we consider in this paper we have a $\SU(1,1)$ subalgebra of the bosonic part of the algebra. This $\SU(1,1)$ is the non-compact part of the algebra, and the $\SU(1,1)$ representations that we have are infinite dimensional. For this reason, one can find a local representation of the states that we have with respect to their $\SU(1,1)$ representations. We shall do this below for the $\SU(1,1|1)$ SMT by considering just the free Hamiltonian. Subsequently we include the interactions: in Section \ref{sec:localsu11} we start from the $\SU(1,1)$ bosonic SMT, we then consider in Section \ref{sec:localsu111} the full $\SU(1,1|1)$ SMT and we finally comment on the $\mathrm{P}\SU(1,1|2)$ case in Section \ref{sec:localpsu112}.

In the $\SU(1,1|1)$ SMT limit of $\CN=4$ SYM the surviving states are $|d_1^n Z\rangle$ and $|d_1^n\chi_1\rangle$ with $n\geq 0$ \cite{Harmark:2007px}. The goal is here to find a local representation of the $\SU(1,1)$ representation of these states. $\SU(1,1)$ has three generators $L_0$, $L_+$ and $L_-$ with algebra
\begin{equation}
\label{su11_algebra}
[L_0,L_\pm ] = \pm L_\pm \spa [L_-,L_+]=2L_0 \,.
\end{equation}
Acting with the three generators on the surviving states one finds
\beq
\label{su111_action}
\begin{array}{ll}
L_0 | d_1^n Z \rangle = \le n+ \frac{1}{2} \ri | d_1^n Z \rangle \, , & \quad
L_0 | d_1^n \chi_1 \rangle = (n+1) | d_1^n \chi_1 \rangle \, ,
\\
L_+ | d_1^n Z \rangle = (n+1) | d_1^{n+1} Z \rangle \, ,&\quad
L_+ | d_1^n \chi_1 \rangle = \sqrt{(n+1)(n+2)} | d_1^{n+1} \chi_1 \rangle \, ,
\\
L_- | d_1^n Z \rangle = n | d_1^{n-1} Z \rangle \, , &\quad
L_- | d_1^n \chi_1 \rangle = \sqrt{n(n+1)} | d_1^{n-1} \chi_1 \rangle \, ,
\end{array}
\eeq
with $n \geq 0$. One can compare this to a general spin $j$ representation of $\SU(1,1)$ 
\begin{equation}
\begin{array}{l}
\ds L_0 | j,j+n\rangle = (j+n)  | j,j+n\rangle\,, \\
\ds L_+ | j,j+n\rangle = \sqrt{(n+1)(n+2j)} | j,j+n+1\rangle\,, \\
\ds L_- | j,j+n\rangle = \sqrt{n(n+2j-1)} | j,j+n-1\rangle\,.
\end{array}
\end{equation}
This shows that the bosonic states $| d_1^n Z \rangle$ are in the $j=1/2$ representation while the fermionic states $| d_1^n \chi_1 \rangle$ are in the $j=1$ representation of $\SU(1,1)$.

The idea is now to find a representation of $L_0$ and $L_\pm$ in terms of differential operators on a local field. Since we have only one quantum number $m$ in the $j=1/2$ and $j=1$ representations it should be a one-dimensional spatial direction. Moreover, since it's quantized, one should put it on a circle.
Hence we introduce the spatial coordinate $x$, periodic with period $2\pi$, to parametrize this circle.

Consider first the bosonic states $| d_1^n Z \rangle$. We introduce the bosonic complex field
\begin{equation}
\label{Phi_field}
\Phi(t,x) = \sum_{n=0}^\infty \Phi_n(t) e^{i (n+\frac{1}{2})x} \,.
\end{equation}
Note that we are in the Heisenberg picture. Identifying $L_0 = n + \frac{1}{2}$ one sees that $L_0 = - i \partial_x$. This means that as a quantum operator $\Phi_n^\dagger$  acting on the vacuum creates the state $|d_1^n Z\rangle$. Note that $\Phi$ is antiperiodic on the circle due to the half-integer momentum on the circle. As we shall see below, $\Phi$ shares some features with $\beta$-$\gamma$ ghost fields, and thus has a mixture of bosonic and fermionic characteristics. A consistent representation on $\Phi(t,x)$ of the $\SU(1,1)$ generators is
\begin{equation}
\label{su11_alg_pos}
L_0 = - i \partial_x \spa L_\pm = e^{\pm i(x-t)} \le - i  \partial_x \pm R \ri \,,
\end{equation}
 since this reproduces the algebra \eqref{su11_algebra}. 
Here $R$  is the $\U(1)$ charge which is $R=\frac{1}{2}$ for bosons and $R=0$ for fermions.
The time-dependence in \eqref{su11_alg_pos} will be addressed below.
An important question is the normalization of the mode $\Phi_n$. The states $|d_1^n Z\rangle= | \frac{1}{2},\frac{1}{2}+n\rangle$ are normalized. Hence one can read off this normalization from the action with $L_\pm$ in \eqref{su111_action}. This shows that $\Phi_n$ is normalized 
and we have
\begin{equation}
[\Phi_m , \Phi^{\dagger}_n ] = i \delta_{m n} \,.
\end{equation}
One can now evaluate the equal-time commutators of $\Phi(t,x)$ giving the result
\begin{equation}
[\Phi(t,x),\Phi(t,x')] = 0 \spa
[\Phi(t,x),\Phi(t,x')^\dagger ] = i S_{\frac{1}{2}}(x-x') \,,
\label{eq:ET_commutators_scalar}
\end{equation}
where 
\begin{equation}
S_j(x) = \sum_{n=0}^\infty e^{i (n+j) x}
\end{equation}
This points to the fact that $\Phi(t,x)$ does not have the standard behavior of a local field. In particular, even if one can find a quantum state that is an eigenstate of the momentum along the circle, one cannot find a quantum state which is an eigenstate of the position along $x$.

Turning to the fermionic states $|d_1^n \chi_1 \rangle$ we introduce the Grassmann-valued complex field 
\begin{equation}
\label{psi_field}
\psi (t,x) = \sum_{n=0}^\infty \frac{1}{\sqrt{n+1}} \psi_n(t) e^{i(n+1)x} \,.
\end{equation}
Identifying $L_0=n+1$ we have again the representation \eqref{su11_alg_pos} of the $\SU(1,1)$ algebra on $\psi(t,x)$. The field $\psi(t,x)$ is periodic in the $x$ direction. Upon quantization, the mode $\psi_n^\dagger$ acting on the vacuum gives the state $|d_1^n\chi_1 \rangle$. Again, one should check the normalization of $\psi_n$ with the action of $L_\pm$ in \eqref{su11_algebra} and \eqref{su111_action}. With the $1/\sqrt{n+1}$ factor in \eqref{psi_field} one gets that $\psi_n$ is normalized, hence in the quantized theory we have
\begin{equation}
\{ \psi_m , \psi_n^\dagger \} = i \delta_{mn} \,.
\end{equation}
With this, the equal-time anti-commutators of $\psi(t,x)$ are
\begin{equation}
\{\psi(t,x),\psi(t,x')\} = 0 \spa
\{\psi(t,x), i \partial_{x'} \psi(t,x')^\dagger \} = i S_{1}(x-x') \,,
\label{eq:ET_commutators_fermion}
\end{equation}
Again, this is not a standard anti-commutator for a local fermionic quantum field.

\subsection{Local formulation of bosonic $\SU(1,1)$ SMT}
\label{sec:localsu11}

We start discussing the basic procedure to build a QFT description for the simplest near-BPS limit, {\sl i.e.} the $\SU{(1,1)}$ bosonic sector.
The main task is to reproduce the interacting Hamiltonian in Eq.~\eqref{fullH_bossu11}, which is given in momentum space, in terms of a local field theory containing the complex scalar field \eqref{Phi_field} satisfying the equal time commutator \eqref{eq:ET_commutators_scalar}.
The presence of the singlet constraint in the Hamiltonian implies that the $\SU{(N)}$ remains gauged. Moreover, we need to integrate in an additional auxiliary field in order to reproduce the interactions.
We can interpret this step as the position space version of the mediation given by the non-dynamical gauge field in the sphere reduction procedure described in Section \ref{sect-sphere_reduction}.

Consider the following (1+1)-dimensional field theory on a circle of unit radius parametrized by the spatial coordinate $x$ with periodic indentification $x \sim x + 2 \pi$
\beq
S= \int dt dx \, \tr \le  i \Phi^{\dagger} (\partial_0 + \partial_x) \Phi + i A^{\dagger} \partial_x A 
+ \tilde{g} \le A^{\dagger} j + A j^{\dagger} \ri \ri \, ,
\label{eq:local_action_su11bos}
\eeq
where $j(t,x)$ is the charge density associated to the $\SU{(N)}$ symmetry defined by
\beq
j(t,x) \equiv [\Phi^{\dagger}(t,x) , \Phi(t,x)] \, .
\eeq
We show that the previous local action gives rise to Eq.~\eqref{fullH_bossu11} with an appopriate decomposition of the auxiliary field in momentum space.
We require
\beq
A(t,x) = \sum_{n=0}^{\infty} A_n (t) e^{inx} \, .
\label{A_field}
\eeq
Combining this expansion with the scalar one in Eq.~\eqref{Phi_field} we obtain
\beq
S= \sum_{n=0}^{\infty} \int dt \, \tr \le i \Phi^{\dagger}_n \partial_t \Phi_n +  \le n+\frac{1}{2} \ri \Phi^{\dagger}_n \Phi_n 
+ n A^{\dagger}_n A_n + \tilde{g} \le A^{\dagger}_n j_n + A_n j^{\dagger}_n \ri  \ri  \, ,
\eeq
where the modes of the charge density $j_n (t)$ are given by Eq.~\eqref{eq:charge_density_scalar_su11bos}.
Since $A(t,x)$ is non-dynamical, its equations of motion give rise to the constraint 
\beq
n A_n (t) + \tilde{g} j_n (t) = 0 \, .
\eeq
For $n=0,$ this coincides with the $\SU{(N)}$ singlet constraint $j_0 = 0.$
When $n>0,$ the constraint can be solved and inserted into the action to get
\beq
S=\int dt \, \tr  \left[  \sum_{n=0}^{\infty}   i \Phi^{\dagger}_n \partial_t \Phi_n +  \sum_{n=0}^{\infty}  \le n+\frac{1}{2} \ri \Phi^{\dagger}_n \Phi_n 
 - \tilde{g}^2 \sum_{n=1}^{\infty} \frac{1}{n} j^{\dagger}_n j_n + \tilde{g} (A_0 + A_0^{\dagger}) j_0 \right]  \, .
\eeq
The corresponding Hamiltonian is easily derived via the Legendre transform and corresponds exactly to Eq.~\eqref{fullH_bossu11}.

We conclude the analysis of the $\SU{(1,1)}$ bosonic limit with some comments on the action \eqref{eq:local_action_su11bos}.
The form of the kinetic term is unusual, being linear in both the time and space derivatives.
In the standard relativistic case the Klein-Gordon operator is quadratic, while in the Schroedinger-invariant case the action is linear in the time derivative, but quadratic along the spatial directions.
Instead, this kinetic term corresponds to an ultra-relativistic dispersion relation between energy and momentum $E=P,$ typical of Carrollian theories \cite{Levy-Leblond}.
In this case, however, there is the non-trivial constraint $P>0,$ which makes the theory non-relativistic.
From this perspective, we see that the momentum constraint and the non-standard Dirac brackets become necessary to get a non-relativistic interpretation of the result.

Finally, we remark that the scalar must necessarily be complex, otherwise the kinetic term would be a total derivative.
In this connection, it is amusing to note that upon introducing two real scalar fields $(\beta, \gamma)$ as
\beq
\Phi = \beta + i \gamma  \,,
\eeq
the kinetic term $\mathcal{L}_0 =   i \Phi^{\dagger} (\partial_0 + \partial_x) \Phi$ of the action \eqref{eq:local_action_su11bos} becomes
\beq
\label{betagammaL}
\mathcal{L}_0 = - 2 \beta (\partial_0 +\partial_x) \gamma  \, .
\eeq
This shows that the bosonic part of the action can be viewed as a $\beta$-$\gamma$ CFT, which is a theory with negative central charge. Exploring this intriguing connection further is a matter for future work.

\subsection{Local formulation of $\SU(1,1|1)$ SMT}
\label{sec:localsu111}

We extend the QFT description of the Section \ref{sec:localsu11} to include to the $\SU{(1,1|1)},$ which contains a fermionic partner for the scalar field.
In particular, we present the result in a manifestly supersymmetric way by giving the position space version of the superspace formulation introduced in Section \ref{sect-momentum_space_superfields}, and then we will comment on the result in terms of component fields.

Having identified the free part of the Hamiltonian with $L_0 = -i \partial_x, $ we need to search for a representation of the supercharge such that $\lbrace \CQ, \CQ^{\dagger} \rbrace = - i \partial_x . $
The presence of a single complex supercharge implies that superspace is composed by one complex Grassmann variable $\theta,$ a common feature with the momentum space description.
It is simple to check that the following representation satisfies the correct anticommutator:
\beq
\CQ = \frac{\partial}{\partial \theta} - \frac{i}{2} \theta^{\dagger} \partial_x  \, , \qquad
\CQ^{\dagger} = \frac{\partial}{\partial \theta^{\dagger}} - \frac{i}{2} \theta \partial_x \, .
\eeq
Consistency between the left and right multiplication in defining superspace implies that we can define the supersymmetric covariant derivatives as
\beq
 D = i \frac{\partial}{\partial \theta} - \frac{1}{2} \theta^{\dagger} \partial_x  \, , \qquad
D^{\dagger} = - i \frac{\partial}{\partial \theta^{\dagger}} + \frac{1}{2} \theta \partial_x \, , 
\eeq
satisfying the commutators
\beq
\lbrace D, \CQ \rbrace = \lbrace D^{\dagger} , \CQ^{\dagger} \rbrace  = \lbrace D, \CQ^{\dagger} \rbrace = \lbrace D^{\dagger} , \CQ \rbrace = 0 \, , \qquad
  \lbrace D, D^{\dagger} \rbrace = i \partial_x = -L_0 \, .
\eeq
These expressions correspond in momentum space to Eq.~\eqref{eq:supercharge_momentum_space} and \eqref{eq:susy_covariant_derivatives_momentum_space}.

The annihilation under the action of the covariant derivatives of a generic superfield allows to define (anti)chiral superfields.
We direcly introduce the quantities that are sufficient to build an action in superspace formalism: the (anti)chiral fermionic superfields containing the dynamical modes of the $\SU{(1,1|1)}$ sector
\bea
& \Psi (t,x,\theta, \theta^{\dagger}) = \psi (t,x) + \theta \Phi(t,x) + \frac{i}{2} \theta \theta^{\dagger} \partial_x \psi (t,x) \, , & \\
& \Psi^{\dagger} (t, x, \theta, \theta^{\dagger}) = \psi^{\dagger}(t,x) + \theta^{\dagger} \Phi^{\dagger} (t,x) - \frac{i}{2} \theta \theta^{\dagger} \partial_x \psi^{\dagger} (t,x) \, , &
\label{eq:chiral_fermionic_superfield_su111}
\eea
and the bosonic (anti)chiral superfield containing the auxiliary fields
\bea
\label{eq:bosonic_chiral_gauge_superfield_su111}
& \mathcal{A} (t,x, \theta, \theta^{\dagger}) = A (t,x) + \theta \lambda (t,x) + \frac{i}{2} \theta \theta^{\dagger} \partial_x A (t,x) \, , & \\
& \mathcal{A}^{\dagger} (t,x, \theta,\theta^{\dagger}) = A^{\dagger} (t,x) - \theta^{\dagger} \lambda^{\dagger} (t,x) - \frac{i}{2} \theta \theta^{\dagger} \partial_x A^{\dagger} (t,x) \, . &
\eea
When expanding in modes the component fields, we obtain Eq.~\eqref{eq:Fourier_modes_superfields_su111}, \eqref{eq:Fourier_modes2_superfields_su111}, \eqref{eq:momentum_space_gauge_superfield1} and \eqref{eq:momentum_space_gauge_superfield2}.

The gauge superfield \eqref{eq:bosonic_chiral_gauge_superfield_su111} is composed by fields that we will call a gauge field $A(t,x)$ and a gaugino $\lambda (t,x)$ in the sense that they play the role of mediators of other interactions, and they are remnants of the original gauge invariance of the $\mathcal{N}=4$ SYM action before imposing Coulomb gauge and performing the sphere reduction.
We can further push on this interpretation by defining derivatives covariant with respect to the gauge superfield
\beq
\mathcal{D}_0 \equiv \partial_0 \, , \qquad
\mathcal{D}_x \equiv \partial_x - i \tilde{g} \mathcal{A} - i \tilde{g} \mathcal{A}^{\dagger} \, .
\eeq 
When applied on the fermionic superfield, it acts as
\beq
\mathcal{D}_x \Psi \equiv \partial_x \Psi - i \tilde{g} [\mathcal{A} , \Psi] - i \tilde{g} [\mathcal{A}^{\dagger}, \Psi] \, ,
\eeq
where in component formalism the brackets are commutators or anticommutators depending from the statistics of the specific field they are acting on.

In this way we obtain a compact expression for the action describing the effective field theory of the $ \SU{(1,1|1)}$ sector
\beq
S = \int dt dx \int d \theta^{\dagger} d \theta \, \tr \le  -i \Psi^{\dagger} (\mathcal{D}_0 + \mathcal{D}_x) \Psi + \mathcal{A}^{\dagger} \mathcal{A}  \ri  \, .
\label{eq:full_QFT_action_su111_superspace}
\eeq
This proposal is very natural: the matter part is a simple generalization of the two-dimensional Dirac action, with the Dirac spinor replaced by a fermionic superfield.
The coupling with the auxiliary field is also straightforward: there is a minimal coupling via the introduction of a covariant derivative containing the real part of the gauge superfield, while the kinetic term is standard for a chiral bosonic superfield.
Notice that while in standard cases, e.g. for the relativistic $\mathcal{N}=1$ chiral superfield in (3+1)-dimensions, a kinetic term of kind $\mathcal{A}^{\dagger} \mathcal{A}$ is dynamical, here the specific expansion in superspace \eqref{eq:bosonic_chiral_gauge_superfield_su111} shows that no time derivative appears, {\sl i.e.} the gauge field and the gaugino are non-dynamical.
We notice that the set of interactions built in this way are quite general, since the Grassmannian nature of the fermionic superfield implies
\beq
\Psi^2 = 0 \, , \qquad
(\Psi^{\dagger})^2 = 0 \, ,
\eeq
so that higher-order (anti)holomorphic terms are forbidden.

We further comment on the supersymmetry invariance of the action \eqref{eq:full_QFT_action_su111_superspace}.
The interacting part of the action is manifestly invariant under supersymmetry because it is built only using the superfield formulation, and it is the non-trivial content of the $\SU(1,1|1)$ sector.
On the other hand, the kinetic term is not supersymmetric invariant: it is given by $L_0$ and is defined using a derivative which is covariant with respect to the gauge superfield \eqref{eq:bosonic_chiral_gauge_superfield_su111}, but not under supersymmetry.
Since the free Hamiltonian given by $H_0$ in Eq.~\eqref{eq:free_Hamiltonian_sphere_red_su111} is instead supersymmetric invariant and it differs from $L_0$ by the conserved charge $\hat{N}$ in \eqref{numberop_su111}, it is easy to obtain a manifestly supersymmetric kinetic term by simply adding a mass shift of 1/2 in the differential operator $ -i(\mathcal{D}_0 + \mathcal{D}_x) .$

It is instructive to decompose the action \eqref{eq:full_QFT_action_su111_superspace} in component fields.
Since it turns out that the gaugino $\lambda (t,x)$ appears simply as a Lagrange multiplier, we immediately solve the corresponding constraint to integrate it out.
We find
\beq
S=  \int dt dx \, \tr \left\lbrace i \Phi^{\dagger} (\partial_0 + \partial_x) \Phi - \partial_x \psi^{\dagger} (\partial_0 + \partial_x) \psi + i A^{\dagger} \partial_x A  + \tilde{g} A j + \tilde{g} A^{\dagger} j^{\dagger} + g^2_0 [\Phi, \psi^{\dagger}] [\psi , \Phi^{\dagger}]  \right\rbrace \, ,
\label{eq:full_component_field_QFT_action_su111}
\eeq
where the scalar field is defined in Eq.~\eqref{Phi_field}, the fermionic field in \eqref{psi_field}, the gauge field in \eqref{A_field} and the current is now given by
\beq
j (t,x) = i \lbrace \partial_x \psi^{\dagger} , \psi \rbrace + [\Phi^{\dagger}, \Phi] \, .
\eeq
When expanding it in momentum space, we obtain the charge density $\hat{q}_n = q_n + \tilde{q}_n $ defined from Eq.~\eqref{eq:charge_density_scalar_su11bos} and \eqref{eq:fermsu11charge}.
Putting the decomposition of all the fields in momentum space inside the component field action \eqref{eq:full_component_field_QFT_action_su111}, we get the Legendre transform of the Hamiltonian \eqref{eq:final_ham_su111}, with interactions \eqref{eq:final_interaction_sphere_reduction_su111}.

Looking at the fermionic kinetic term in the action \eqref{eq:full_component_field_QFT_action_su111}, we notice that while the quadratic part in the spatial derivatives is standard e.g. in Schroedinger-invariant theories, instead the product of one time and one spatial derivative is peculiar.
Notice that had we taken the fermionic field to be real, the kinetic term would have been a total derivative. Note further that the structure of the fermionic kinetic term is in complete agreement with that of a complex chiral boson (see e.g. \cite{Sonnenschein:1988ug}), only that the field is Grassmann valued. Again, this unveils a curious correspondence with ghost fields with nonstandard statistics.

It is also interesting to observe that we obtain a natural superfield description of the model with action \eqref{eq:full_QFT_action_su111_superspace} by defining a gauge superfield, which requires the inclusion of a fermionic partner for the gauge field.
However $\lambda(t,x)$ turns out to be completely auxiliary, and in fact it is not necessary to introduce it when considering a component field formulation.
In this sense, it plays the same role of the auxiliary field $F$ entering the relativistic $\mathcal{N}=1$ bosonic chiral superfield in 3+1 dimensions.

\subsection{Local formulation of $\PSU(1,1|2)$ SMT}
\label{sec:localpsu112}

It is straightforward to extend the QFT description of Section \ref{sec:localsu111} in order to obtain the full $\mathrm{P}\SU{(1,1|2)}$ sector.
We work in component field formulation and require the following decomposition of the fields: 
\bea
& \Phi_a (t,x)= \sum_{n=0}^{\infty} (\Phi_a)_n (t) \, e^{i(n+\frac{1}{2})x} \, , \qquad
\psi_a (t,x) = \sum_{n=0}^{\infty} \frac{1}{\sqrt{n+1}} (\psi_a)_n (t) \, e^{i(n+1)x} \, , & \\
& A(t,x) = \sum_{n=0}^{\infty} A_n (t) \, e^{i n x} \, ,
\qquad  B_{ab} (t,x) = \sum_{n=0}^{\infty} (B_{ab})_n (t) \, e^{i n x} \, .  &
\eea
In addition to the doublet structure of bosons and fermions under $\SU{(2)},$ we introduced another bosonic field $B_{ab} (t,x)$ which will mediate the interactions; the difference with $A(t,x)$ is that it will give rise to single trace structures, while the latter will contribute to double trace interactions.

We then consider the total action
\beq
\begin{aligned}
S =  \int dt dx \, \tr & \left\lbrace   i \Phi^{\dagger}_a \le \partial_0 + \partial_x  \ri \Phi_a - \partial_x \psi_a^{\dagger} \le \partial_0 + \partial_x  \ri \psi_a  - i A^{\dagger} \partial_x A   \right. \\
& \left. - i B^{\dagger}_{ab} \partial_x B_{ab} - \tilde{g} A^{\dagger} j - \tilde{g} A j^{\dagger} - \tilde{g} B_{ab} j^{\dagger}_{ab} - \tilde{g} j_{ab} B^{\dagger}_{ab} - \tilde{g}^2 |[\Phi_a , \psi^{\dagger}_a]|^2   \right\rbrace \, ,
\end{aligned}
\label{eq:total_local_action_su112}
\eeq
with currents
\beq
j (x) =  i \lbrace \partial_x \psi_a^{\dagger} (x) , \psi_a (x) \rbrace + [\Phi^{\dagger}_a , \Phi_a (x)] \, , \qquad
j_{ab} (x) = - i \lbrace \partial_x \psi_a (x) , \psi_b (x) \rbrace + [\Phi_a (x) , \Phi_b (x)] \, .
\eeq
The matching of the kinetic terms in \eqref{eq:full_Hamiltonian_su112} with the double trace  interactions in \eqref{eq:sphere_reduction_su112_result_refined} is straightforward and works as in Section \ref{sec:localsu111}.
We briefly show how the matching of the single trace structure works: the current $j_{ab} (t,x)$ in Fourier space reads
\beq
(j_{ab})_{s} =
 \sum_{m, n=0}^{\infty} \le  \sqrt{\frac{m+1}{n+1}} \lbrace (\psi_a)_{m}, (\psi_b)_{n}  \rbrace \delta (m + n +2 -s)  + [(\Phi_a)_{m} , (\Phi_b)_{n}] \delta (m + n +1 -s) \ri \, .
\eeq
This is exactly the expression \eqref{eq:current_su112}.
Notice that the mode expansion for the dynamical bosonic and fermionic fields is shifted by $R,$ which takes the values $\frac{1}{2}, 0$ respectively.
This behaviour is responsible for the different support of the delta functions and is a consequence of the fields belonging to the representations $j=\frac{1}{2},0$ of the $\SU(1,1)$ group.
The equations of motion for the non-dynamical field $B_{ab}$ in Fourier space are
\beq
s (B_{ab})_s - \tilde{g} (j_{ab})_s = 0 \, ,
\eeq
and integrating out this field we get the interaction
\beq
 \sum_{s=1}^{\infty} \frac{1}{s} \tr \left[  (j^{\dagger}_{ab})_s (j_{ab})_s \right] \, ,
\eeq
which is exactly Eq.~\eqref{eq:sphere_reduction_su112_result_refined}.


\section{Conclusions and outlook}
\label{sect-conclusions-and-outlook}

In this paper we have shown how to take the near-BPS limits \eqref{BPSlimit} directly of the classical formulation of $\CN=4$ SYM on a three-sphere, following \cite{Harmark:2019zkn}. The BPS bounds we considered were all of the form \eqref{BPSbound} giving a surviving $\SU(1,1)$ global symmetry along with a $\U(1)$-symmetry corresponding to conservation of the number operator. In the $\SU(1,1|1)$ and $\PSU(1,1|2)$ near-BPS theories the $\SU(1,1)$ symmetry is a subgroup of a larger symmetry.
The techniques used for taking the limits include the spherical reduction of $\CN=4$ SYM on a three-sphere, following \cite{Harmark:2019zkn}, as well as integrating out non-dynamical fields that in some cases contribute to the interaction of the surviving modes. 

We have shown explicitly how to quantize the near-BPS theories, and shown that the result is equivalent to taking the near-BPS limit directly of the quantized $\CN=4$ SYM. This means the quantized near-BPS theories corresponds to the Spin Matrix theories \cite{Harmark:2014mpa}. 

Finally, we found a superfield formulation of $\SU(1,1|1)$ in momentum space, and we have explored a way to represent the near-BPS/Spin Matrix theories as local non-relativistic quantum field theories. This has revealed interesting and surprisingly elegant structures for the interactions, in particular in the $\SU(1,1|1)$ case and its two $\SU(1,1)$ subcases. The quantum fields we found are not fully local and have ghost-like features in that the bosonic fields have fermionic features, and the fermionic fields have bosonic features.

As explained in the introduction, near-BPS/Spin Matrix theories are interesting in view of their possible realization of the holographic principle, and, related to this, also as a possible way to access new regimes in the AdS/CFT correspondence that have yet to be explored. For this reason, it is highly interesting to observe that the form of the interactions in the $\SU(1,1|1)$ theory makes it possible to access the strong coupling. Indeed, we find a classical Hamiltonian of the form (see Section \ref{sec:su111}) 
\begin{equation}
\label{su111_posdef}
H_{\rm limit} = L_0 + \frac{\tilde{g}^2}{2N} \left(  \sum_{l=1}^\infty  \frac{1}{l} \tr \le \hat{q}_{l}^{\dagger} \hat{q}_{l} \ri  
 +  \sum_{l = 0}^{\infty}  \tr ( F_l^\dagger F_l ) \right) \, .
\end{equation}
We see that since the terms in the interaction are positive definite, we notice that in the strong coupling limit $\tilde{g}\rightarrow \infty$ the leading contribution is given by $\hat{q}_{l+1}=F_l=0$ for $l \geq 0$. 
Thus, one can solve the strong coupling limit in this way, at least in the semi-classical limit. Indeed, this was also noticed in \cite{Berkooz:2014uwa} for the fermionic $\SU(1,1)$ case. 

A similar argument can be applied to the Hamiltonian of the $\mathrm{P}\SU(1,1|2)$ sector
\beq
	H_{\rm limit}  = L_0 + \frac{\tilde{g}^2}{2N} \left[  \sum_{l=1}^{\infty}  \frac{1}{l} \tr \le \hat{q}_l^{\dagger} \, \hat{q}_l  \ri
	+  \sum_{l = 0}^{\infty}  \tr ( (F_{ab})_l^\dagger (F_{ab})_l )
+ 	\sum_{l=1}^{\infty} 
 \tr ( (G_{ab})_l^\dagger (G_{ab})_l ) \right]
	 \, ,
\label{su112_posdef}
\eeq
suggesting that the strong coupling limit $\tilde{g} \rightarrow \infty$ corresponds to a leading contribution where $\hat{q}_{l+1}=(F_{ab})_l=(G_{ab})_{l+1}=0$ for $l \geq 0 . $  
It will be illuminating to understand better the $\SU(1,1|1)$-algebraic structure of the interactions in \eqref{su111_posdef}, and the analog problem for the maximal case in Eq.~\eqref{su112_posdef}.

Since we can access the strong coupling limit of these near-BPS theories it is interesting to ask what holographic dual one should compare this to. In the planar limit, the answer is presumably the string theory duals of \cite{Harmark:2017rpg,Harmark:2018cdl,Harmark:2019upf}. For instance, in the bosonic $\SU(1,1)$ case one finds a dual $\U(1)$-Galilean geometry which is basically $\mathbb{R}$ times a cigar-geometry, where $\mathbb{R}$ is the time-direction. It will be vital to explore this further, also in view of the non-standard features of the $\SU(1,1)$ theory formulated as local quantum field theories.
Moreover, beyond the planar limit, the BPS bounds \eqref{BPSbound} examined in this paper are related to limits of black holes in $\mbox{AdS}_5\times S^5$ with vanishing entropy.  There is also the intriguing possibility that one can observe the emergence of dual D-branes, in the form of Giant Gravitons, similarly to what was found in \cite{Harmark:2016cjq}.

We found an interesting connection to the $\beta$-$\gamma$ ghost CFT for the kinetic term of the scalar fields, see Eq.~\eqref{betagammaL}. This would be interesting to explore further as it possibly could provide another view point on the $P > 0$ constraint. 

Other representations of $\SU(1,1)$ may allow for a more natural field theoretic formulation of our near-BPS theories. We note that $\SU(1,1)$ representations are also realized on $\mbox{AdS}_2$ and on the hyperbolic plane. 

Finally, we would like to advertise our companion paper \cite{companion} in which we explore a class of near-BPS limits with a $\SU(1,2)$ global symmetry present. Among these near-BPS theories are the theory with $\PSU(1,2|3)$ symmetry that captures the behavior of $\CN=4$ SYM near the BPS bound $E \geq S_1+S_2+J_1+J_2+J_3$, a bound saturated by the supersymmetric black hole in $\mbox{AdS}_5\times S^5$ \cite{Gutowski:2004yv}.

\section*{Acknowledgements}

We thank Yang Lei for many interesting discussions and useful comments on the draft of this paper. We acknowledge support from the Independent Research Fund Denmark grant number DFF-6108-00340 “Towards a deeper understanding of black holes with non-relativistic holography”.


\appendix

\section{Spherical harmonics on $S^3$}
\label{app-spher_harmonics_S3}

In this Appendix we review the decomposition of fields on $\mathbb{R} \times S^3$ into a basis of spherical harmonics following \cite{Ishiki:2006rt}.

Any field on this background can be factorized into a part depending only from the time direction, and another term living on the three-sphere.
We focus on the latter factor.

The three-sphere has isometry group $G=SO(4)$ and local rotational invariance under $H=SO(3),$ hence it can be defined by the coset $G/H = SO(4)/SO(3) .$
For convenience, we use the local isometry $SO(4) \simeq SU(2) \times SU(2)$ to split the irreducible representations of $G$
into products of the irreducible representations of $\SU(2),$ which are labelled by integer and half-integer spins $J, \tilde{J} .$
A basis for such a representation will be denoted by $|J, m \rangle  |\tilde{J}, \tilde{m} \rangle ,$ with $|m| \leq J$ and $|\tilde{m}| \leq \tilde{J}.$
For the local invariance, we denote the spin of the irreducible representation as $L$ and the states as $|L n \rangle,$ with the constraint $|n| \leq L .$

If we denote the generators of $G$ with $J_i , \tilde{J}_i$ and the generators of $H$ with $L_i$ (in both cases $i \in \lbrace 1,2,3 \rbrace$), they are related by $L_i = J_i + \tilde{J}_i .$
In this way, we can introduce $\SU(2)$ Clebsch-Gordan coefficients to obtain the representations of $H$ from the sum of the two representations $\SU(2)$ composing $G,$ yielding the expression
\beq
|Ln; J \tilde{J} \rangle = 
\sum_{m, \tilde{m}} C^{Ln}_{Jm; \tilde{J} \tilde{m}} | J m \rangle | \tilde{J} \tilde{m} \rangle \, ,
\eeq
with the triangle inequalities
\beq
|J - \tilde{J}| \leq L \leq J + \tilde{J} \, .
\eeq
Spherical harmonics on $S^3$ are defined starting from these basis and from the choice of a representative element of $G/H.$ While most of the results do not depend from the specific choice of this representative, the rotation charges will.
For this reason, we specify that we parametrize the unit three-sphere with coordinates
\beq
d \Omega_3^2 = d \psi^2 + \cos^2 \psi \, d \phi_1^2 
+ \sin^2 \psi \, d \phi_2^2 \, .
\eeq
In this way we find the group element 
\beq
\Upsilon (\Omega) = e^{-i \phi_1 (J_3 - \tilde{J}_3)} e^{i \phi_2 (J_3 + \tilde{J}_3)} e^{-i \psi (J_1 - \tilde{J}_1)} \, ,
\label{eq:appA_group_element_three_sphere}
\eeq
and the corresponding inverse
\beq
\Upsilon^{-1} (\Omega) = e^{i \psi (J_1 - \tilde{J}_1)} e^{i \phi_1 (J_3 - \tilde{J}_3)} e^{- i \phi_2 (J_3 + \tilde{J}_3)} \, .
\eeq
The spherical harmonics are defined as
\beq
\mathcal{Y}^{Ln}_{Jm;\tilde{J} \tilde{m}} (\Omega) = \sqrt{\frac{(2J+1)(2 \tilde{J}+1)}{2L+1}} \langle Ln; J \tilde{J} | \Upsilon^{-1} (\Omega) | Jm; \tilde{J} \tilde{m} \rangle \, ,
\eeq
with $m, \tilde{m}$ being the eigenvalues of the generators $J_3, \tilde{J}_3 ,$ respectively.

At this point, we specify the previous decomposition for the fields of interest on the three-sphere: scalars, fermions and gauge fields.
Since a scalar field is a singlet under the local rotations $SO(3),$ its spin is $L=0$ and this immediately implies that $J=\tilde{J} .$
The decomposition is easily given by
\beq
\Phi^a (t,\Omega)= \sum_{J,M} \Phi^a_{JM} (t) \mathcal{Y}^{JM} (\Omega) \, , \qquad
 \Phi^{\dagger}_a (t,\Omega) = \sum_{JM} \Phi^{\dagger}_{JM,a} (t) \bar{\mathcal{Y}}^{JM} (\Omega) \, ,
 \label{eq:app_expansions_spherical_harmonics_scalars}
\eeq
where we denoted
\beq
\mathcal{Y}^{JM}   \equiv \mathcal{Y}^{L=0,n=0}_{J,m;J,\tilde{m}} \, .
\eeq
In the previous sums we collected the eigenvalues of the momenta as $M=(m, \tilde{m}),$ both running from $-J$ to $J,$ while $J$ itself runs over positive integers and half-integers.

Spinor fields have $L=\frac{1}{2},$ which allows for the possibilities to take momenta $(J+\frac{1}{2}, J)$ or viceversa, {\sl i.e.} $(J, J + \frac{1}{2}).$
In this case the mode expansion reads
 \beq
 \psi^A_{\alpha} (t, \Omega) = \sum_{\kappa = \pm 1} \sum_{JM} \psi^A_{JM,\kappa} (t) \mathcal{Y}^{\kappa}_{JM,\alpha} (\Omega) \, , \qquad
\psd_{\dot{\alpha},A} (t, \Omega) = \sum_{\kappa = \pm 1} \sum_{JM} \psd_{JM,\kappa,A} (t) \bar{\mathcal{Y}}^{\kappa}_{JM,\dot{\alpha}} (\Omega) \, ,
 \label{eq:app_expansions_spherical_harmonics_fermions}
 \eeq
where we defined
\beq
\mathcal{Y}^{\kappa=1}_{JM,\alpha} \equiv \mathcal{Y}^{L=\frac{1}{2},\alpha}_{J+\frac{1}{2},m;J,\tilde{m}} \, , \qquad
\mathcal{Y}^{\kappa=-1}_{JM,\alpha}  \equiv \mathcal{Y}^{L=\frac{1}{2},\alpha}_{J,m;J+\frac{1}{2},\tilde{m}} \, .
\eeq
While $J$ runs again over positive integers and half-integers, now the momenta $(m, \tilde{m})$ are summed from $-U$ to $U$ and $-\tilde{U}$ to $\tilde{U},$ respectively, with $U= J + \frac{1+\kappa}{4}$ and $\tilde{U}=J + \frac{1-\kappa}{4}.$

The gauge fields are vectors, and then they have $L=1.$
This allows for even more possibilities, {\sl i.e.} we can take $(J+1,J), (J,J) $ or $(J,J+1).$
Then their decomposition is
\beq
 A_i(t,\Omega) = \sum_{\rho=-1,0,1}\sum_{JM} A_{(\rho)}^{JM}(t){\cal Y}_{JM,i}^\rho (\Omega) \, ,
\eeq
with
\beq
\mathcal{Y}^{\rho=1}_{JM,i} \equiv i \mathcal{Y}^{L=1,i}_{J+1,m;J,\tilde{m}} \, , \qquad
\mathcal{Y}^{\rho=0}_{JM,i}  \equiv \mathcal{Y}^{L=1,i}_{J,m;J,\tilde{m}} \, , \qquad
\mathcal{Y}^{\rho=-1}_{JM,i}  \equiv -i \mathcal{Y}^{L=1,i}_{J,m;J+1,\tilde{m}} \, .
 \label{eq:app_expansions_spherical_harmonics_gauge_fields}
\eeq
In this case $(m, \tilde{m})$ run from $-Q$ to $Q$ and $-\tilde{Q}$ to $\tilde{Q},$ respectively, with $Q= J + \frac{\rho(1+\rho)}{2}$ and $\tilde{Q}=J - \frac{\rho(1-\rho)}{2},$ and $J$ is summed over positive integers and half-integers.

Notice that while the components along the three-sphere of the gauge field $A_i$ are vectors, instead the temporal component $A_0$ behaves as a scalar.
Consequently, its decomposition only involves the harmonics with $L=n=0,$ and reads
\beq
A_0 (t, \Omega) \equiv \chi(t,\Omega) = \sum_{J,M} \chi^{JM}(t) {\cal Y}_{JM}(\Omega)\,.
\eeq
In view of the manipulations with the Hamiltonian formalism, we also introduce the mode expansion of the momenta, which we denote as $\Pi^{(F)},$ being $F \in \lbrace \Phi, \psi, A \rbrace$ the field whose the specific momentum is associated to.
Since the orthonormality of the basis involves an inner product between spherical harmonics and their complex conjugate, it is convenient to choose\footnote{In principle we can expand the fields on the three-sphere in terms of the spherical harmonics $\mathcal{Y}$ or in terms of their complex conjugate $\bar{\mathcal{Y}}.$ The difference between them amounts to a phase and some change of the labels, as shown in \eqref{eq:app_complex_conjugate_spher_harmonics}. Therefore, the two choices correspond to a redefinition of the modes defining the expansion of the field.} a decomposition of spherical harmonics for the corresponding canonical momenta given by
\beq
 \Pi^{a (\Phi)} (t,\Omega)= \sum_{J,M} \Pi^{a (\Phi)}_{JM} (t) \bar{\mathcal{Y}}^{JM} (\Omega) \, , \qquad
 \Pi^{\dagger (\Phi)}_a (t,\Omega) = \sum_{JM} \Pi^{\dagger (\Phi)}_{JM,a} (t) \mathcal{Y}^{JM} (\Omega) \, , 
 \eeq
 \beq
\Pi^{A (\psi)}_{\alpha} (t, \Omega) = \sum_{\kappa = \pm 1} \sum_{JM} \Pi^{A (\psi)}_{JM,\kappa} (t) \bar{\mathcal{Y}}^{\kappa}_{JM,\alpha} (\Omega) \, , \qquad	
 \Pi_i^{(A)} (t,\Omega) = \sum_{\rho=-1,0,1} \sum_{JM} \Pi_{(\rho)}^{JM (A)}(t) \bar{{\cal Y}}_{JM,i}^\rho (\Omega) \, .
 \eeq
We also specify the mode expansion of the current and the Lagrange multiplier entering Eq.~\eqref{eq:Hamiltonian_gauge_field_plus_currents}, where we apply the same convenient choice:
\beq
 j_0 (t,\Omega)= \sum_{J,M} j_0^{\dagger JM} (t) \bar{\mathcal{Y}}^{JM} (\Omega) \, , \qquad
  j_i (t,\Omega)= \sum_{J,M,\rho} j_{(\rho))}^{\dagger JM} (t) \bar{\mathcal{Y}}^{\rho}_{JM,i} (\Omega) \, , 
\eeq
\beq
 \eta (t,\Omega)= \sum_{J,M} \eta^{\dagger JM} (t) \bar{\mathcal{Y}}^{JM} (\Omega) \, .
\eeq

\section{Hamiltonian and conserved charges of ${\cal N} = 4$ on $S^3$}
\label{app-quadratic_Hamiltonian_conserved_charges}

In view of the evaluation of BPS limits for the various sectors of $\mathcal{N}=4$ SYM, we collect the conventions about the relevant rotational and internal charges on the three-sphere $S^3$.
We mostly follow the same notation as \cite{Ishiki:2006rt}.

The free part of the action on $\mathbb{R} \times S^3$  is given by
\beq
S_0 = \int_{\mathbb{R}\times S^3} \sqrt{-\mathrm{det} \, g_{\mu\nu}} \tr \left\lbrace  
- (\nabla_{\mu} \Phi_a)^{\dagger} \nabla^{\mu} \Phi^a - \Phi^{\dagger}_a \Phi^a - i \psd_A \bar{\sigma}^{\mu} \nabla_{\mu} \psi^A - \frac{1}{2} F^{\mu\nu} F_{\mu\nu} 
\right\rbrace \, ,
\label{eq:free_N4_SYM_action}
\eeq
where we remind that $\nabla$ denotes the covariant derivative containing only the gravity contributions, without the minimal coupling with the gauge fields.
We stress that excluding the gauge coupling from the free action is not restrictive to compute the conserved charges. 
In fact the Noether currents are defined up to total derivatives and we can always consider the canonical current associated to a particular symmetry, which is independent from the gauge coupling.

The canonical momenta associated to scalars $\Phi,$ Weyl fermions $\psi$ and gauge fields $A$ are respectively given by 
\beq
\begin{aligned}
	& \Pi^{(\Phi)}_a = \frac{1}{\sqrt{-\mathrm{det} \, g_{\mu\nu}}} \frac{\delta S}{\delta \dot{\Phi}^a} = \dot{\Phi}_a^{\dagger} \, , \qquad
	\Pi_{(\Phi)}^{a \dagger} = \frac{1}{\sqrt{-\mathrm{det} \, g_{\mu\nu}}} \frac{\delta S}{\delta \dot{\Phi}^{\dagger}_a} = \dot{\Phi}^a \, , \\
 &	\Pi_A^{(\psi)} = \frac{1}{\sqrt{-\mathrm{det} \, g_{\mu\nu}}} \frac{\delta S}{\delta \dot{\psi}^A} = i \psd_A  \, , \qquad & \\
	& \Pi_0^{(A)} = \frac{1}{\sqrt{-\mathrm{det} \, g_{\mu\nu}}} \frac{\delta S}{\delta \dot{A}_0} = 0 \, , \qquad
	\Pi_i^{(A)} = \frac{1}{\sqrt{-\mathrm{det} \, g_{\mu\nu}}} \frac{\delta S}{\delta \dot{A}_i} = F_{0 i} \, . &
\end{aligned}
\eeq
From now on, we will avoid specifying the field $F$ associated to the momentum $\Pi^{(F)},$ since this will be clear from the context.
Notice that the first order nature of the fermionic action is responsible for obtaining a proportionality between the Weyl fermion and the hermitian conjugate of the corresponding momentum.
This allows to choose if we want to express the Hamiltonian and the charges of interest in terms of squares of the fields or of their momenta, or if we want mixed products of them.

These momenta allow to compute the free Hamiltonian of the system by means of the Legendre transform, giving
\beq
H_0 = \int_{\mathbb{R} \times S^3} \sqrt{-\mathrm{det} \, g_{\mu\nu}} \, \tr \left\lbrace  |\Pi_a|^2 + |\nabla_i \Pi_a|^2
- i \psi^{\dagger}_A \sigma^i \nabla_i \psi^A
 + \Pi_i^2+ \frac{1}{2} F_{ij}^2  \right\rbrace \, .
\eeq
The issues related to imposing the Coulomb gauge and the corresponding constraints, which require to introduce the Dirac brackets, are discussed in section \ref{sect-sphere_reduction}.
Here we report the result of such discussion: the $\rho=0$ mode of the gauge field $A_i$ is vanishing, and the temporal component $A_0$ can be integrated out.

In this way, after using the mode expansions \eqref{eq:app_expansions_spherical_harmonics_scalars}, \eqref{eq:app_expansions_spherical_harmonics_fermions} and \eqref{eq:app_expansions_spherical_harmonics_gauge_fields}, we find the free Hamiltonian
\beq
\begin{aligned}
\label{H0_sphere}
H_0 =  \sum_{J,m,\tilde{m}} & \tr \left\lbrace  |\Pi_a^{Jm\tilde{m}}|^2 + \omega_J^2 |\Phi_a^{Jm\tilde{m}}|^2 
 - \sum_{\kappa=\pm 1}  \kappa \omega_J^{\psi} \, \psd_{JM,\kappa, A} \psi^A_{JM,\kappa}    \right.\\
& \left.+ \sum_{\rho = \pm1}\left(|\Pi_{(\rho)}^{Jm\tilde{m}}|^2 + \omega_{A,J}^2 |A_{(\rho)}^{Jm\tilde{m}}|^2\right) \right\rbrace \, ,
\end{aligned}
\eeq
where
\beq
\omega_J \equiv 2J+1 \, , \qquad
\omega_J^{\psi} \equiv 2J + \frac{3}{2} \, , \qquad
\omega_{A,J} \equiv 2J + 2 \, . 
\eeq
A peculiarity of this free Hamiltonian is that while the scalar and gauge terms are manifestly positive-definite, instead the fermionic part is apparently negative-definite when $\kappa=1,$ {\sl i.e.} we have
\beq
H_{0}^{(\psi)} =  \sum_{JM} \tr \le  -  \omega_J^{\psi} \psi^{A \dagger}_{JM,\kappa=1} \psi^A_{JM,\kappa=1} + \omega_J^{\psi} \psi^{A \dagger}_{JM,\kappa=-1} \psi^A_{JM,\kappa=-1}  \ri   \, .
\eeq
The reason for this apparent negativity of the fermionic term arises from the conventions in \cite{Ishiki:2006rt}, because after quantization it is required that the two chiralities of the fermions are decomposed as follows:
\beq
\psi^A_{JM,\kappa=1} = d^{A \dagger}_{J,-M} e^{i \omega_J^{\psi}} \, , \qquad
\psi^A_{JM,\kappa=-1} = b^A_{JM} e^{- i \omega_J^{\psi}} \, .
\eeq
Since one polarization acts as a creation operator and the other one as an annihilation operator, in the end the Hamiltonian is positive definite.

We find a manifestly positive definite expression even at the level of the classical action if we redefine
\beq
\psi^A_{JM,\kappa=1} \rightarrow \psi^{A \dagger}_{J,-M,\kappa=1} \, , \qquad
\psi^{A \dagger}_{JM,\kappa=1} \rightarrow \psi^{A}_{J,-M,\kappa=1} \, .
\label{eq:redefinition_fermions}
\eeq 
It is important to notice that the redefinition also involves a change of sign for the orbital momentum eigenvalue $M=(m, \tilde{m}).$
This change can be easily understood if we think that the creation of a particle with momentum $M$ is now interpreted as the annihilation of an antiparticle of momentum $-M,$ ad viceversa.

In this way, using the Grassmannian nature of the fermions, we find in terms of the redefined quantities that
\beq
\begin{aligned}
	H^{(\psi)}_{0} & =  \sum_{JM} \tr \le  -  \omega_J^{\psi} \psi^{A }_{J,-M,\kappa=1} \psi^{A \dagger}_{J,-M,\kappa=1} + \omega_J^{\psi} \psi^{A \dagger}_{JM,\kappa=-1} \psi^A_{JM,\kappa=-1}  \ri  = \\
	& =  \sum_{JM} \tr \le   \omega_J^{\psi} \psi^{A \dagger}_{JM,\kappa=1} \psi^{A }_{JM,\kappa=1} + \omega_J^{\psi} \psi^{A \dagger}_{JM,\kappa=-1} \psi^A_{JM,\kappa=-1}  \ri = \sum_{\kappa=\pm 1} \sum_{JM} \tr \le  \omega_J^{\psi} \psd_{JM,\kappa, A} \psi^A_{JM,\kappa}  \ri 
	\, .
\end{aligned}
\eeq
In the first step we sent the index $M \rightarrow -M$ due to the symmetry of the range of summation.
We observe that the sign given by the factor $- \kappa$ in the free Hamiltonian disappears, and the map from the previous notation to the new conventions implies
\beq
\sum_{JM} \sum_{\kappa= \pm 1} \psd_{JM,\kappa, A} \psi^A_{JM,\kappa}  \rightarrow
\sum_{JM} \sum_{\kappa= \pm 1} - \kappa \psd_{JM,\kappa, A} \psi^A_{JM,\kappa} \, .
\eeq
From now on, all the quantities involving fermionic fields will be computed after applying the prescription \eqref{eq:redefinition_fermions}.
We will add some additional comments on these terms only when computing the Cartan charges associated to rotation and R-symmetry.

Now we compute the relevant currents corresponding to the symmetries of the action \eqref{eq:free_N4_SYM_action}.
We start with the canonical energy-momentum tensor
\beq
T_{\mu\nu} \equiv T_{\mu\nu}^{(\Phi)} + T_{\mu\nu}^{(\psi)} + T_{\mu\nu}^{(A)} + \frac{g_{\mu\nu}}{\sqrt{-g}} \mathcal{L} \, ,
\eeq
where $\mathcal{L}$ is the Lagrangian density and
\bea
& T_{\mu\nu}^{(\Phi)} = (\partial_{\mu} \Phi_a)^{\dagger} \partial_{\nu} \Phi^a + (\partial_{\mu} \Phi_a)^{\dagger} \partial_{\nu} \Phi^a \, , & \\
&  T_{\mu\nu}^{(\psi)} = - \frac{i}{2} \psd_A \bar{\sigma}_{\mu} (\nabla_{\nu} \psi^A ) + \frac{i}{2} (\nabla_{\nu} \psi_A)^{\dagger} \bar{\sigma}_{\nu} \psi^A \, , & \\
& T_{\mu\nu}^{(A)} = F_{\mu}^{\sigma} F_{\nu \sigma} \, . &
\eea
The conserved charges corresponding to the commuting rotation generators $S_1, S_2$ on the three-sphere are defined as
\beq
S_i = \int_{S^3} d \Omega \, T^0_{\,\,\ i} \, ,
\eeq
being $d \Omega$ the volume form on the three-sphere.
Introducing the decomposition of the fields into spherical harmonics from Appendix \ref{app-spher_harmonics_S3}, we get 
\beq
S_i \equiv S_i^{(\Phi)} + S_i^{(\psi)} + S_i^{(A)} \, , 
\label{eq:appB_rotation_charges}
\eeq 
where
\bea
&  S_1^{(\Phi)} = \sum_{J,M} i (\tilde{m} - m) \tr \le \Phi^{JM}\Pi^{JM} - \Phi^{\dagger\,JM}\Pi_\phi^{\dagger\,JM} \ri \, , & 
\\
& S_1^{(\psi)} = \sum_{JM} \sum_{\kappa= \pm 1}  (\tilde{m}-m) \tr \le  \psd_{JM,\kappa, A} \psi^A_{JM,\kappa}  \ri \, ,  &  \\
& S_1^{(A)} = \sum_{J,m,\tilde{m}} \sum_{\rho=-1,1} \frac{i}{2} (\tilde{m} - m)   \tr \left( {A}_{(\rho)}^{Jm\tilde{m}}\Pi_{(\rho)}^{Jm\tilde{m}} - A_{(\rho)}^{\dagger\,Jm\tilde{m}}\Pi_{(\rho)}^{\dagger\,Jm\tilde{m}}\right) \,, & 
\eea
and 
\bea
&  S_2^{(\Phi)} = \sum_{J,M} i (m+\tilde{m}) \tr \le \Phi^{JM}\Pi^{JM} - \Phi^{\dagger\,JM}\Pi_\phi^{\dagger\,JM} \ri \, , & 
\\
& S_2^{(\psi)} =  \sum_{JM} \sum_{\kappa= \pm 1}  (m+\tilde{m}) \tr \le  \psd_{JM,\kappa, A} \psi^A_{JM,\kappa}  \ri \, ,  &  \\
& S_2^{(A)} =  \sum_{J,m,\tilde{m}} \sum_{\rho=-1,1} \frac{i}{2} (m+\tilde{m})   \tr \left( {A}_{(\rho)}^{Jm\tilde{m}}\Pi_{(\rho)}^{Jm\tilde{m}} - A_{(\rho)}^{\dagger\,Jm\tilde{m}}\Pi_{(\rho)}^{\dagger\,Jm\tilde{m}}\right) \, . & 
\eea
In order to derive the previous action of the derivatives on the spherical harmonics, it is crucial to use the specific group elemen on the three-sphere $G/H=SO(4)/SO(3)$ in Eq.~\eqref{eq:appA_group_element_three_sphere}, since derivatives along the angular directions are required.

Notice that the expression for the fermionic charge is exactly the same before and after the redefinition \eqref{eq:redefinition_fermions}, as can be seen by direct evaluation.
For this equivalence to hold it is crucial to use the flipping of $M.$

The other relevant charges are associated to the Cartan subalgebra of the global R-symmetry of the action.
They can be written as 
\beq
Q_a = Q_a^{(\Phi)} + Q_a^{(\psi)} \, ,
\eeq
where
\bea
\label{eq:R_charges_general_scalars}
& Q_a^{(\Phi)} = i \sum_{J,M} \tr(\Phi^{JM}_a\Pi^{JM}_a - \Phi_a^{\dagger\,JM}\Pi_a^{\dagger,JM})\,, & \\
& Q_a^{(\psi)} =  \sum_{\kappa=\pm 1} \sum_{JM} \kappa \tr \le \psd_{JM, \kappa,A} (T_a)^A_{\,\,\, B} \psi^B_{JM,\kappa} \ri \, . &
\label{eq:R_charges_general_fermions}
\eea
The matrices of the Cartan subalgebra in the fundamental representation of $\SU(4)$ are
\beq
\label{eq:cartans_fund}
T_1 \equiv \frac{1}{2} \, \mathrm{diag} \lbrace 1,-1,-1,1  \rbrace \, , \qquad
T_2 \equiv \frac{1}{2} \, \mathrm{diag} \lbrace 1,-1,1,-1  \rbrace \, , \qquad
T_3 \equiv \frac{1}{2} \, \mathrm{diag} \lbrace 1,1,-1,-1  \rbrace \, .
\eeq
Notice that in the end all the explicit $\kappa$ dependence in the Hamiltonian and the other conserved quantities is only isolated to the R-charges.

\subsection*{Weights}

In order to sistematically explore the near-BPS limits of $\mathcal{N}=4$ SYM on $\mathbb{R} \times S^3,$ it is convenient to list the set of letters of the theory, which is composed by 6 complex scalars, 16 complex Grassmannian fields and 6 independent gauge field strength components, plus the descendants obtained by acting with the 4 components of the covariant derivatives.
We assign weights under the subgroups $\mathrm{SO}(4)$ (rotations) and $\SU{(4)}$ (R-symmetry) of $ \mathrm{P}\SU{(2,2|4)},$ following the conventions of reference \cite{Harmark:2007px}.  
We start with the weights under $\SU{(4)},$ which are reported in Table \ref{tab:su4_weights_scalars_Troels_paper}, \ref{tab:su4_weights_fermions1_Troels_paper} and \ref{tab:su4_weights_fermions2_Troels_paper}.
The field strength and the covariant derivatives $d_1 , d_2, \bar{d}_1, \bar{d}_2$ are uncharged under this symmetry.

\begin{table}\centering
	\ra{1.3}
	\begin{tabular}{@{}cccccc@{}}\toprule[0.1em]
		$Z$ & $X$ & $W$ & $ \bar{Z} $ & $ \bar{X}$ & $\bar{W}$ \\ \midrule[0.05em]
		$\left(1,0,0\right)$ & $\left(0,1,0\right)$ & $\left(0,0,1\right)$ & $\left(-1,0,0\right)$ & $\left(0,-1,0\right)$ & $\left(0,0,-1\right)$ \\
		\bottomrule[0.1em]
	\end{tabular}
	\caption{Scalar $\SU(4)$ weights in the notation of \cite{Harmark:2007px}}
	\label{tab:su4_weights_scalars_Troels_paper}
\end{table}

\begin{table}\centering
	\ra{1.3}
	\begin{tabular}{@{}cccc@{}}\toprule[0.1em]
		$\chi_1, \chi_2$ & $\chi_3, \chi_4$ & $\chi_5, \chi_6$ & $ \chi_7 , \chi_8 $ \\ \midrule[0.05em]
		$\left(\frac{1}{2},\frac{1}{2},\frac{1}{2}\right)$ & $\left(\frac{1}{2},-\frac{1}{2},-\frac{1}{2}\right)$ & $\left(-\frac{1}{2},\frac{1}{2},-\frac{1}{2}\right)$ & $\left(-\frac{1}{2},-\frac{1}{2},\frac{1}{2}\right)$  \\
		\bottomrule[0.1em]
	\end{tabular}
	\caption{Fermionic $\SU(4)$ weights for fermions $\chi$ in the notation of \cite{Harmark:2007px}}
	\label{tab:su4_weights_fermions1_Troels_paper}
\end{table}

\begin{table}\centering
	\ra{1.3}
	\begin{tabular}{@{}cccc@{}}\toprule[0.1em]
		$ \bar{\chi}_1, \bar{\chi}_2 $ & $ \bar{\chi}_3, \bar{\chi}_4$ & $ \bar{\chi}_5, \bar{\chi}_6$ & $ \bar{\chi}_7, \bar{\chi}_8$\\ \midrule[0.05em]
		 $\left(-\frac{1}{2},-\frac{1}{2},-\frac{1}{2}\right)$ & $\left(-\frac{1}{2},\frac{1}{2},\frac{1}{2}\right)$ & $\left(\frac{1}{2},-\frac{1}{2},\frac{1}{2}\right)$ & $\left(\frac{1}{2},\frac{1}{2},-\frac{1}{2}\right)$ \\
		\bottomrule[0.1em]
	\end{tabular}
	\caption{Fermionic $\SU(4)$ weights for fermions $\bar{\chi}$ in the notation of \cite{Harmark:2007px}}
	\label{tab:su4_weights_fermions2_Troels_paper}
\end{table}

In order to make contact with the notation used in this work, we read off the $\SU(4)$ R-symmetry weights of all fields using \eqref{eq:R_charges_general_scalars}, \eqref{eq:R_charges_general_fermions} and \eqref{eq:cartans_fund}. 
We list them in tables \ref{tab:su4_weights_scalars} and \ref{tab:su4_weights_fermions}, respectively.
By looking at the dynamical modes that we describe in the sectors of section \ref{sect-sphere_reduction}, we verify that the results are consistent with the list of surviving field in the limits given in reference \cite{Harmark:2007px}, see subsection below.

\begin{table}\centering
	\ra{1.3}
	\begin{tabular}{@{}ccc@{}}\toprule[0.1em]
		$\Phi_1$ & $\Phi_2$ & $\Phi_3$ \\ \midrule[0.05em]
		$\left(1,0,0\right)$ & $\left(0,1,0\right)$ & $\left(0,0,1\right)$ \\
		\bottomrule[0.1em]
	\end{tabular}
	\caption{Scalar $\SU(4)$ weights}
	\label{tab:su4_weights_scalars}
\end{table}

\begin{table}\centering
	\ra{1.3}
	\begin{tabular}{@{}ccccc@{}}\toprule[0.1em]
		& $\psi_1$ & $\psi_2$ & $\psi_3$ & $\psi_4$ \\ \midrule[0.05em]
		$\kappa = 1$  & $\left(\frac{1}{2},\frac{1}{2},\frac{1}{2}\right)$ & $\left(-\frac{1}{2},-\frac{1}{2},\frac{1}{2}\right)$ & $\left(-\frac{1}{2},\frac{1}{2},-\frac{1}{2}\right)$ & $\left(\frac{1}{2},-\frac{1}{2},-\frac{1}{2}\right)$ \\
		$\kappa = -1$ & $\left(-\frac{1}{2},-\frac{1}{2},-\frac{1}{2}\right)$ & $\left(\frac{1}{2},\frac{1}{2},-\frac{1}{2}\right)$ & $\left(\frac{1}{2},-\frac{1}{2},\frac{1}{2}\right)$ & $\left(-\frac{1}{2},\frac{1}{2},\frac{1}{2}\right)$ \\
		\bottomrule[0.1em]
	\end{tabular}
	\caption{Fermionic $\SU(4)$ weights}
	\label{tab:su4_weights_fermions}
\end{table}

It is also convenient to compare this notation with the conventions of the paper \cite{Ishiki:2006rt}, where the antisymmetric representation $\mathbf{4}$ is instead used for the scalar fields.
In Eq.~(2.19) of reference \cite{Ishiki:2006rt} the transformation properties of the antisymmetric tensor and of the fermions under R-symmetry are reported:
\begin{equation}
	\delta_R X^{AB} = i T^A{}_CX^{CB} + T^B{}_CX^{AC}\,, \qquad
	\delta_R \psi^A = i T^A{}_B \psi^B\,.
	\end{equation}
The weights of scalars can be immediately deduced from this rule and from the basis for the generators \eqref{eq:cartans_fund}.
The transformation of the fermionic fields is the same used here to derive the expression \eqref{eq:R_charges_general_fermions}, and then they agree.
The comparison is consistent if we choose
\begin{equation}
\boxed{	\Phi_1 = X_{14} \,,\quad\Phi_2 = X^{\dagger}_{24}\,,\quad\Phi_3 = X^{\dagger}_{34}\,. }
	\label{eq:final_mapping_scalars}
	\end{equation}
This is the dictionary that we will use throughout all the computations in the present work.

In addition, the fields also carry charge under $SO(4)$ rotations, except for the scalars.
We list their quantum number in the subsection below, referring to the specific sectors where we take the limits.

\subsection*{Charges for the specific near-BPS limits}

From Eq.~\eqref{eq:appB_rotation_charges} applied to the specific cases, we identify
\begin{equation}
S_1 = - m + \tilde{m} \spa S_2 = m+\tilde{m}
\end{equation}
Here we list the rotation and the R-symmetry charges for all the limits considered in this work.
\begin{itemize}
\item In the bosonic $\SU(1,1)$ sector we have a surviving dynamical scalar with derivatives $d_1^n Z.$ 
The associated momenta and charges are
\bea
& -m = \tilde{m} = J 
\spa  
S_1 = 2J \spa S_2 = 0 \, , &
\\
& (Q_1,Q_2,Q_3) = (1,0,0) \,, &
\eea
with $2J \in \mathbb{N}$ and $n=2J$.

\item In the fermionic $\SU(1,1)$ limit we have the fermion with derivatives $d_1^n \chi_1$ with quantum numbers
\bea
& \kappa= 1 \spa
m = - J - \frac{1}{2} \spa \tilde{m} = - J\spa
S_1 = 2J + \frac{1}{2} \spa S_2 = -\frac{1}{2} \,,
& \\
& (Q_1,Q_2,Q_3) = \le \frac{1}{2},\frac{1}{2},\frac{1}{2} \ri \,, &
\eea
with $2J \in \mathbb{N}$ and $n=2J$.

\item The $\SU(1,1|1)$ sector simply contains the union of the degrees of freedom in the $\SU(1,1)$ bosonic and fermionic sectors.
Then, the field content and the quantum numbers are the same reported above.

\item In the $\PSU(1,1|2)$ sector, in addition to the abovementioned fields $d_1^n Z , d_1^n \chi_1 ,$ there are one more scalar field with derivatives $d_1^n X$ and one more fermion with derivatives $d_1^n \bar{\chi}_7 . $

The additional scalar has the same quantum numbers as $d_1^n Z,$ {\sl i.e.}
\bea
& -m = \tilde{m} = J 
\spa 
S_1 = 2J \spa S_2 = 0 \, ,
& \\
&
(Q_1,Q_2,Q_3) = (1,0,0) \,, &
\eea
with $2J \in \mathbb{N}$ and $n=2J$.

The fermion $d_1^n \bar{\chi}_7$ has instead different momenta and charges, given by
\bea
& \kappa=-1  \spa
m = - J  \spa \tilde{m} = J + \frac{1}{2} \spa
S_1 = 2J + \frac{1}{2} \spa S_2 = -\frac{1}{2} \,,
&
 \\
&
(Q_1,Q_2,Q_3) = \le \frac{1}{2},\frac{1}{2},-\frac{1}{2} \ri \,, &
\eea
with $2J \in \mathbb{N}$ and $n=2J$.

\end{itemize}

\subsection*{Interacting Hamiltonian}

Using the decomposition into spherical harmonics on the three-sphere, we derive the interacting Hamiltonian of $\mathcal{N}=4$ SYM on $\mathbb{R} \times S^3 .$
The entire expression can be found in \cite{Ishiki:2006rt}, but we report here the result using our notation.
Due to the redefinition of fermions \eqref{eq:redefinition_fermions} and the dictionary \eqref{eq:final_mapping_scalars}, it is convenient to introduce the notations
\beq
(Z_a)_{JM}  \equiv \begin{pmatrix} (\Phi_1)_{JM} \\
 (-1)^{m-\tilde{m}} (\Phi^\dagger_2)_{J,-M} \\
  (-1)^{m-\tilde{m}} (\Phi^\dagger_3)_{J,-M} 
  \end{pmatrix} \, ,
  \label{eq:mapping_Z_scalars}
\eeq
\beq
(\Psi_A)_{J,M,\kappa=1} \equiv   (\psi^{\dagger}_A)_{J,-M,\kappa=1} \, , \qquad
(\Psi_A)_{J,M,\kappa=-1} \equiv (\psi_A)_{J,M, \kappa=-1} \, .
\label{eq:mapping_Psi_fermions}
\eeq 
Notice that these definitions account precisely for the different interpretation of scalars and fermions with respect to reference \cite{Ishiki:2006rt}, see {\sl i.e.} the action of complex conjugation on spherical harmonics described in Eq.~\eqref{eq:app_complex_conjugate_spher_harmonics}. 

The result is:
\beq
\begin{aligned}
H_{\rm int}   = \sum_{J_i, M_i, \kappa_i, \rho_i} & \tr  \left\lbrace i g {\cal C}^{J_2M_2}_{J_1M_1;JM} \, \chi_{JM} \left([(Z_a^\dagger)_{J_2M_2},(\Pi^{(\Phi)\dagger}_a)_{J_1M_1}] + [Z^a_{J_1M_1},\Pi^{a(\Phi)}_{J_2M_2}]\right)  \right. \\
& \left. - 4g\sqrt{J_1(J_1+1)} {\cal D}^{J_2M_2}_{J_1M_1 0; JM\rho} \, A_{(\rho)}^{JM}
[Z^a_{J_1M_1},(Z_a^\dagger)_{J_2M_2}]  \right. \\
& \left. +g \mathcal{F}^{J_1 M_1 \kappa_1}_{J_2 M_2 \kappa_2; JM}  \, \chi_{JM} 
\lbrace (\Psi_A^{\dagger})_{J_1 M_1 \kappa_1} , \Psi^A_{J_2 M_2 \kappa_2} \rbrace \right. \\
& \left. + g \mathcal{G}^{J_1 M_1 \kappa_1}_{J_2 M_2 \kappa_2; JM \rho} \, A_{(\rho)}^{JM}
\lbrace (\Psi^{\dagger}_A)_{J_1 M_1 \kappa_1} , \Psi^A_{J_2 M_2 \kappa_2} \rbrace   
  \right. \\
  & \left. + \frac{g^2}{2} {\cal C}^{J_2 M_2}_{J_1 M_1;J M}{\cal C}^{J_3 M_3}_{J_4 M_4;J M}[Z^a_{J_1 M_1},(Z^{\dagger}_a)_{J_2 M_2}][Z^b_{J_3 M_3},(Z_b^{\dagger})_{J_4 M_4}] \right. \\
 & \left.   - \sqrt{2} ig (-1)^{-m_1+\tilde{m}_1+\frac{\kappa_1}{2}} \mathcal{F}^{J_1,-M_1,\kappa_1}_{J_2 M_2 \kappa_2; J M} 
 \psi^4_{J_2 M_2 \kappa_2} [(Z_a)^{JM} , \Psi^a_{J_1 M_1 \kappa_1}]  \right. \\
 & \left.   + \sqrt{2} ig (-1)^{-m_1+\tilde{m}_1+\frac{\kappa_1}{2}} \mathcal{F}^{J_1,-M_1,\kappa_1}_{J_2 M_2 \kappa_2; J M} 
\, \epsilon_{abc} \Psi^a_{J_1 M_1 \kappa_1} [(Z_b^{\dagger})^{JM} , \Psi^c_{J_2 M_2 \kappa_2}]  \right. \\
  & \left.   + \sqrt{2} ig (-1)^{m_2-\tilde{m}_2+\frac{\kappa_2}{2}} \mathcal{F}^{J_1 M_1 \kappa_1}_{J_2, -M_2, \kappa_2; J M} 
 (\Psi^{\dagger}_4)_{J_2 M_2 \kappa_2} [(Z^{\dagger}_a)^{JM} , (\Psi^{\dagger}_a)_{J_1 M_1 \kappa_1}]  \right. \\
 & \left.   - \sqrt{2} ig (-1)^{m_2-\tilde{m}_2+\frac{\kappa_2}{2}} \mathcal{F}^{J_1 M_1 \kappa_1}_{J_2, -M_2, \kappa_2; J M} 
\, \epsilon_{abc} (\Psi^{\dagger}_a)_{J_1 M_1 \kappa_1} [(Z_b)^{JM} , (\Psi^{\dagger}_c)_{J_2 M_2 \kappa_2}]  \right. \\
& \left. + i g {\cal D}^{JM}_{J_1M_1\rho_1;J_2M_2\rho_2} \, \chi_{JM} [\Pi^{J_1M_1}_{(\rho_1)},A^{J_2M_2}_{(\rho_2)}] \right. \\
& \left. + g^2 \mathcal{C}^{JM}_{J_2 M_2; J_4, -M_4} \mathcal{D}_{JM; J_1 M_1 \rho_1; J_3 M_3 \rho_3} [A_{(\rho_1)}^{J_1 M_1} , Z^a_{J_2 M_2}] [A_{(\rho_3)}^{J_3 M_3} , (Z^{\dagger}_a)_{J_4 M_4}]   \right. \\
& \left. + 2 i g \rho_1(J_1 + 1) {\cal E}_{J_1M_1\rho_1;J_2M_2\rho_2;J_3M_3\rho_3}A^{J_1M_1}_{(\rho_1)}[A^{J_2M_2}_{(\rho_2)},A^{J_3M_3}_{(\rho_3)}] \right. \\
& \left. - \frac{g^2}{2} {\cal D}^{JM}_{J_1M_1\rho_1; J_3M_3\rho_3} {\cal D}_{JM;J_2M_2\rho_2; J_4 M_4 \rho_4}[A^{J_1M_1}_{(\rho_1)},A^{J_2M_2}_{(\rho_2)}][A^{J_3M_3}_{(\rho_3)},A^{J_4M_4}_{(\rho_4)}] \right. \\
& \left. -2g \sqrt{J_1(J_1+1)} \mathcal{D}_{J_2 M_2; J_1 M_1 0; J M \rho} \, \chi_{J_1 M_1} [\chi_{J_2 M_2} , A_{(\rho)}^{JM}] \right. \\
& \left. + \frac{g^2}{2} \mathcal{C}^{JM}_{J_1 M_1; J_3 M_3} \mathcal{D}_{JM; J_2 M_2 \rho_2; J_4 M_4 \rho_4} [\chi_{J_1 M_1} , A_{(\rho_2)}^{J_2 M_2}] [\chi_{J_3 M_3} , A_{(\rho_4)}^{J_4 M_4}]  \right. \\
& \left. + g^2 \mathcal{C}^{JM}_{J_1 M_1; J_2 M_2} \mathcal{C}_{JM; J_3 M_3; J_4 M_4} [\chi_{J_1 M_1} , Z^a_{J_2M_2}] [\chi_{J_3 M_3} , (Z_a^{\dagger})_{J_4 M_4}]
 \right\rbrace \, .
\end{aligned}
\label{eq:app_full_interacting_N=4SYM_Hamiltonian}
\eeq
The notation used is the following. The initial sum represents a summation over all contracted indices: momenta $(J,M),$ labels for the spherical harmonics involving fermions $\kappa$ and gauge fields $\rho,$ and indices of the fields under $\SU{(4)}$ R-symmetry. 
The Yukawa term contains sums of the spinors over only the subset $a \in \lbrace 1,2,3 \rbrace$ and the Levi-Civita symbol $\epsilon_{abc}$ is defined in such a way that $\epsilon_{123}=1.$
In order to avoid confusion, we specified that $\Pi^{\Phi}_a$ are the canonical momenta associated to the scalar fields $\Phi_a,$ while $\Pi_{(\rho)}$ is the symplectic partner of the gauge field $A_{(\rho)}.$

The terms involving the gauge fields, except for the terms contributing to the bosonic and fermionic currents, are not needed for the near-BPS limits included in this work, but are put for completeness.
In order to derive from this expression the relevant contributions to the interacting Hamiltonians in section \ref{sect-sphere_reduction}, few simplifications still need to be performed, in order to obtain the fields $\Phi_a, \psi_A$ from the variables \eqref{eq:mapping_Z_scalars} and \eqref{eq:mapping_Psi_fermions}.
The simplified expressions are written explicitly in section \ref{sect-sphere_reduction} for each case considered.


\section{Properties of spherical harmonics and Clebsch-Gordan coefficients}
\label{app-properties_Clebsch_Gordan_coeff}

\subsection*{Definition of the Clebsch-Gordan coefficients}

We give the explicit definitions of the Clebsch-Gordan coefficients which are used to compute the interacting Hamiltonians in section \ref{sect-sphere_reduction}.
They were previously given e.g. in \cite{Ishiki:2006rt}.
\beq
\mathcal{C}^{J_1 M_1}_{J_2 M_2; JM} = 
\sqrt{\frac{(2J+1)(2J_2+1)}{2J_1+1}}
C^{J_1 m_1}_{J_2 m_2; J m} C^{J_1 \tilde{m}_1}_{J_2 \tilde{m}_2; J \tilde{m}} \, ,
\label{eq:app_definitionClebsch_C}
\eeq
\beq
\begin{aligned}
\mathcal{D}^{J_1 M_1}_{J_2 M_2 \rho_2; JM \rho} & = 
(-1)^{\frac{\rho_2+\rho}{2} +1}
\sqrt{3(2J_2+1)(2J_2+2 \rho_2^2 +1)(2J+1)(2J+2 \rho^2 +1)} \\
& \times C^{J_1,m_1}_{Q_2,m_2; Q,m} C^{J_1, \tilde{m}_1}_{\tilde{Q}_2, \tilde{m}_2 ; \tilde{Q},\tilde{m}}
\begin{Bmatrix}
Q_2 & \tilde{Q}_2 & 1 \\
Q & \tilde{Q} & 1 \\
J_1 & J_1 & 0  
\end{Bmatrix} \, ,
\label{eq:app_definitionClebsch_D}
 \end{aligned}
\eeq
\beq
\begin{aligned}
 \mathcal{E}_{J_1 M_1 \rho_1; J_2 M_2 \rho_2; JM \rho} & =  \sqrt{6(2J_1+1)(2J_1 + 2 \rho_1^2 +1)(2J_2+1)(2J_2+2 \rho_2^2 +1)(2J+1)(2J+2 \rho^2 +1)} \\
& \times (-1)^{-\frac{\rho_1+\rho_2+\rho+1}{2}} 
\begin{Bmatrix}
Q_1 & \tilde{Q}_1 & 1 \\
Q_2 & \tilde{Q}_2 & 1 \\
Q & \tilde{Q} & 1 
\end{Bmatrix} 
\begin{pmatrix}
Q_1 & Q_2 & Q \\
m_1 & m_2 & m
\end{pmatrix}
\begin{pmatrix}
\tilde{Q}_1 & \tilde{Q}_2 & \tilde{Q} \\
\tilde{m}_1 & \tilde{m}_2 & \tilde{m}
\end{pmatrix} \, ,
\label{eq:app_definitionClebsch_E}
\end{aligned}
\eeq
\beq
\begin{aligned}
\mathcal{F}^{J_1 M_1 \kappa_1}_{J_2 M_2 \kappa_2; JM} = & (-1)^{\tilde{U}_1+U_2+J+\frac{1}{2}} \sqrt{(2J+1)(2J_2+1)(2J_2+2)}  \\
& \times C^{U_1, m_1}_{U_2, m_2; J,m} C^{\tilde{U}_1,\tilde{m}_1}_{\tilde{U}_2,\tilde{m}_2; J,\tilde{m}}
\begin{Bmatrix}
U_1 & \tilde{U}_1 & \frac{1}{2} \\
\tilde{U}_2 & U_2 & J
\end{Bmatrix} \, ,
\label{eq:app_definitionClebsch_F}
\end{aligned}
\eeq
\beq
\begin{aligned}
\mathcal{G}^{J_1 M_1 \kappa_1}_{J_2 M_2 \kappa_2; JM \rho} = & (-1)^{\frac{\rho}{2}} \sqrt{6(2J_2+1)(2J_2+2)(2J+1)(2J+2 \rho^2 +1)} \\
& \times C^{U_1,m_1}_{U_2,m_2; Q,m} C^{\tilde{U}_1, \tilde{m}_1}_{\tilde{U}_2, \tilde{m}_2 ; \tilde{Q},\tilde{m}}
\begin{Bmatrix}
U_1 & \tilde{U}_1 & \frac{1}{2} \\
U_2 & \tilde{U}_2 & \frac{1}{2} \\
Q & \tilde{Q} & 1 
\end{Bmatrix} \, ,
\label{eq:app_definitionClebsch_G}
\end{aligned}
\eeq
where we defined the quantities
\beq
U \equiv J + \frac{\kappa+1}{4} \spa
\tilde{U} \equiv J + \frac{1-\kappa}{4} \spa
Q \equiv J+ \frac{\rho(\rho+1)}{2} \spa
\tilde{Q} \equiv J + \frac{\rho(\rho-1)}{2} \, .
\label{eq:app_labels_harmonics}
\eeq
Properties of 9-j and 6-j Wigner symbols were used to write the coefficient $\mathcal{F}$ in this form, but the expression is still completely general.

In view of the crossing relations that we will derive, it is also useful to record the integral representation of the previous Clebsch-Gordan coefficients as products of spherical harmonics on the three-sphere.
Precisely, they are given by
\beq
\mathcal{C}^{J_1 M_1}_{J_2 M_2; JM} = \int_{S^3} d \Omega \, \bar{\mathcal{Y}}_{J_1 M_1} \mathcal{Y}_{J_2 M_2} \mathcal{Y}_{JM} \, , 
\label{eq:app_integral_repr_C}
\eeq
\beq
\mathcal{D}^{J_1 M_1}_{J_2 M_2 \rho_2; JM \rho} = \int_{S^3} d \Omega \, \bar{\mathcal{Y}}_{J_1 M_1} \mathcal{Y}^{\rho_2}_{J_2 M_2 i} \mathcal{Y}^{\rho}_{JM i} \, , 
\label{eq:app_integral_repr_D}
\eeq
\beq
\mathcal{E}_{J_1 M_1 \rho_1; J_2 M_2 \rho_2; JM \rho} = \int_{S^3} d \Omega \, \epsilon_{ijk} \mathcal{Y}^{\rho_1}_{J_1 M_1 i} \mathcal{Y}^{\rho_2}_{J_2 M_2 j} \mathcal{Y}^{\rho}_{JM k} \, , 
\label{eq:app_integral_repr_E}
\eeq
\beq
\mathcal{F}^{J_1 M_1 \kappa_1}_{J_2 M_2 \kappa_2; JM} = \int_{S^3} d \Omega \, \bar{\mathcal{Y}}^{\kappa_1}_{J_1 M_1 \alpha} \mathcal{Y}^{\kappa_2}_{J_2 M_2 \alpha} \mathcal{Y}_{JM} \, , 
\label{eq:app_integral_repr_F}
\eeq
\beq
\mathcal{G}^{J_1 M_1 \kappa_1}_{J_2 M_2 \kappa_2; JM \rho} = \int_{S^3} d \Omega \, \sigma^i_{\alpha \beta} \bar{\mathcal{Y}}^{\kappa_1}_{J_1 M_1 \alpha} \mathcal{Y}^{\kappa_2}_{J_2 M_2 \beta} \mathcal{Y}^{\rho}_{JMi} \, ,
\label{eq:app_integral_repr_G}
\eeq
where $\bar{\mathcal{Y}}$ denotes the complex conjugate of the harmonics $\mathcal{Y}.$
We reported here the coefficient $\mathcal{E}$ for completeness, but it will never enter in any interacting Hamiltonian for the near-BPS limits considered in this work because it only couple terms containing dynamical gauge fields, while in the $\SU(1,1)$ sector and its generalizations the gauge field always decouples.

At this point we start specializing these definitions to the cases of interest for the near-BPS limits.
The crossing relations between them will allow to analitically solve the sums over intermediate momenta $J$ appearing in the computation of the interacting Hamiltonian.

We start from $\mathcal{C},$ which enters all the computations of the various sectors only via the prescription of momenta \eqref{eq:short-hand_notation_momenta}.
Using the definition \eqref{eq:app_definitionClebsch_C} and specializing to this case, we easily obtain by direct computation
\beq
\mathcal{C}^{\CJ_1}_{\CJ_2;JM} \equiv \mathcal{C}^{J_1,-J_1,J_1}_{J_2,-J_2,J_2; J m \tilde{m}} =
(-1)^{J-J_1+J_2} \sqrt{\frac{(2J+1)(2J_2+1)}{2J_1+1}} \, \frac{(2J_1+1)!(2J_2)!}{(J_1+J_2-J)!(J+J_1+J_2+1)!} \, .
\eeq

\subsection*{Crossing relations at saturated angular momenta -- ${\cal C}$ and ${\cal D}$}
\label{app-crossing_CD}

We consider the definition \eqref{eq:app_definitionClebsch_D} and we specialize the momenta to the assignments in Eq.~\eqref{eq:short-hand_notation_momenta} with $\rho= \pm 1.$
In fact, these are the only two cases of interest for the computation of the interacting part of the Hamiltonian mediated by the non-dynamical gauge field.
The explicit expressions are
\beq
\begin{aligned}
\mathcal{D}^{\CJ_1}_{\CJ_2; Jm \tilde{m}, \rho=1} = & - i (-1)^{J-J_1+J_2} \sqrt{\frac{(2J_1+1)(J+\Delta J+1)(J-\Delta J+1)}{J_2(J+1)(J_2+1)(2J_2+1)}} \\ 
& \times  \frac{(2J_1)!(2J_2+1)!}{2(J+1+J_1+J_2)!(J_1+J_2-J-1)!} \, ,
\end{aligned}
\eeq
where $\Delta J= J_1 - J_2 ,$ and
\beq
\mathcal{D}^{\CJ_1}_{\CJ_2; Jm \tilde{m}, \rho=-1} = - \mathcal{D}^{\CJ_1}_{\CJ_2; Jm \tilde{m}, \rho=1}  \, .
\label{eq:app_relations_between_twoprescriptions_D}
\eeq
It is convenient for the following manipulations to factorize from this formula appropriate factors of the Clebsch-Gordan coefficient $\mathcal{C}$ computed above.
We find
\beq
\begin{aligned}
\mathcal{D}^{\CJ_1}_{\CJ_2; Jm \tilde{m}, \rho=1} & = \frac{i}{2} (J-J_1-J_2) \sqrt{\frac{(J+\Delta J +1)(J - \Delta J +1)}{J_2(J+1)(2J+1)(J_2+1)}}
\mathcal{C}^{\CJ_1}_{\CJ_2; Jm \tilde{m}} = \\
& = \frac{i}{2} (J+J_1+J_2+2) \sqrt{\frac{(J+\Delta J+1)(J - \Delta J+1)}{J_2(J+1)(2J+3)(J_2+1)}}
\mathcal{C}^{\CJ_1}_{\CJ_2; J+1,m \tilde{m}} \, .
\end{aligned}
\eeq
At this point, we consider appropriate quadratic combinations of the Clebsch-Gordan coefficients $\mathcal{C}, \mathcal{D}$ in view of finding simplifications which allow to solve the sum over intermediate momenta $J.$
We define the quantities
\begin{equation}
\label{eq:ACDdef}
{\cal A}_{\CJ_1,\CJ_4;Jm\tilde{m}}^{\CJ_2,\CJ_3} = \bigg(
1
+ \frac{( \omega_{J_1}+\omega_{J_2})( \omega_{J_3}+\omega_{J_4})}{4J(J+1)}\bigg) {\cal C}^{\CJ_2}_{\CJ_1;Jm\tilde{m}}{\cal C}^{\CJ_3}_{\CJ_4;Jm\tilde{m}}
\end{equation}
and
\begin{equation}
{\cal B}_{\CJ_1,\CJ_4;Jm\tilde{m}\rho}^{\CJ_2,\CJ_3} = \frac{16}{\omega_{A,J}^2 - (m-\tilde{m})^2}\sqrt{J_1(J_1+1)J_4(J_4+1)}{\cal D}^{\CJ_2}_{\CJ_1; Jm\tilde{m}\rho}{\cal \bar{D}}^{\CJ_3}_{\CJ_4; Jm\tilde{m}\rho}\,,
\end{equation}
which for $\rho = \pm 1$ reads
\begin{align}
 \label{eq:Bp1CDdef}
{\cal B}_{\CJ_1,\CJ_4;Jm\tilde{m}, \rho=1}^{\CJ_2,\CJ_3} &= 
\frac{ (2 + J + J_1 + J_2)(2 + J + J_3 + J_4)
}{(J+1)(2J+3)}
{\cal C}^{\CJ_2}_{\CJ_1,J+1\, m \tilde{m}}
{\cal C}^{\CJ_3}_{\CJ_4,J+1\, m \tilde{m}} \,, \\
\label{eq:Bm1CDdef}
{\cal B}_{\CJ_1,\CJ_4;Jm\tilde{m}, \rho=-1}^{\CJ_2,\CJ_3} &= 
\frac{ (J_1+J_2-J)(J_3+J_4-J)
}{(J+1)(2J+1) } {\cal C}^{\CJ_2}_{\CJ_1,J m \tilde{m}}
 {\cal C}^{\CJ_3}_{\CJ_4,J m \tilde{m}}\,.
\end{align}
Simple algebraic manipulations now give rise to the relation
\begin{equation}
\label{eq:CDrelation}
{\cal B}_{\CJ_1,\CJ_4;Jm\tilde{m},\rho=-1}^{\CJ_2,\CJ_3} + {\cal B}_{\CJ_1,\CJ_4;J-1\,m\tilde{m},\rho=1}^{\CJ_2,\CJ_3}
= {\cal A}_{\CJ_1,\CJ_4;Jm\tilde{m}}^{\CJ_2,\CJ_3}\,,
\end{equation}
valid for $J \geq 1$.

As a particular application, we find
\begin{align}
\sum_{Jm\tilde{m}} \Bigg(&\bigg(
1
+ \frac{( \omega_{J_1}+\omega_{J_2})( \omega_{J_3}+\omega_{J_4})}{4J(J+1)}\bigg) {\cal C}^{\CJ_2}_{\CJ_1,Jm\tilde{m}}{\cal C}^{\CJ_3}_{\CJ_4,Jm\tilde{m}}\nonumber\\
&\hspace{2em}-\sum_{\rho = \pm 1} \frac{16}{\omega_{A,J}^2 - (m-\tilde{m})^2}\sqrt{J_1(J_1+1)J_4(J_4+1)}{\cal D}^{\CJ_2}_{\CJ_1; Jm\tilde{m}\rho}{\cal \bar{D}}^{\CJ_3}_{\CJ_4; Jm\tilde{m}\rho}
\Bigg)\nonumber\\
\label{eq:app_Hphi}
&= \sum_{J \geq J_\text{min}(\rho)} \left(
{\cal A}_{\CJ_1,\CJ_4;J,-\Delta J, \Delta J}^{\CJ_2,\CJ_3} - {\cal B}_{\CJ_1,\CJ_4;J,-\Delta J, \Delta J,\rho=-1}^{\CJ_2,\CJ_3} - {\cal B}_{\CJ_1,\CJ_4;J,-\Delta J, \Delta J,\rho=1}^{\CJ_2,\CJ_3}
\right)
\,,
\end{align}
where here $\Delta J = J_2 - J_1 = J_3 - J_4.$ 
In general, if we also define $\Delta m = m_2 - m_1 = m_3 - m_4$ (and similarly for the $\tilde{m}$), we should be careful in distinguishing the cases $|\Delta{J}| < |\Delta{m}|$ and $|\Delta{J}| \geq |\Delta{m}|,$ because the triangle inequalities and the constraints on the eigenvalues of momenta imply that the lower bound of summation for $J$ changes.

In this case, however, all the momenta are fixed and we notice that $ \Delta m=- \Delta \tilde{m} = - \Delta J,$ so there is only one case to consider.
The eqs.\eqref{eq:ACDdef} - \eqref{eq:Bp1CDdef} imply that $J_\text{min} = |\Delta{J}|$ for all the terms in \eqref{eq:app_Hphi}. Shifting $J \to J-1$ thus only cancels the terms in the sum with $J > |\Delta{J}|$, leaving the final expression
\beq
\hspace{-\leftmargin}
\sum_{J \geq |\Delta{J}|} \left(
{\cal A}_{\CJ_1,\CJ_4;J, -\Delta J, \Delta J}^{\CJ_2,\CJ_3} - {\cal B}_{\CJ_1,\CJ_4;J,-\Delta J, \Delta J,\rho=-1}^{\CJ_2,\CJ_3} - {\cal B}_{\CJ_1,\CJ_4;J,-\Delta J, \Delta J,\rho=1}^{\CJ_2,\CJ_3}
\right) = 
{\cal B}_{\CJ_1,\CJ_4;|\Delta J|-1\, , -\Delta J, \Delta J,\rho=1}^{\CJ_2,\CJ_3}\,,
\eeq
where we have used Eq.~\eqref{eq:CDrelation} to simplify the result.

Explicit evaluation yields
\beq
 \hspace{-4mm} {\cal B}_{\CJ_1,\CJ_4;|\Delta J|-1,-\Delta J, \Delta J,\rho=1}^{\CJ_2,\CJ_3} \\
= \frac{ (1 +|\Delta J| + J_1 + J_2)(1 + |\Delta J| + J_3 + J_4)
}{|\Delta J|(2|\Delta J|+1)}
{\cal C}^{\CJ_2}_{\CJ_1,|\Delta J|\, , -\Delta{J}, \Delta{J}}
{\cal C}^{\CJ_3}_{\CJ_4,|\Delta J|\, , -\Delta{J}, \Delta{J}}\,.
\eeq

\subsection*{Crossing relations at saturated angular momenta -- ${\cal C}$ and ${\cal F}$}

We start by deriving a couple of properties which relate Clebsch-Gordan coefficients $\mathcal{F}$ with different assignments of momenta, useful to obtain the simplification in Eq.~\eqref{yukawa_terms}:
\bea
\label{eq:property_complex_conjugation_F_coefficients}
& \mathcal{F}^{J_1,M_1,\kappa_1}_{J_2,-M_2,\kappa_2 ; JM} = (-1)^{m- \tilde{m}} \mathcal{F}^{J_2,-M_2,\kappa_2}_{J_1,M_1,\kappa_1 ; J,-M} \, , & \\
& \mathcal{F}^{J_1,M_1,\kappa_1}_{J_2,M_2,\kappa_2 ; JM} =
(-1)^{m_1 - \tilde{m}_1 + m_2 - \tilde{m}_2 + \frac{\kappa_1 + \kappa_2}{2}} \mathcal{F}^{J_2,-M_2,\kappa_2}_{J_1,-M_1,\kappa_1 ; JM} \, .  &
\label{eq:properties_F_coefficients}
\eea
These identities can be easily derived by using the integral representation \eqref{eq:app_integral_repr_F} combined with the properties of spherical harmonics under complex conjugation, {\sl i.e.}
\beq
\bar{\mathcal{Y}}_{JM} = (-1)^{m - \tilde{m}} \mathcal{Y}_{J, -M} \, ,
\qquad
\bar{\mathcal{Y}}^{\kappa}_{JM \alpha} = (-1)^{m - \tilde{m}+ \kappa \alpha +1} \mathcal{Y}^{\kappa}_{J, -M, - \alpha} \, .
\label{eq:app_complex_conjugate_spher_harmonics}
\eeq
Now we focalize instead to the specific cases of interest for the computation of the Hamiltonian in the near-BPS bounds of interest.
Contrarily to the bosonic case, the Clebsch-Gordan coefficients involving spherical harmonics of fermions appear in the sectors with two different possibilities, see the analysis of the $\PSU(1,1|2)$ sector: with chirality $\kappa=1$ and momenta $(m, \tilde{m})=(-J-\frac{1}{2},J)$ or with $\kappa=-1$ and momenta $(-J,J+\frac{1}{2}).$
Due to the redefinition of the former in Eq.~\eqref{eq:redefinition_fermions}, we get the two quantities
\beq
\mathcal{F}^{J_1, J_1+\frac{1}{2},-J_1,\kappa_1=1}_{J_2,J_2+\frac{1}{2},-J_2,\kappa=1; Jm \tilde{m}} \, , \qquad
\mathcal{F}^{J_1,-J_1, J_1+\frac{1}{2},\kappa_1=-1}_{J_2,-J_2,J_2+\frac{1}{2},\kappa=-1; Jm \tilde{m}} \, .
\label{eq:app_two_assignments_momenta_ClebschF}
\eeq
Looking at the definition \eqref{eq:app_definitionClebsch_F}, we notice that the two expressions are related.
In fact, the two factors of $\SU(2)$ Clebsch-Gordan coefficients $C$ entering the definition of $\mathcal{F}$ are simply exchanged in the two cases, while the Wigner 6-j symbols are the same due to the symmetry under the interchange of two elements of a line with the other one
\beq
\begin{Bmatrix}
J_1 + \frac{1}{2} & J_1 & \frac{1}{2} \\
J_2 & J_2 + \frac{1}{2} & J 
\end{Bmatrix} =
\begin{Bmatrix}
J_1 & J_1+ \frac{1}{2} & \frac{1}{2} \\
J_2 + \frac{1}{2} & J_2  & J 
\end{Bmatrix} \, .
\eeq
Finally, the triangle inequalities and the conditions on integer sums of momenta coincide in the two cases, thus we conclude that
\beq
\mathcal{F}^{J_1, J_1+\frac{1}{2},-J_1,\kappa_1=1}_{J_2,J_2+\frac{1}{2},-J_2,\kappa=1; Jm \tilde{m}} =
\mathcal{F}^{J_1,-J_1, J_1+\frac{1}{2},\kappa_1=-1}_{J_2,-J_2,J_2+\frac{1}{2},\kappa=-1; Jm \tilde{m}} \, .
\label{eq:app_relation_between_twoprescriptions_F}
\eeq
More generally, the interactions of the $\PSU(1,1|2)$ sector also involve terms where the chiralities $\kappa_1, \kappa_2$ entering the definition of $\mathcal{F}$ assume both values $\pm 1.$   
In such case, it is convenient to find additional crossing relations between the coefficients with various assignments of momenta and chiralities.

We follow the same steps depicted above, but now we notice that when $\kappa_1, \kappa_2$ are not fixed the triangle conditions on momenta imply
$J+J_1 + J_2 \in \mathbb{Z},$ while when $\kappa_1 = - \kappa_2$ we have $J+J_1+J_2+\frac{1}{2} \in \mathbb{Z}.$
This gives only two possibilities for the overall sign, which are
\beq
\mathcal{F}^{J_1 m_1 \tilde{m}_1 \kappa_1}_{J_2 m_2 \tilde{m}_2 \kappa_2 ; J m \tilde{m}} =
\begin{cases}
	\mathcal{F}^{J_1, \tilde{m}_1, m_1, -\kappa_1}_{J_2, \tilde{m}_2, m_2, -\kappa_2 ; J \tilde{m} m}  & $if$ \,\, \kappa_1=\kappa_2  \\
	- \mathcal{F}^{J_1, \tilde{m}_1, m_1, -\kappa_1}_{J_2, \tilde{m}_2, m_2, -\kappa_2 ; J \tilde{m} m}  & $if$ \,\, \kappa_1 = -\kappa_2 \, .
	\label{eq:property_F_exchanging_m_and_mtilde}
\end{cases}
\eeq
In fact, the relation \eqref{eq:app_relation_between_twoprescriptions_F} corresponds to the first case with $\kappa_1 = \kappa_2 = 1. $

We observe that under the exchange $m \leftrightarrow \tilde{m}, \kappa \rightarrow - \kappa ,$ the following identity holds:
\beq
\kappa \omega^{\psi}_J - (m- \tilde{m}) \rightarrow - \le \kappa \omega^{\psi}_J - (m- \tilde{m})  \ri \, .
\eeq
This expression appears at the denominator of a relevant interaction in the $\mathrm{PSU}(1,1|2)$ sector. We thus study in details the following expression when $\kappa_1 = \kappa_4$
\beq
\sum_{\kappa= \pm1} \frac{\mathcal{F}^{J_1,-m_1,-\tilde{m}_1,\kappa_1}_{J m \tilde{m} \kappa; J_2 m_2 \tilde{m}_2} \mathcal{F}^{J_4,-m_4,-\tilde{m}_4,\kappa_4}_{J m \tilde{m} \kappa; J_3 m_3 \tilde{m}_3}}{\kappa \omega^{\psi}_J - (m - \tilde{m})} =
	- \sum_{\kappa=\pm1} \frac{\mathcal{F}^{J_1,-\tilde{m}_1,-m_1,-\kappa_1}_{J \tilde{m} m \kappa; J_2 \tilde{m}_2 m_2} \mathcal{F}^{J_4,-\tilde{m}_4, -m_4,-\kappa_4}_{J \tilde{m} m \kappa; J_3 \tilde{m}_3 m_3}}{\kappa \omega^{\psi}_J - (m - \tilde{m})}  \, ,
\label{eq:app_products_F_su112_part1}
\eeq
and when $\kappa_1 = - \kappa_4$ 
\beq
\sum_{\kappa= \pm1} \frac{\mathcal{F}^{J_1,-m_1,-\tilde{m}_1,\kappa_1}_{J m \tilde{m} \kappa; J_2 m_2 \tilde{m}_2} \mathcal{F}^{J_4,-m_4,-\tilde{m}_4,\kappa_4}_{J m \tilde{m} \kappa; J_3 m_3 \tilde{m}_3}}{\kappa \omega^{\psi}_J - (m - \tilde{m})} =	\sum_{\kappa=\pm1} \frac{\mathcal{F}^{J_1,-\tilde{m}_1,-m_1,-\kappa_1}_{J \tilde{m} m \kappa; J_2 \tilde{m}_2 m_2} \mathcal{F}^{J_4,-\tilde{m}_4, -m_4,-\kappa_4}_{J \tilde{m} m \kappa; J_3 \tilde{m}_3 m_3}}{\kappa \omega^{\psi}_J - (m - \tilde{m})}  \, .
\label{eq:app_products_F_su112_part2}
\eeq
Summarizing, we found that there is only one independent assignment of momenta which is relevant to derive the interacting Hamiltonians of section \ref{sect-sphere_reduction}, identified by the short-hand notation \eqref{FGcoef_modes}.
We use the definition \eqref{eq:app_definitionClebsch_F} to find
\beq
\mathcal{F}^{\BCJ_1}_{\BCJ_2; J m \tilde{m}} = \mathcal{C}^{\CJ_1}_{\CJ_2; J m \tilde{m}} \, .
\label{eq:app_relation_CF_saturated_momenta}
\eeq
Surprisingly, when momenta are saturated in this way, we find an equivalence between the Clebsch-Gordan coefficients involving only scalar harmonics, and this one involving mixed products between scalar and spinorial harmonics.

\subsection*{Crossing relations at saturated angular momenta -- ${\cal C}$ and ${\cal G}$}

The Clebsch-Gordan coefficient $\mathcal{G}$ only appears in the computation of terms involving the fermionic current.
This restricts the assignments on momenta of interest to the two cases
\beq
\mathcal{G}^{J_1, J_1+\frac{1}{2},-J_1,\kappa_1=1}_{J_2,J_2+\frac{1}{2},-J_2,\kappa=1; Jm \tilde{m}\rho} \, , \qquad
\mathcal{G}^{J_1,-J_1, J_1+\frac{1}{2},\kappa_1=-1}_{J_2,-J_2,J_2+\frac{1}{2},\kappa=-1; Jm \tilde{m}\rho} \, .
\label{eq:app_two_assignments_momenta_ClebschG}
\eeq
The strategy is to consider the general definition \eqref{eq:app_definitionClebsch_G} and apply the symmetry properties of $\SU(2)$ Clebsch-Gordan and Wigner 9-j symbols to find a relation between the two possibilities.
Then, putting the assignments of momenta, we will relate the specific cases with $\rho= \pm 1$ to the scalar coefficient $\mathcal{C}.$

For convenience, we write here the two explicit expressions:
\beq
\begin{aligned}
\mathcal{G}^{J_1, J_1+\frac{1}{2},-J_1,\kappa_1=1}_{J_2,J_2+\frac{1}{2},-J_2,\kappa_2=1 ; JM \rho} & =  (-1)^{\frac{\rho}{2}} \sqrt{6(2J_2+1)(2J_2+2)(2J+1)(2J+3)} \\
& \times
 C^{J_1+\frac{1}{2}, J_1+\frac{1}{2}}_{J_2+\frac{1}{2}, J_1+\frac{1}{2} ; Q, J_1-J_2} 
 C^{J_1,-J_1}_{J_2,-J_2; \tilde{Q}, J_2-J_1}
\begin{Bmatrix}
J_1 + \frac{1}{2} & J_1 & \frac{1}{2} \\
J_2 + \frac{1}{2} & J_2 & \frac{1}{2} \\
Q & \tilde{Q} & 1 
\end{Bmatrix} \, ,
\end{aligned}
\label{eq:definition_G_appendix_1}
\eeq
and the other one
\beq
\begin{aligned}
\mathcal{G}^{J_1,-J_1, J_1+\frac{1}{2},\kappa_1=-1}_{J_2,-J_2,J_2+\frac{1}{2},\kappa=-1; Jm \tilde{m}\rho} & =  (-1)^{\frac{\rho}{2}} \sqrt{6(2J_2+1)(2J_2+2)(2J+1)(2J+3)} \\
& \times
 C^{J_1,-J_1}_{J_2,-J_2; Q, J_2-J_1}
 C^{J_1+\frac{1}{2}, J_1+\frac{1}{2}}_{J_2+\frac{1}{2}, J_1+\frac{1}{2} ; \tilde{Q}, J_1-J_2} 
\begin{Bmatrix}
J_1 & J_1 + \frac{1}{2} & \frac{1}{2} \\
J_2  & J_2+ \frac{1}{2} & \frac{1}{2} \\
Q & \tilde{Q} & 1 
\end{Bmatrix} \, ,
\end{aligned}
\label{eq:definition_G_appendix_2}
\eeq
where in both cases we used the labels \eqref{eq:app_labels_harmonics}.

Interestingly, the two expressions are almost the same: the prefactors coincide and the 9-j symbol satisfies the property
\beq
\begin{Bmatrix}
J_1 + \frac{1}{2} & J_1 & \frac{1}{2} \\
J_2 + \frac{1}{2} & J_2 & \frac{1}{2} \\
Q & \tilde{Q} & 1 
\end{Bmatrix}  =(-1)^{2(J+J_1+J_2)}
\begin{Bmatrix}
J_1 & J_1 + \frac{1}{2} & \frac{1}{2} \\
J_2  & J_2+ \frac{1}{2} & \frac{1}{2} \\
Q & \tilde{Q} & 1 
\end{Bmatrix} \, .
\eeq
Since for all the admitted choices of $\rho$ we have $J+J_1+J_2 \in \mathbb{Z},$ the prefactor is 1 and the two expressions coincide.
Thus the only difference between the two cases is due to the $\SU(2)$ Clebsch-Gordan coefficients, which however are simply exchanged if we also send $Q \leftrightarrow \tilde{Q}.$

Since the interacting Hamiltonian contains only these quantities with $\rho= \pm 1,$ we have that $Q, \tilde{Q}$ assume in the two cases the values $J+1, J.$
This implies the simple relation
\beq
\mathcal{G}^{J_1, J_1+\frac{1}{2},-J_1,\kappa_1=1}_{J_2,J_2+\frac{1}{2},-J_2,\kappa_2=1; Jm \tilde{m}\rho} =
- \mathcal{G}^{J_1,-J_1, J_1+\frac{1}{2},\kappa_1=-1}_{J_2,-J_2,J_2+\frac{1}{2},\kappa_2=-1; Jm \tilde{m} -\rho} \, .
\label{eq:relation_between_two_cases_G}
\eeq
The different sign arises due to the factor $(-1)^{\frac{\rho}{2}},$ which gives an opposite sign to the imaginary unit when considering $\rho= \pm 1.$

Then we can simply focus on one specific choice of the momenta and compute explicitly the coefficient $\mathcal{G},$ e.g. in the case in Eq.~\eqref{FGcoef_modes}.
We find 
\beq
 \mathcal{G}^{\BCJ_1}_{\BCJ_2; J m \tilde{m}, \rho=1} =
 i \sqrt{\frac{(J+\Delta J+1) (J-\Delta J+1)}{(J+1) (2 J+1)}}
   \mathcal{C}^{\CJ_1}_{\CJ_2; J m} \, ,
   \label{eq:app_relation_CGp_saturated_momenta}
\eeq
\beq
 \mathcal{G}^{\BCJ_1}_{\BCJ_2; J m \tilde{m}, \rho=-1} =
 -i \sqrt{\frac{(J+\Delta J+1) (J-\Delta J+1)}{(J+1) (2 J+3)}}
   \mathcal{C}^{\CJ_1}_{\CJ_2; J+1, m} \, .
   \label{eq:app_relation_CGm_saturated_momenta}
\eeq

\subsection*{Crossing relations at saturated angular momenta -- ${\cal F}$ and ${\cal G}$}

In this section we put together the identities between $\mathcal{F},\mathcal{G}$ with $\mathcal{C}$ in order to obtain a simplification for the terms involving fermions mediated by the non-dynamical gauge fields.
The interaction of interest is
\beq
\sum_{J_i,J,m,\tilde{m}} \le \frac{1}{8J(J+1)} \mathcal{F}^{\BCJ_1}_{\BCJ_2, J m \tilde{m}} \mathcal{F}^{\BCJ_4}_{\BCJ_3, J m \tilde{m}}
- \sum_{\rho = \pm 1} \frac{1}{2(\omega^2_{A,J}- (m-\tilde{m})^2)} \mathcal{G}^{\BCJ_1}_{\BCJ_2; Jm \tilde{m} \rho} \bar{\mathcal{G}}^{\BCJ_4}_{\BCJ_3; J m \tilde{m} \rho}   \ri \, .
\eeq
Using the relations \eqref{eq:app_relation_CF_saturated_momenta}, \eqref{eq:app_relation_CGp_saturated_momenta}, \eqref{eq:app_relation_CGm_saturated_momenta} and splitting the sum over $\rho= \pm 1$ we obtain
\beq
\begin{aligned}
& \sum_{J,m,\tilde{m}}  \le \frac{1}{8J(J+1)} \mathcal{C}^{\CJ_1}_{\CJ_2; J m \tilde{m}} \mathcal{C}^{\CJ_4}_{\CJ_3; J m \tilde{m}} -  \frac{1}{8(J+1)(2J+3)} \mathcal{C}^{\CJ_1}_{\CJ_2; J+1, m, \tilde{m}} \mathcal{C}^{\CJ_4}_{\CJ_3; J+1, m, \tilde{m}} \right. \\
& \left. 
-  \frac{1}{8(J+1)(2J+1)} \mathcal{C}^{\CJ_1}_{\CJ_2; J m \tilde{m}} \mathcal{C}^{\CJ_4}_{\CJ_3; J m \tilde{m}}  \ri \, .
\end{aligned}
\eeq
Now we define the convenient quantities
\bea
& \mathcal{P}^{\CJ_1;\CJ_4}_{\CJ_2; \CJ_3; J m \tilde{m}} = \frac{1}{8J(J+1)} \mathcal{C}^{\CJ_1}_{\CJ_2; J m \tilde{m}} \mathcal{C}^{\CJ_4}_{\CJ_3; J m \tilde{m}} \, ,  & \\
& \mathcal{Q}^{\CJ_1;\CJ_4}_{\CJ_2; \CJ_3; J m \tilde{m}, \rho=-1} = -  \frac{1}{8(J+1)(2J+3)} \mathcal{C}^{\CJ_1}_{\CJ_2; J+1, m, \tilde{m}} \mathcal{C}^{\CJ_4}_{\CJ_3; J+1, m, \tilde{m}} \, , & \\
& \mathcal{Q}^{\CJ_1;\CJ_4}_{\CJ_2; \CJ_3; J m \tilde{m}, \rho=1} =  -  \frac{1}{8(J+1)(2J+1)} \mathcal{C}^{\CJ_1}_{\CJ_2; J m \tilde{m}} \mathcal{C}^{\CJ_4}_{\CJ_3; J m \tilde{m}} \, . &
\eea
It can be shown that
\beq
\mathcal{P}^{\CJ_1;\CJ_4}_{\CJ_2; \CJ_3; J m \tilde{m}}  +
 \mathcal{Q}^{\CJ_1;\CJ_4}_{\CJ_2; \CJ_3; J-1, m, \tilde{m}, \rho=-1} +
  \mathcal{Q}^{\CJ_1;\CJ_4}_{\CJ_2; \CJ_3; J m \tilde{m}, \rho=1} = 0 \, .
  \label{eq:FGrelation_saturated_momenta}
\eeq
The sum (over an appropriate interval) of the previous quantities exactly gives the term mediated by the non-dynamical gauge field for the $\SU(1,1)$ fermionic sector, see Eq.~\eqref{eq:interacting_Hamiltonian_fermSU(1,1)}:
\beq
\sum_{J \geq J_{\rm min} (\rho)}  \le \mathcal{P}^{\CJ_1;\CJ_4}_{\CJ_2; \CJ_3; J, -\Delta J, \Delta J}  +
 \mathcal{Q}^{\CJ_1;\CJ_4}_{\CJ_2; \CJ_3; J, -\Delta J, \Delta J, \rho=-1} +
  \mathcal{Q}^{\CJ_1;\CJ_4}_{\CJ_2; \CJ_3; J, -\Delta J, \Delta J, \rho=1} \ri \, ,
\eeq
where we defined
\beq
\Delta J = J_1 - J_2 = J_4 - J_3 \, , \qquad
\Delta m = m_1 - m_2 = m_4 - m_3  \, .
\eeq
The lower extremum of summation plays a crucial role for the simplifications below. 
As in the previous cases considered in this Appendix, the assignments of momenta completely fix $\Delta m = - \Delta \tilde{m} = - \Delta J,$ which implies that we need to consider only one possibility for the endpoints of summation.

Indeed, all the sums start from the same value $J_{\rm min}= |\Delta J|,$ and the shift $J \rightarrow J-1$ in the term $\mathcal{Q}_{\rho=1}$ changes the lower endpoint of its summation to $J_{\rm min}= |\Delta J|-1.$
In this way, using Eq.~\eqref{eq:FGrelation_saturated_momenta}, we get a remarkable simplification which only leaves a non-vanishing term coming from the boundary of summation
\beq
\begin{aligned}
& \sum_{J \geq |\Delta J|}  \le \mathcal{P}^{\CJ_1;\CJ_4}_{\CJ_2; \CJ_3; J, -\Delta J, \Delta J}  +
 \mathcal{Q}^{\CJ_1;\CJ_4}_{\CJ_2; \CJ_3; J, -\Delta J, \Delta J, \rho=-1} +
  \mathcal{Q}^{\CJ_1;\CJ_4}_{\CJ_2; \CJ_3; J, -\Delta J, \Delta J, \rho=1} \ri = \\
  & = - \mathcal{Q}^{\CJ_1;\CJ_4}_{\CJ_2; \CJ_3; \Delta J-1, -\Delta J, \Delta J, \rho=-1} \, .
\end{aligned}
\eeq
In particular, we can explicitly evaluate this last term to obtain an expression in terms of the coefficient $\mathcal{C},$ {\sl i.e.} we obtain
\beq
- \mathcal{Q}^{\CJ_1;\CJ_4}_{\CJ_2; \CJ_3; \Delta J-1, -\Delta J, \Delta J, \rho=-1} =
 \frac{1}{8 |\Delta J| (2 |\Delta J|+1)} \mathcal{C}^{\CJ_1}_{\CJ_2; |\Delta J|, -\Delta J, \Delta J} \mathcal{C}^{\CJ_4}_{\CJ_3; |\Delta J|, -\Delta J, \Delta J} \, .
\label{eq:result_sum_J_FGterm}
\eeq
Due to Eq.~\eqref{eq:relation_between_two_cases_G}, it is clear that the same procedure can be applied to the case with a dynamical fermion having $\kappa=-1.$
The difference in such case is that the shift $J \rightarrow J-1$ and the surviving terms come from the Clebsch-Gordan coefficients with $\rho=1.$
The symmetry of the problem guarantees that the final result is the same, as it is explained for the sphere reduction in the $\PSU(1,1|2)$ sector.

\subsection*{Crossing relations at saturated angular momenta --  products of  ${\cal C},{\cal D},{\cal F}$ and ${\cal G}$}

In this subsection we show another remarkable simplification for the sum over $J$ of the mixed bosonic-fermionic term mediated by the non-dynamical gauge field.

We refer to Eq.~\eqref{eq:interaction_gauge_current_su111}, where however due to symmetry reasons it is sufficient to consider only half of the terms.
We thus define
\bea
&  \mathcal{S}^{\CJ_1,\CJ_4}_{\CJ_2,\CJ_3;JM} \equiv \frac{J_1+J_2+1}{4J(J+1)} \mathcal{C}^{\CJ_1}_{\CJ_2; JM} \mathcal{F}^{\BCJ_3}_{\BCJ_4; JM}  &   \\
& \mathcal{T}^{\CJ_1,\CJ_4}_{\CJ_2,\CJ_3;JM\rho} \equiv \frac{\sqrt{J_2(J_2+1)}}{\omega^2_{A,J}- (m - \tilde{m})^2} (\bar{\mathcal{D}}^{\CJ_1}_{\CJ_2; JM \rho} \mathcal{G}^{\BCJ_3}_{\BCJ_4, JM \rho}  + \mathcal{D}^{\CJ_1}_{\CJ_2, JM \rho} \bar{\mathcal{G}}^{\BCJ_3}_{\BCJ_4; JM \rho}  ) \, .&   
\eea
We can write these combinations only in terms of the Clebsch-Gordan coefficient $\mathcal{C}$ by means of the crossing relations proved in this Appendix.
The result is
\bea
& \mathcal{S}^{\CJ_1,\CJ_4}_{\CJ_2,\CJ_3;JM} = \frac{J_1+J_2+1}{4J(J+1)} \mathcal{C}^{\CJ_1}_{\CJ_2; JM} \mathcal{C}^{\CJ_3}_{\CJ_4; JM} \, , & \\
& \mathcal{T}^{\CJ_1,\CJ_4}_{\CJ_2,\CJ_3;JM\rho=-1} =  
- \frac{J+J_1+J_2+2}{4(J+1)(2J+3)}
\mathcal{C}^{\CJ_1}_{\CJ_2; J+1, M}  \mathcal{C}^{\CJ_3}_{\CJ_4; J+1, M}   \, ,  & \\
& \mathcal{T}^{\CJ_1,\CJ_4}_{\CJ_2,\CJ_3; JM\rho=1} = 
- \frac{J_1+J_2-J}{4(J+1)(2J+1)}
 \mathcal{C}^{\CJ_1}_{\CJ_2; J M}  \mathcal{C}^{\CJ_3}_{\CJ_4; J M}   \, .  &
\eea
As we learnt from previous example, the strategy is to send $J \rightarrow J-1$ in the term with $\rho=-1$ and observe that the following relation holds:
\beq
\mathcal{S}^{\CJ_1,\CJ_4}_{\CJ_2,\CJ_3;JM}  +
\mathcal{T}^{\CJ_1,\CJ_4}_{\CJ_2,\CJ_3;J-1, M\rho=-1} +
\mathcal{T}^{\CJ_1,\CJ_4}_{\CJ_2,\CJ_3;JM\rho=-1} = 0 \, .
\label{eq:app_crossing_relation_gauge_current_su111}
\eeq
In this way, we find that the sum over $J$ required in Eq.~\eqref{eq:interaction_gauge_current_su111} reduces to a boundary term.
Indeed, we obtain
\beq
\begin{aligned}
& \sum_{J \geq |\Delta J|}  \le \mathcal{S}^{\CJ_1;\CJ_4}_{J_2; J_3; J, -\Delta J, \Delta J}  +
 \mathcal{T}^{\CJ_1;\CJ_4}_{\CJ_2; \CJ_3; J, -\Delta J, \rho=-1} +
  \mathcal{T}^{\CJ_1;\CJ_4}_{\CJ_2; \CJ_3 m_3; J, -\Delta J, \rho=1} \ri = \\
  & = - \mathcal{T}^{\CJ_1;\CJ_4}_{\CJ_2; \CJ_3; \Delta J-1, -\Delta J, \rho=-1} =
\frac{J_1 + J_2 + \Delta J}{4|\Delta J| (2 |\Delta J|+1)}  
 \mathcal{C}^{\CJ_1}_{\CJ_2; |\Delta J|}  \mathcal{C}^{\CJ_4}_{\CJ_3;|\Delta J|}    \, .
\end{aligned}
\eeq
The same result also applies to the analog term involving fermions with $\kappa=-1$ relevant for the $\PSU(1,1|2)$ sector, 
as can be seen by applying \eqref{eq:app_relations_between_twoprescriptions_D}, \eqref{eq:app_relation_between_twoprescriptions_F} and \eqref{eq:relation_between_two_cases_G}.
The only difference for the fermions with opposite chirality is that we need instead to shift the terms with $\rho=1 ,$ but the procedre is formally the same, as well as the final result that we obtain.

\section{Algebra and oscillator representation}
\label{app-algebra_osc_repr}

The oscillator representation \cite{Beisert:2004ry} is a convenient way to represent the set of letters of $\mathcal{N}=4$ SYM and its superconformal algebra $u(2,2|4) .$ 
In this Appendix we review such a representation, following the conventions of \cite{Harmark:2007px}.

We consider two sets of bosonic oscillators $\mathbf{a}^{\alpha}, \mathbf{b}^{\dot{\alpha}}$ with four dimensional spinorial indices $\alpha, \dot{\alpha} \in \lbrace 1,2 \rbrace$ and one fermionic oscillator $\mathbf{c}^a$ with $a \in \lbrace 1,2,3,4 \rbrace$ satisfying the canonical commutation relations
\beq
\left[\mathbf{a}^{\alpha} , \mathbf{a}^{\dagger}_{\beta}  \right] = \delta^{\alpha}_{\beta} \, , \quad
\left[ \mbf{b}^{\dot{\alpha}} , \mbf{b}^{\dagger}_{\dot{\beta}} \right] = \delta^{\dot{\alpha}}_{\dot{\beta}} \, , \quad
\lbrace \mbf{c}^a , \mbf{c}_b^{\dagger} \rbrace = \delta^a_b \, .
\eeq
We conveniently introduce the following notation to denote the number operators:
\beq
a^{\alpha} \equiv \mbf{a}^{\dagger}_{\alpha} \mbf{a}^{\alpha} \, , \quad
b^{\dot{\alpha}} \equiv \mbf{b}^{\dagger}_{\dot{\alpha}} \mbf{b}^{\dot{\alpha}} \, , \quad
c^a \equiv \mbf{c}^{\dagger}_a \mbf{c}^a \, ,
\eeq
with no sum over the indices.

These oscillators can be combined in order to define the generators of the algebra and the physical states.
We have the 6 generators of the $so(4)$ subalgebra
\beq
\mbf{L}^{\alpha}_{\,\, \beta} = \mbf{a}^{\dagger}_{\beta} \mbf{a}^{\alpha} - \frac{a^1+a^2}{2} \delta^{\alpha}_{\beta} \, , \qquad
\dot{\mbf{L}}^{\dot{\alpha}}_{\,\, \dot{\beta}} = \mbf{b}^{\dagger}_{\dot{\beta}} \mbf{b}^{\dot{\alpha}} - \frac{b^1+b^2}{2} \delta^{\dot{\alpha}}_{\dot{\beta}} \, ,
\eeq
and 15 generators for the $su(4)$ subalgebra
\beq
\mbf{R}^a_{\,\, b} = \mbf{c}^{\dagger}_b \mbf{c}^a - \frac{1}{4} \delta^a_b \sum_{d=1}^4 c^d \, .
\eeq
For the purposes of this work, we need to take BPS bounds given by combinations of charges in the Cartan subalgebra of $u(2,2|4).$ 
Among the previous set, they are given by the rotation ones
\beq
S_1 = \frac{1}{2} \le a^1 - a^2 +  b^1 - b^2 \ri \, , \qquad
S_2 = \frac{1}{2} \le -a^1 + a^2 + b^1 - b^2 \ri \, ,
\eeq
and of the $su(4)$ Cartan charges\footnote{In this Appendix, we call $J_i$ the $su(4)$ Cartan generators instead of the notation $Q_i$ used in the main text to avoid confusion with the supercharges.}
\beq
J_1 = \frac{1}{2} \le -c^1 - c^2 + c^3 + c^4 \ri \, ,  \quad
 J_2 = \frac{1}{2} \le -c^1 +c^2 -c^3 + c^4 \ri \, , \quad
J_3 = \frac{1}{2} \le c^1 -c^2 -c^3 +c^4 \ri \, .
\eeq
In addition, the $u(2,2|4)$ algebra contains three $u(1)$ charges: the bare dilatation operator $D_0,$ the central charge $C$ and the hypercharge $B$, given by
\beq
D_0 = 1 + \frac{1}{2} \le a^1 + a^2 + b^1 + b^2 \ri \, ,
\eeq
\beq
C= 1 - \frac{1}{2} \le -a^1-a^2+b^1+b^2-c^1-c^2-c^3-c^4 \ri \, ,
\eeq
\beq
B= \frac{1}{2} (a^1 + a^2 - b^1 - b^2) \, .
\eeq
All the letters of $\mathcal{N}=4$ SYM  satisfy $C=0,$ then they correspond to a representation of $psu(2,2|4).$

The complete algebra also contains the generators for translations and boosts
\beq
\mbf{P}_{\alpha \dot{\beta}} = \mbf{a}^{\dagger}_{\alpha} \mbf{b}^{\dagger}_{\dot{\beta}} \, , \qquad
\mbf{K}^{\alpha \dot{\beta}} = \mbf{a}^{\alpha} \mbf{b}^{\dot{\beta}} \, ,
\eeq
and the fermionic generators for supersymmetry plus the superconformal partners:
\bea
& \mbf{Q}^a_{\,\, \alpha} = \mbf{a}^{\dagger}_{\alpha} \mbf{c}^{a} \, , \qquad
\dot{\mbf{Q}}_{\dot{\alpha} a} = \mbf{b}^{\dagger}_{\dot{\alpha}} \mbf{c}^{\dagger}_a \, ,  & \\
& \mbf{S}^{\alpha}_{\,\,a} = \mbf{c}^{\dagger}_a \mbf{a}^{\alpha} \, , \qquad
\dot{\mbf{S}}^{\dot{\alpha} a} = \mbf{b}^{\dot{\alpha}} \mbf{c}^a \, . &
\eea
Among the entire set of generators, an important role is played by the $su(1,1)$ subalgebra spanned by
\begin{equation}
\label{oscrepSU11}
L_0 = \frac{1}{2} (1 + a^1 + b^1 ) \spa L_+ = a_1^\dagger b_1^\dagger \spa L_- = a_1 b_1 \, ,
\end{equation}
which is common to all the near-BPS limits considered in Section \ref{sect-sphere_reduction}.
They satisfy the commutation relations
\begin{equation}
[L_0 ,L_\pm ]= \pm L_\pm \spa [L_-,L_+]=2L_0
\end{equation}
The letters of $\mathcal{N}=4$ SYM are composed by the bosonic and fermionic fields listed in Table \ref{tab:su4_weights_scalars_Troels_paper}, \ref{tab:su4_weights_fermions1_Troels_paper} and \ref{tab:su4_weights_fermions2_Troels_paper}, plus the gauge fields strengths and the covariant derivatives.
Schematically, they are given by 
\bea
& \Phi : (\mbf{c}^{\dagger})^2 |0 \rangle \, , \qquad
\chi : \mbf{a}^{\dagger} \mbf{c}^{\dagger} | 0 \rangle \, , \qquad
\bar{\chi} : \mbf{b}^{\dagger} (\mbf{c}^{\dagger})^3 | 0 \rangle \, , & \\
& F : (\mbf{a}^{\dagger})^2 | 0 \rangle \, , \qquad
\bar{F} : (\mbf{b}^{\dagger})^2 (\mbf{c}^{\dagger})^2 |0 \rangle \, , \qquad
d : \mbf{a}^{\dagger} \mbf{b}^{\dagger} |0 \rangle \, . &
\eea
Normalization factors are omitted, while the precise labelling of the indices depends from the specific letter; this can be easily found by considering the Cartan generators and the fields surviving the various near-BPS limits.
In this Appendix we focus on the main cases considered in this paper: the $su{(1,1|1)}$ and $psu{(1,1|2)}$ algebras.

\subsection*{$\mbf{su(1,1|1)}$ algebra}

The BPS limit in the $\SU(1,1|1)$ sector reads
\beq
D_0 - \le S_1 + J_1 + \frac{1}{2} J_2 + \frac{1}{2} J_3 \ri =  0 \, ,
\eeq
which implies
\beq
a^2 = b^2 = 0 \spa c^1 = c^2 = 0 \spa c^4 = 1 \, .
\label{eq:conditions_number_operators_su111}
\eeq
Moreover, the vanishing of the central charge gives the additional condition
\beq
c^3 = 1-a^1 +b^1 \, .
\eeq
The letters in this sector are
\begin{equation}
\label{oscrepstates}
| d_1^n Z \rangle = \frac{1}{n!} (a_1^\dagger b_1^\dagger)^n c_3^\dagger c_4^\dagger |0\rangle\spa
| d_1^n \chi_1 \rangle = \frac{1}{\sqrt{n! (n+1)!}} (a_1^\dagger b_1^\dagger)^n a_1^\dagger c_4^\dagger |0\rangle\spa
\end{equation}
where the factors are chosen to achieve unity normalization:
\beq
\langle d_1^m Z | d_1^n Z \rangle = \delta_{mn} \, , \qquad
\langle d_1^m \chi_1 | d_1^n \chi_1 \rangle = \delta_{mn} \, .
\eeq
The generators of the algebra are the following: there are four bosonic generators, generating the algebra $su(1,1)\times u(1),$ given by the set \eqref{oscrepSU11} plus the additional $u(1)$ generator
\begin{equation}
R = \frac{1}{2} c^3 \, .
\end{equation}
Then we have four fermionic generators, that we collect using the notation
\begin{equation}
Q = a_1 c_3^\dagger \spa Q^\dagger = a_1^\dagger c_3 \spa S = b_1 c_3 \spa S^\dagger = b_1^\dagger c_3^\dagger \, .
\end{equation}
The generators of this sector satisfy the following commutation relations:
\bea
\label{QQdagger}
& \{ Q,Q^\dagger \}  = L_0 + R  \, , \quad
 \{ S,S^\dagger\} =  L_0 - R \, , \quad
 \{ S^\dagger , Q^\dagger \} = L_+  \, , \quad
 \{ S , Q \} = L_- \, , & \\
& [L_0,Q]= -\frac{1}{2} Q \, , \quad
 [L_0,Q^\dagger]=\frac{1}{2} Q^\dagger \, , \quad
 [L_0,S]=-\frac{1}{2} S \, , \quad
  [L_0,S^\dagger]=\frac{1}{2} S^\dagger \, , & \\
 & [Q,L_+]=S^\dagger \, , \quad
 [Q,L_-]=0 \, , \quad
[Q^\dagger,L_+]=0 \, , \quad
 [Q^\dagger,L_-]=-S \, , & \\
& [S,L_+]=Q^\dagger \, , \quad
 [S,L_-] = 0 \, , \quad
[S^\dagger , L_+]=0 \, , \quad
 [S^\dagger,L_-]=-Q \, .
\eea
A typical feature of supersymmetric-invariant theories is that the anticommutator of the supercharges closes on the free Hamiltonian.
Looking at Eq.~\eqref{QQdagger}, this points towards the identification
\beq
H_0 = L_0 + R = S_1 + J_1 \, .
\eeq
On the other hand, we remark in Section \ref{sect-sphere_reduction} that a more natural choice for all the near-BPS limits is to take $L_0$ to be the free part of the Hamiltonian, see e.g. Eq.~\eqref{eq:final_ham_su111}.
Indeed, following the near-BPS limit \eqref{H0_su111}, the free Hamiltonian of the system would naturally be
\beq
S_1 + J_1 + \frac{1}{2} J_2 + \frac{1}{2} J_3  = L_0 + \frac{1}{2} \, ,
\eeq
and then the choice of take instead $L_0$ to represent the free part simply corresponds to a convenent mass shift.

In order to follow this interpretation, we introduce a linear combination of the original supercharges which closes on $L_0$ instead of the combination $S_1 + J_1.$
We define
\beq
\mathcal{Q} = \frac{1}{\sqrt{2}} (Q+S) \, , \qquad
\mathcal{Q}^{\dagger} = \frac{1}{\sqrt{2}} (Q^{\dagger} + S^{\dagger}) \, .
\eeq
Since 
\beq
\lbrace Q, S^{\dagger} \rbrace = 0 \, , \qquad
\lbrace Q^{\dagger} , S \rbrace = 0 \, , 
\eeq
we obtain
\beq
\lbrace \mathcal{Q} , \mathcal{Q}^{\dagger} \rbrace = L_0 = \le  S_1 + J_1 + \frac{1}{2} J_2 + \frac{1}{2} J_3 \ri  - \frac{1}{2} \, .
\eeq
This representation of the supercharges is used in Section \ref{sect-sphere_reduction} and to build the superfield formulation in Section \ref{sect-momentum_space_superfields}.

\subsection*{$\mathbf{psu(1,1|2)}$ algebra}

The BPS limit in this sector reads
\beq
D_0 - \le S_1 + J_1 + J_2 \ri = 0 \, , 
\eeq
which implies
\beq
a^2 = b^2 = 0 \, , \qquad
c^1 = 0 \, , \qquad
c^4 = 1 \, .
\eeq
The vanishing of the central charge in the representation gives
\beq
c^2 + c^3 = 1 -a^1 + b^1 \, .
\eeq
The set of letters of the sector is\footnote{Notice that we are choosing conventions such that the fermion field $\psi^2_n$ creates the state  $ - | \bar{\chi}_7 \rangle .$}
\bea
& | d_1^n Z \rangle = \frac{1}{n!} (a_1^\dagger b_1^\dagger)^n c_3^\dagger c_4^\dagger |0\rangle\spa
| d_1^n \chi_1 \rangle = \frac{1}{\sqrt{n! (n+1)!}} (a_1^\dagger b_1^\dagger)^n a_1^\dagger c_4^\dagger |0\rangle\spa & \\
& | d_1^n X \rangle = \frac{1}{n!} (a_1^\dagger b_1^\dagger)^n c_2^\dagger c_4^\dagger |0\rangle\spa
| d_1^n \bar{\chi}_7 \rangle = \frac{1}{\sqrt{n! (n+1)!}} (a_1^\dagger b_1^\dagger)^n b_1^\dagger c_2^{\dagger} c_3^{\dagger} c_4^\dagger |0\rangle \spa &
\eea
where the prefactors ensure a unit normaliztion
\bea
& \langle d_1^m Z | d_1^n Z \rangle = \delta_{mn} \, , \qquad
\langle d_1^m \chi_1 | d_1^n \chi_1 \rangle = \delta_{mn} \, , & \\
& \langle d_1^m X | d_1^n X \rangle = \delta_{mn} \, , \qquad
\langle d_1^m \bar{\chi}_7 | d_1^n \bar{\chi}_7 \rangle = \delta_{mn} \, . &
\eea
The bosonic generators of the $su(1,1)$ subalgebra are the same as in the $su(1,1|1)$ sector. 
The R-symmetry generators form now a $su(2)$ subalgebra, given by
\beq
\mbf{R}^2_{\,\,\, 3} = \mbf{c}^{\dagger}_3 \mbf{c}^2  \, , \quad
\mbf{R}^3_{\,\,\, 2} = \mbf{c}^{\dagger}_2 \mbf{c}^3  \, , \quad
R = \frac{1}{2} (c^3 - c^2) \, .
\eeq
The fermionic generators can be collected in the convenient basis
\bea
& Q = a_1 c^{\dagger}_3 \spa Q^{\dagger} = a_1^{\dagger} c_3 \spa
S = b_1 c_3 \spa S^{\dagger} = b_1^{\dagger} c_3^{\dagger} \spa & \\
& \tilde{Q} = a_1 c^{\dagger}_2 \spa \tilde{Q}^{\dagger} = a_1^{\dagger} c_2 \spa
\tilde{S} = b_1 c_2 \spa \tilde{S}^{\dagger} = b_1^{\dagger} c_2^{\dagger} \, . &
\eea
They satisfy the following commutation relations:
\bea
& \{ Q,Q^\dagger \} = L_0 + R  \spa 
\{ S,S^\dagger\} = L_0 - R \spa & \\
&  \{ \tilde{Q},\tilde{Q}^\dagger \} = L_0 - R  \spa 
\{ \tilde{S},\tilde{S}^\dagger\} = L_0 + R \spa & \\
& \{ Q, \tilde{Q}^{\dagger} \} = \mbf{R}^2_{\,\,\, 3} \spa
\{ S, \tilde{S}^{\dagger} \} = - \mbf{R}^3_{\,\,\, 2}  \spa
\{ \tilde{Q}, Q^{\dagger} \} = \mbf{R}^3_{\,\,\, 2}  \spa
 \{ \tilde{S} , S^{\dagger} \} = - \mbf{R}^2_{\,\,\, 3}  \, . &
\eea
Similarly to the $su(1,1|1)$ sector, we would like to identify the free part of the Hamiltonian to be $L_0,$ and define a linear combination of the supercharges such that they close on this generator.
This is also motivated by the fact that $L_0$ differs by the combinations of Cartan charges defining the $\mathrm{P}\SU(1,1|2)$ by a constant:
\beq
S_1 + J_1 + J_2 = L_0 + \frac{1}{2} \, .
\eeq
We then define
\bea
& \mathcal{Q}_1 = \frac{1}{\sqrt{2}} \le Q + S \ri \, , \qquad
\mathcal{Q}_2 = \frac{1}{\sqrt{2}} \le \tilde{Q} + \tilde{S} \ri \, ,  & \\
&  \mathcal{Q}^{\dagger}_1 = \frac{1}{\sqrt{2}} \le Q^{\dagger} + S^{\dagger} \ri \, , \qquad
\mathcal{Q}^{\dagger}_2 = \frac{1}{\sqrt{2}} \le \tilde{Q}^{\dagger} + \tilde{S}^{\dagger} \ri \, ,   &
\eea
which satisfy 
\bea
& \lbrace \mathcal{Q}_1 , \mathcal{Q}_1^{\dagger} \rbrace = \lbrace \mathcal{Q}_2 , \mathcal{Q}_2^{\dagger} \rbrace  = L_0 \spa & \\
&  \lbrace \mathcal{Q}_1 , \mathcal{Q}_2^{\dagger} \rbrace = \frac{1}{2} \le  \mathbf{R}^2_{\,\, 3} - \mathbf{R}^3_{\,\, 2} \ri \, , & \\
& \lbrace \mathcal{Q}_1 , \mathcal{Q}_2 \rbrace = \{ \mathcal{Q}_1^{\dagger} , \mathcal{Q}_2^{\dagger} \}  = 0 \, . &
\eea
They correspond to the supercharges defined in \eqref{eq:CQdef2}.

\addcontentsline{toc}{section}{References}

\bibliography{newbib}
\bibliographystyle{newutphys}

\end{document}